\documentclass[%
 aip,
 amsmath,amssymb,
 reprint,%
]{revtex4-1}

\usepackage{graphicx}
\usepackage{dcolumn}
\usepackage{bm}

\usepackage[utf8]{inputenc}
\usepackage[T1]{fontenc}
\usepackage{mathptmx}
\usepackage{etoolbox}

\makeatletter
\def\@email#1#2{%
 \endgroup
 \patchcmd{\titleblock@produce}
{\frontmatter@RRAPformat}
{\frontmatter@RRAPformat{\produce@RRAP{*#1\href{mailto:#2}{#2}}}\frontmatter@RRAPformat}
{}{}
}%
\makeatother
\begin{document}

\preprint{AIP/123-QED}

 \title{Airborne Ultrasound Focusing Aperture with Binary Amplitude Mask Over Planar Ultrasound Emissions} {Airborne Ultrasound Focusing with Amplitude Mask}

 \author{Masatake Kitano}
 \author{Keisuke Hasegawa}
\email{keisuke\_hasegawa@ipc.i.u-tokyo.ac.jp}
\thanks{Keisuke Hasegawa is the author to whom correspondence should be addressed.}%
 \affiliation{The Department of Mathematical Engineering and Information Physics, Faculty of Engineering, the University of Tokyo, Tokyo, 113-8656, Japan.}






\date{\today} 

\begin{abstract}

 Phased arrays of airborne ultrasound transducers are widely utilized as a key technology to achieve mid-air convergence of intense ultrasound, which is applied to a variety of systems, such as contactless tactile presentation, acoustic-levitation and its application, mid-air-flow acceleration, etc. 
However, it requires considerably precise phase control with temporally severe synchronization between elements, which leads to difficulty in scaling up the entire system beyond the tabletop size as most of the current application systems. 
Here, we propose a much simpler and easier scaling-up method of airborne ultrasound convergence, where a binary amplitude mask that serves as a Fresnel Zone Plate (FZP) is placed on the planar in-phase ultrasound sources. 

We experimentally demonstrate that the FZP-based ultrasound focusing achieved a spatial resolution that is comparable to conventional methods, based on the use of phase-controlled transducers.
The ultrasound foci created using FZPs are sufficiently intense for most application scenarios that are currently in practical use. 
We also determine favorable side effects of our method suppressing grating lobes, which is inevitable with the conventional phase-controlling method. 

The FZPs and planar ultrasound sources are both readily implemented with inexpensive ingredients and components.
The result of our study contributes to upsizing dimensions in which a mid-air convergent ultrasound field is successfully generated.
Accordingly, unprecedented application scenarios that target the entire room as the workspace will be possible.
\end{abstract}


\maketitle
\newpage
\section{Introduction}
\subsection{Prevalent use of phase-controlled airborne ultrasound transducer arrays in nonlinear mid-air ultrasound applications}
The nonlinear acoustic effect of strong mid-air ultrasound has been known \cite{Strutt1884,King1934,Westervelt1951,Westervelt1953} for more than a century. However, its practical applications had mostly been limited within underwater cases. Recently, however, the situation has been altered by the advent of wave emission control devices employed for generating spatially localized intense ultrasound fields in the air, which can be electronically steered. Such strong and localized airborne ultrasonic power fields enable the generation of nonlinear acoustic phenomena at desired locations in space. This has led to the development of various applications of mid-air convergent ultrasound in several fields over the past decade. Examples of such real-world applications include mid-air ultrasound tactile presentation \cite{Hoshi2010, Takahashi2020,Frier2018,Hajas2020}, acoustic levitation systems\cite{Morales2019, Marzo2017, Inoue2019}, mid-air three-dimensional displays \cite{Hirayama2019, Hirayama2022}, utilization of mid-air ultrasound-driven acoustic flows \cite{Hasegawa2017,Hasegawa2018,Hasegawa2019}, etc. To date, new applications have been continuously devised by many researchers.

As aforementioned, most of these applications rely on the technique of creating ultrasound fields with controlled spatial distribution, which is based on the principle of ultrasound source emission that is spatially controlled in a greater emission area than the wavelength. Currently, the most prevalently employed method is the ultrasonic phased array technique, where the coherent ultrasound emission plane is constituted by a large number of ultrasonic transducers, with their individual emission amplitudes and phase delays electronically controlled. By appropriately controlling the driving phases of the transducers, a wide variety of spatial distributions of ultrasound fields are created. The fabrication of the first airborne ultrasound phased array device \cite{Hoshi2010} triggered widespread research on its applications. Development of the airborne ultrasound phased arrays have been continued to date by several research groups and most of the aforementioned application scenarios are based on this technique.

\subsection{Difficulty in upsizing current phased-array-based ultrasound manipulation scenarios}
However, the phased array technique suffers from difficulties in technical implementation \cite{Camps2020}.
Particularly, the need for synchronization and minute phase control of all individual transducers requiring a $\mu$s precision, prevents the workspace of mid-air ultrasound systems from being upscaled.
Therefore, most of the current mid-air ultrasound applications are limited in tabletop systems, in terms of their spatial tract.

As a potential solution that could address this challenge, there is a different approach to manipulate airborne ultrasound.
It is the development of wave manipulating elements that convert the incident ultrasound from a single fixed ultrasound source on its surface into desired spatial distribution of the ultrasound field.
They are passive wave-manipulation elements and upsizing, and such devices are much easier and less expensive than upsizing the phased-array system for most cases.
Among them, many phase controlling element arrays are studied, which serves as a phased array with fixed spatial phase delay profiles in conjunction with wave sources.
One of the most intuitive ones are phase delay elements directly placed above the plane wave source to convert the incident waves inside them into the desired field. They operate in the water \cite{Melde2016}, in vivo \cite{Gambin2020}, and in the air \cite{Memoli2017, Zhao2018}. There are also a number of methods that handle incident waves at a distance from the wave source that comes into the elements \cite{Morelon2014, Li2014}. Several reflective elements that control the phase of incident wave have also been proposed both in water and air \cite{Camps2020, Poly2020, Sallam2021} as well. Such devices do not require electronical synchronizations among individuals once fabricated. 
At the same time, most of these elements do not allow their phase distribution to be changed once they are constructed, apart from some that can be adjusted manually \cite{Memoli2017}. Other examples of related technologies for wave control are the use and fabrication of acoustic metamaterials \cite{zhao2020, Brunet2015, Krodel2016}.

There is another approach for fabricating wave manipulation devices, which does not rely on phase control of incident waves. Instead, such devices control only the amplitude distribution of the wave emissions. There is no need for the phase control of the vibrating surface, which significantly simplifies the system implementation; in addition, the required precision in element fabrication is substantially less than that of phase-controlling alternatives. In addition, it is advantageous in terms of the simplicity of the fabrication in that amplitude-control-based devices do not require acoustic impedance matching to efficiently transmit incident waves, whereas this is indispensable for phase-controlling devices. The main drawback of this approach is that unlike the phase-controlling devices, a portion of radiated wave is weakened by amplitude control. This results in the requirement of a larger ultrasound-emitting aperture in applications that require strong ultrasonic output when compared with the phase-control scheme.
However, fabrication simpleness of such amplitude-controlling devices enables them to be made larger with less cost and effort than that required with upsizing the phase-controlling devices.
There are several examples of acoustic convergence in air and water achieved by concentric amplitude lenses \cite{Schindel1997, Shen2019, Welter2011, Brown2016, Korozlu2018, Tarrazo-Serrano2019} and concentric transducer arrays \cite{Hur2022}. With proper designing, the spatial resolution of the generated sound fields by the amplitude-controlled emissions are not particularly degraded, compared with the phase-controlled cases.

\begin{figure*}[t]
\includegraphics[width=6in]{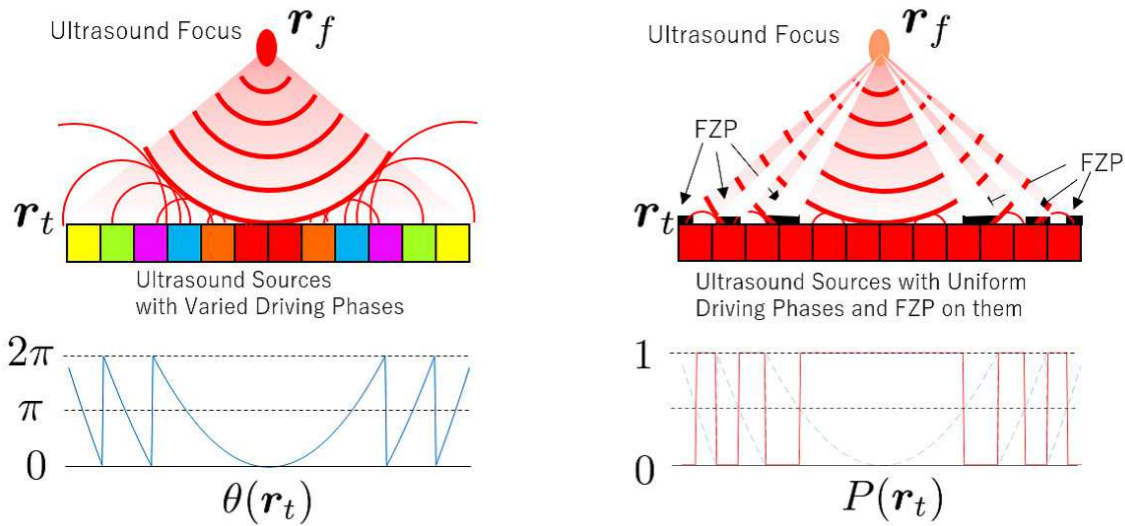}
\caption{\label{fig:calc}Schematic illustration of how the ultrasonic focus is formed by phased array transducers (Left figure) and in-phase planar wave sources under an FZP (right figure).}
\end{figure*}
\subsection{Proposed technique: Large-aperture amplitude mask on planar ultrasonic source for ultrasound manipulation}
The acoustic convergence using concentric amplitude emissions described above is a technique called Fresnel Zone Plate (FZP). Converged sound field is realized by placing the FZP at a certain distance from, or directly on radiating sound wave sources, to block off the ultrasound emissions that are not “in phase” at the desired region in the workspace in the air. Several airborne ultrasonic applications including potential ones are implemented using 40 kHz ultrasonic transducers \cite{Hirayama2022, Allevato2022, Tseng2017, Marzo2017-2, Inoue2019, Frier2018, Hajas2020, Morales2019}. 
As pointed out by a preceding study \cite{Rutsch2015}, one of the reason for many related works choosing 40~kHz is its low attenuation during propagation owing to relatively long wavelength as ultrasound.
In addition, good availability, inexpensiveness, high emission efficiency and requirement of relatively coarse temporal timing control between the transducers would be another benefits of utilizing 40~kHz ultrasound transducers.

However, to the best of the authors’ knowledge, there have been no examples of the use of large-aperture FZPs for convergence of 40 kHz large-amplitude airborne ultrasonic waves in mid-air ultrasound applications.
FZPs can be easily scaled up in size, and the realization of a large-area ultrasonic radiation surface using FZP is expected to significantly expand the range of applications in mid-air ultrasound research, because of the ability to create well-concentrated ultrasound foci at a great distance from the aperture.
For example, a mid-air tactile display will be realized that can present tactile stimuli all over the body of users at several locations in the room, whereas the current system can only stimulate a part of the user’s limb situated in front of the fixed small-apertured phased arrays. It is also expected that a wide range of aerial object manipulation using the entire room as a workspace will be achieved. 

As a fundamental technology to turn these potential applications into reality, we propose a method for realizing a convergent sound field as an alternative to the phased array, by placing a thin, large-area FZP binary amplitude mask on a planar ultrasonic source. More specifically, we demonstrate the generation of an ultrasonic focus with this setup, which is often utilized in various applications including mid-air tactile presentation. The proposed FZP amplitude mask can be made from any material that has acoustic impedance sufficiently different from that of the air and blocks off emissions from the plane wave source. In this study, we utilize an acrylic plate cut with a laser cutting machine to fabricate the FZP and demonstrate that a focus can be generated with it. In the proposed sound field control method, a machining accuracy of millimeters is sufficient for 40 kHz ultrasonic waves (8.5 mm in wavelength in the air), which are commonly utilized in airborne ultrasonic research, and any of such complicated machining processes as required in fabricating metamaterials are unnecessary.
The proposed technique does not have a real-time focus shifting function like the phased array. Nevertheless, the fabricated FZP mask can be larger than the area of the ultrasound radiation surface, and the focus can be shifted by translating it over the fixed radiation surface. Although not as easy as a phased array whose transducers' phases can be electronically controlled, this focus shifting strategy can be achieved with appropriate actuators.

In this study, it is assumed that a large number of ultrasonic transducers forming a large emitting aperture is utilized as the plane wave radiation surface. A reason for this assumption is the difficulty in fabricating a monolithic plane-wave radiation surface that utilizes a single flat plate to perform exclusive normal mode vibration at ultrasonic frequency. It is much easier to construct a planar radiation surface using a large number of separate ultrasonic transducers driving in phase instead. 
Although the fundamental physical principle of FZP-based ultrasound field control is not affected by the source frequency, we focused on the utilization of 40~kHz ultrasound transducer array in combination with FZP amplitude masks.
This is because it is currently the most reasonable solution for construction of large ultrasound emitting aperture thanks to their availability and fabrication readiness and the prevalent use of 40~kHz midair ultrasound in many current practical applications.

In this study, phased arrays that have already been developed were used as a plane wave source in the experiment by driving all their transducers with no phase differences among individuals. For actual applications, we envision the use of transducer arrays in which all elements are driven by a common driving signal. This strategy does not require synchronization control of each element, unlike the case with phased arrays, and thus can be easily applied to large scale application systems.
 
There is another finding in this paper that is concerned with the grating lobe issue, which is the strong and localized radiation of ultrasonic energy in an unintended direction, caused by phased arrays whose element spacing on their radiation plane is wider than half of the wavelength. In contrast, the spatial resolution of the amplitude mask fabricated in this study is finer than half of the wavelength; therefore, it is experimentally demonstrated that the afore-mentioned grating lobes do not occur when the in-phase driven transducer array is covered with the FZP mask. This feature in this study has great practical significance, in that it suppresses people’s unintentional exposure to strong ultrasound in several application scenarios. 

\section{Physical Principles}
\subsection{Ultrasound focusing by phased array systems}
Prior to the description of the formation principle of ultrasound foci by FZPs, we start with brief introduction of focal formation by phased arrays: it has much in common with our method, and thus would bring better comprehension of this research. 
Figure \ref{fig:calc} illustrates how the two strategies, the phase-controlled-transducer-based and FZP-based methods, form an ultrasound focus.

 As aforementioned, an airborne ultrasonic phased array has a radiation surface with many transducers, and the output signal of each transducer can be individually controlled. With a phased array, focus formation is achieved by electronically controlling the phase delay of each transducer so that the sound waves from all transducers are in-phase and yield strong acoustic energy spot at a desired point (the focus).
 Let $\bm{r}_t$ be the position of an element in the array, $\bm{r}_f$ be the focal point, and $k$ be the ultrasonic wave number.
 Then, the driving phase $\theta(\bm{r}_t)$ of the ultrasonic wave emitted by the element at $\bm{r}_t$ should be set to compensate for the phase delay owing to the distance between the element and focal point. Therefore, the driving phase $\theta(\bm{r}_t)$ is expressed as
 \begin{equation}
 \theta(\bm{r}_t) = k||\bm{r}_t - \bm{r}_f|| + \alpha,
 \end{equation}
where $\alpha$ is an arbitrary constant expressing the phase indefiniteness and $||\cdot||$ denotes the Euclidean norm of a vector $\cdot$.

\subsection{Principles and designing procedures of amplitude FZP for ultrasound focusing}
 The FZP converges ultrasound power around a desired position by blocking off a part of the wave emission from an ultrasonic source. Let a plane wave source be constructed by a set of in-phase driven ultrasonic elements and $\bm{r}_t$ be an element position. Then at the focal position, the observed phase delay of the sound wave emitted from each element varies with the distance between the element and focus, as in the case with a phased array. Here, we consider driving the wave source with the rule that only those elements that are in phase or nearly in-phase at the focal point are activated, whereas the rest are deactivated. It is expected that nearly-in-phase addition of sound waves will be realized at the focus. 
\begin{figure*}[t]
\includegraphics[width=2in]{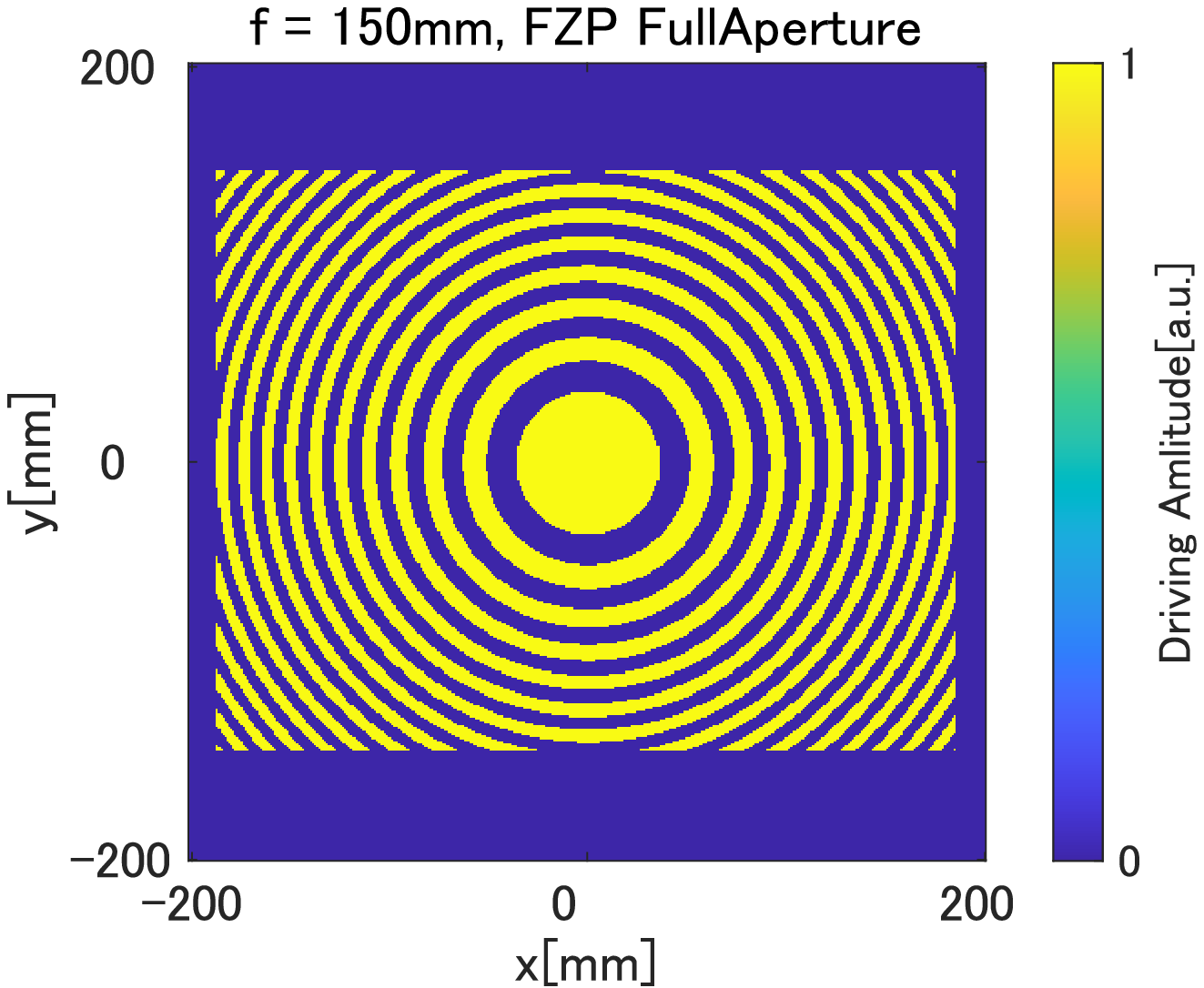}
\includegraphics[width=2in]{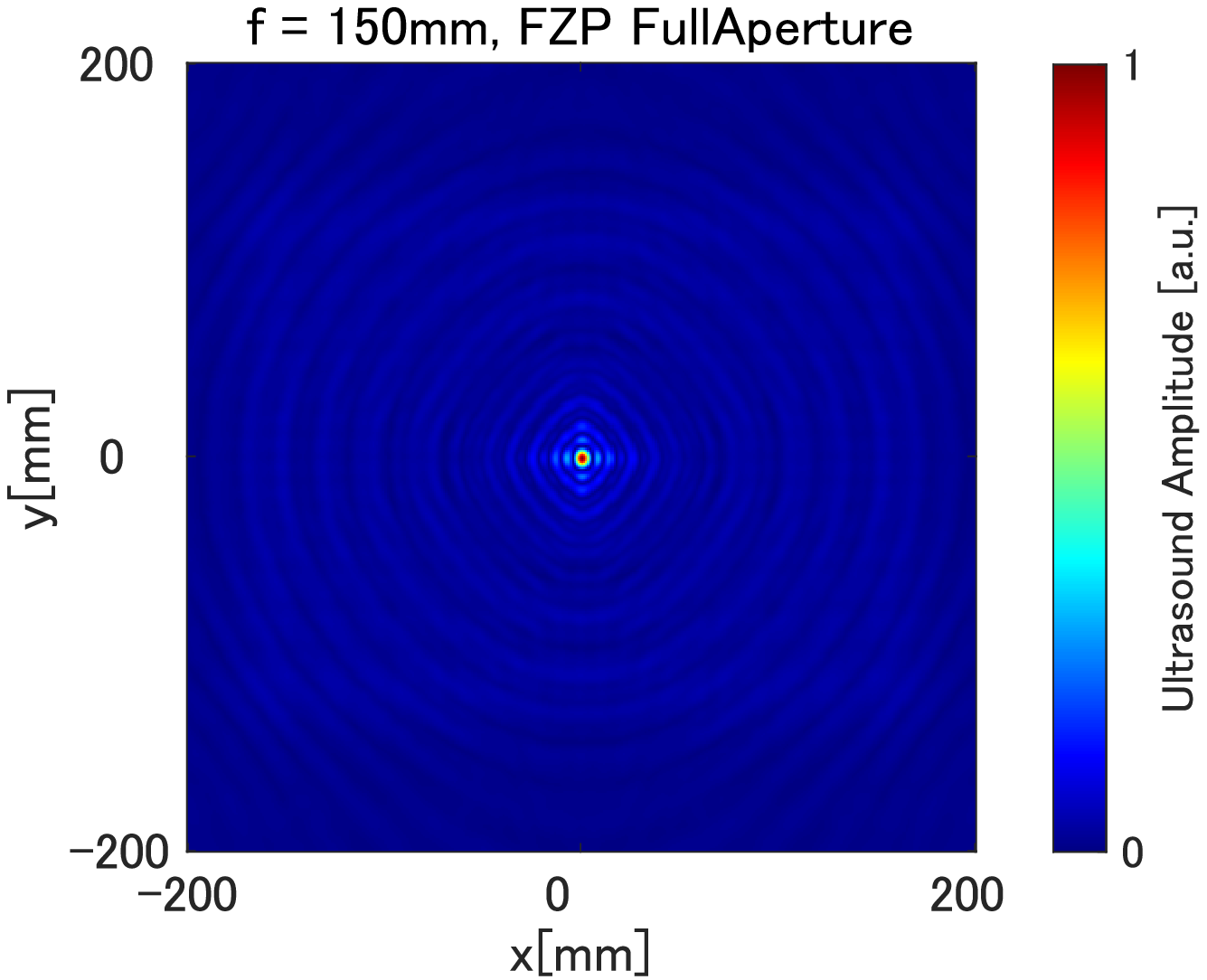}
\includegraphics[width=2in]{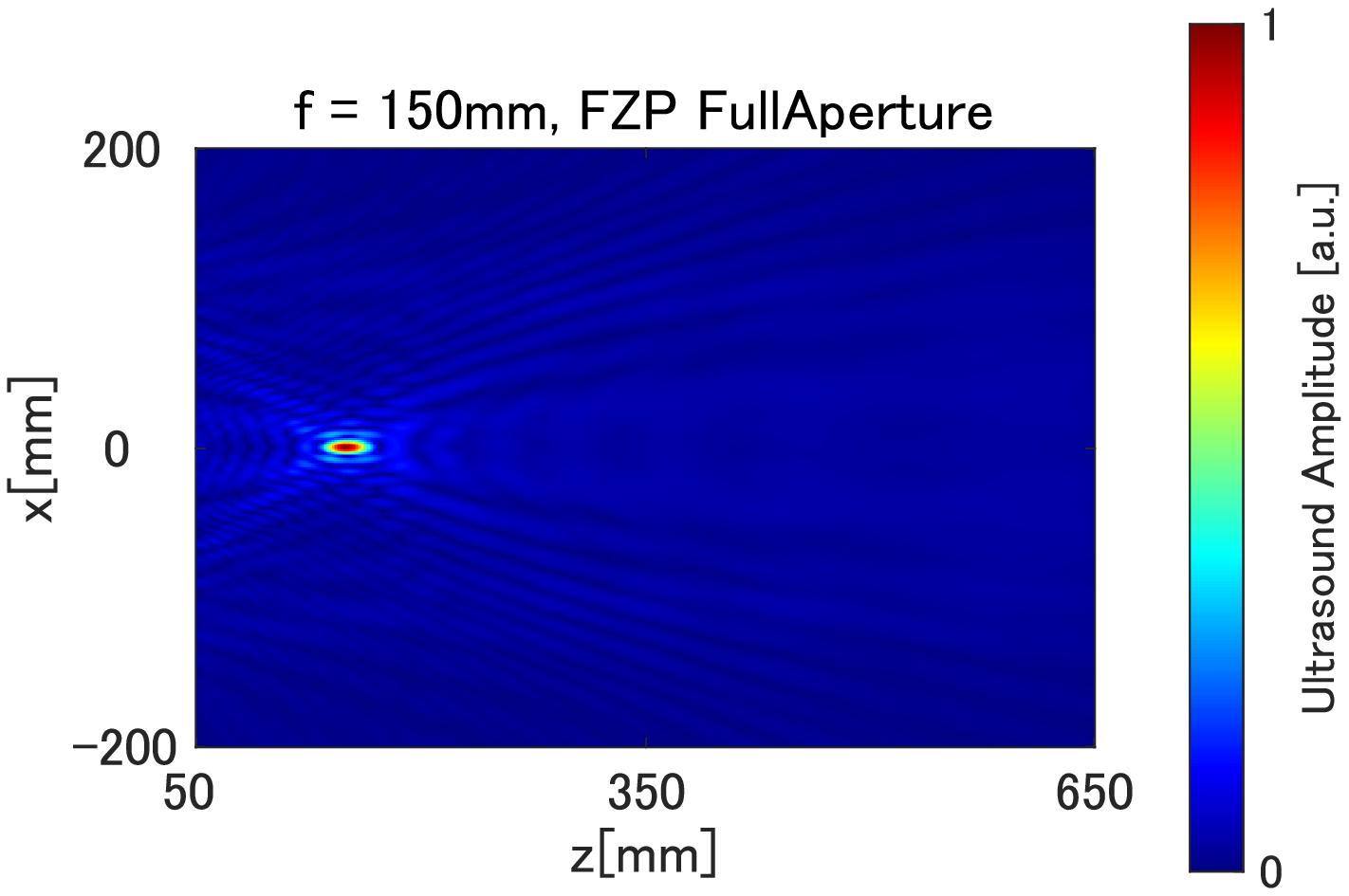}
\includegraphics[width=2in]{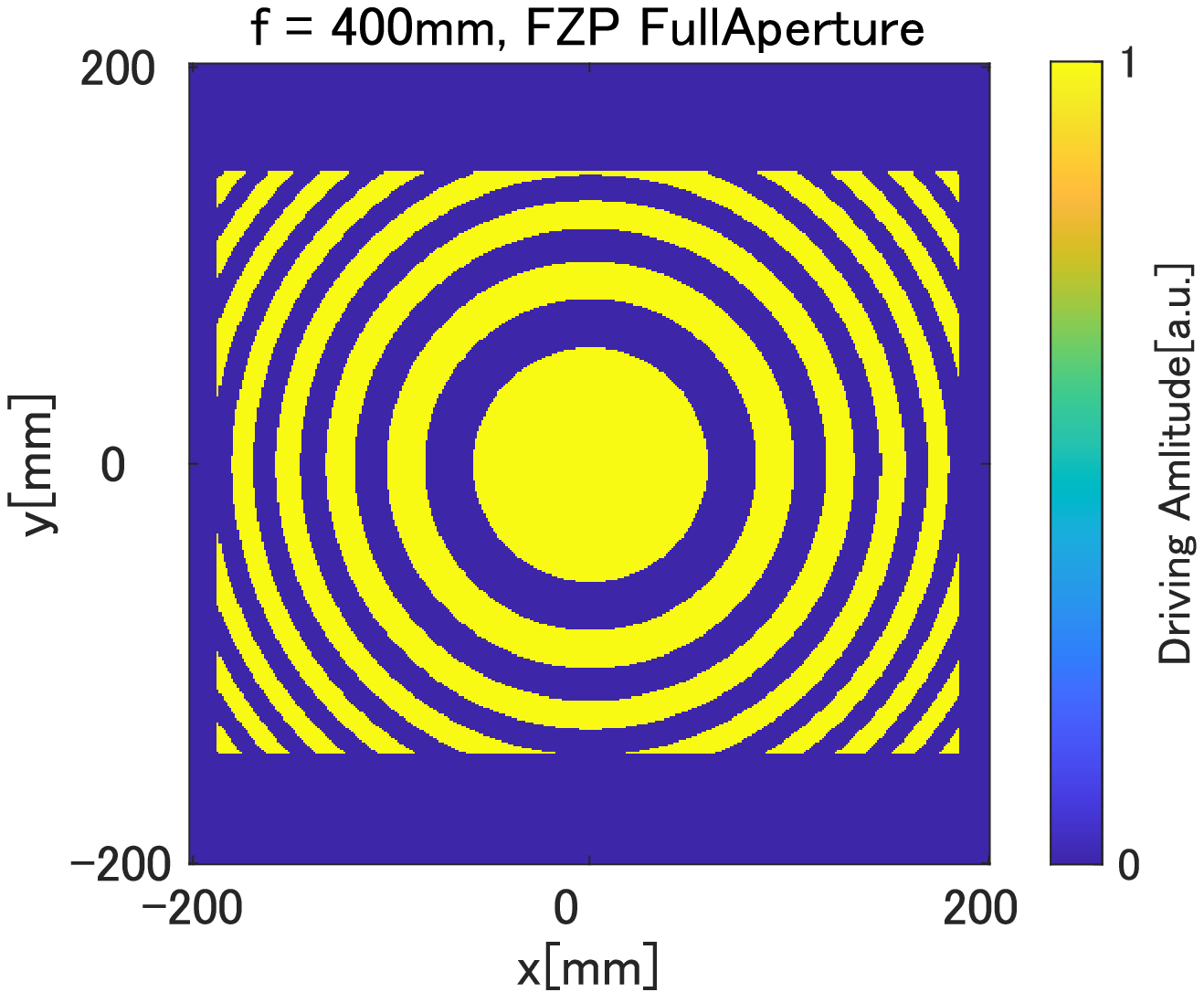}
\includegraphics[width=2in]{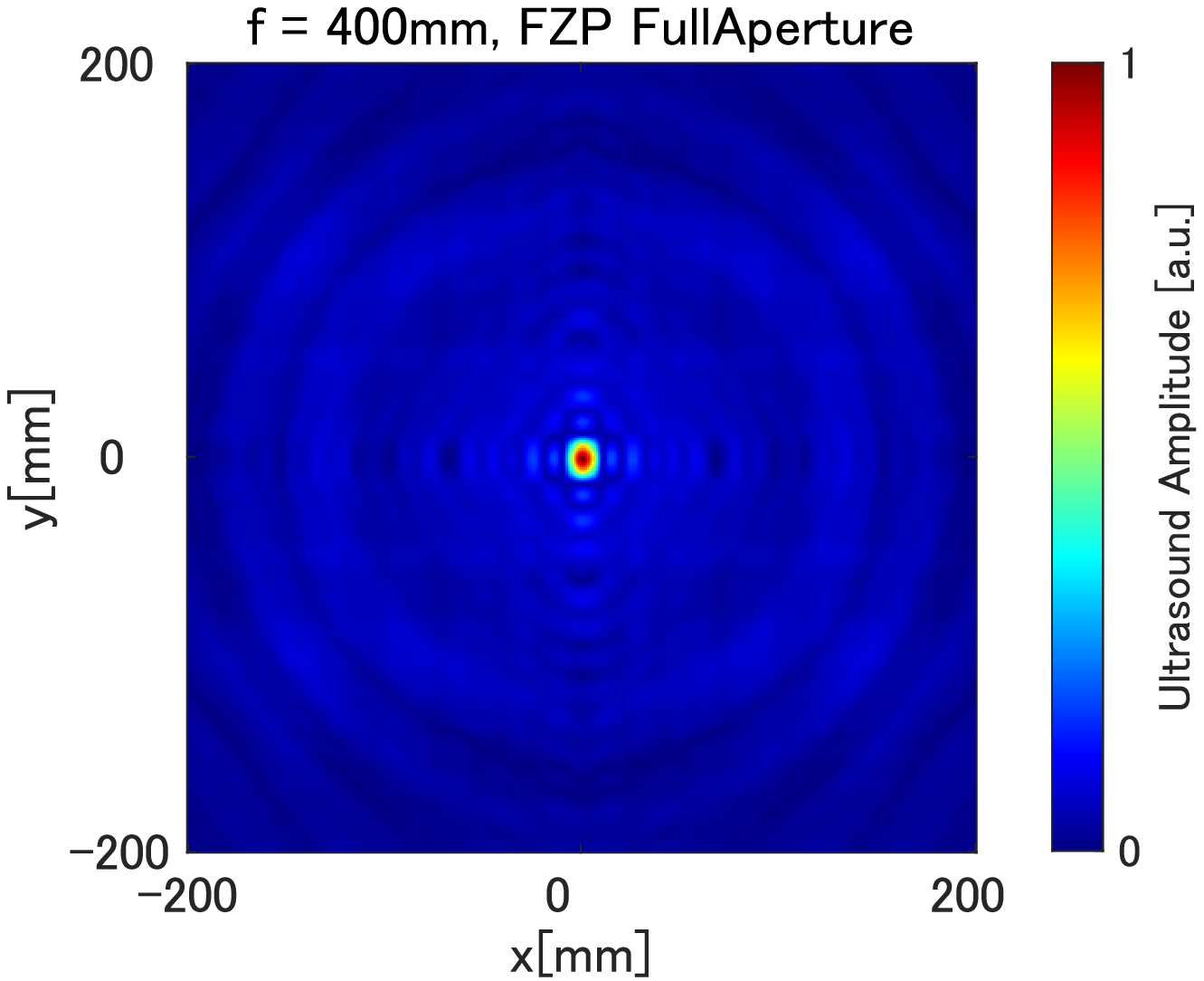}
\includegraphics[width=2in]{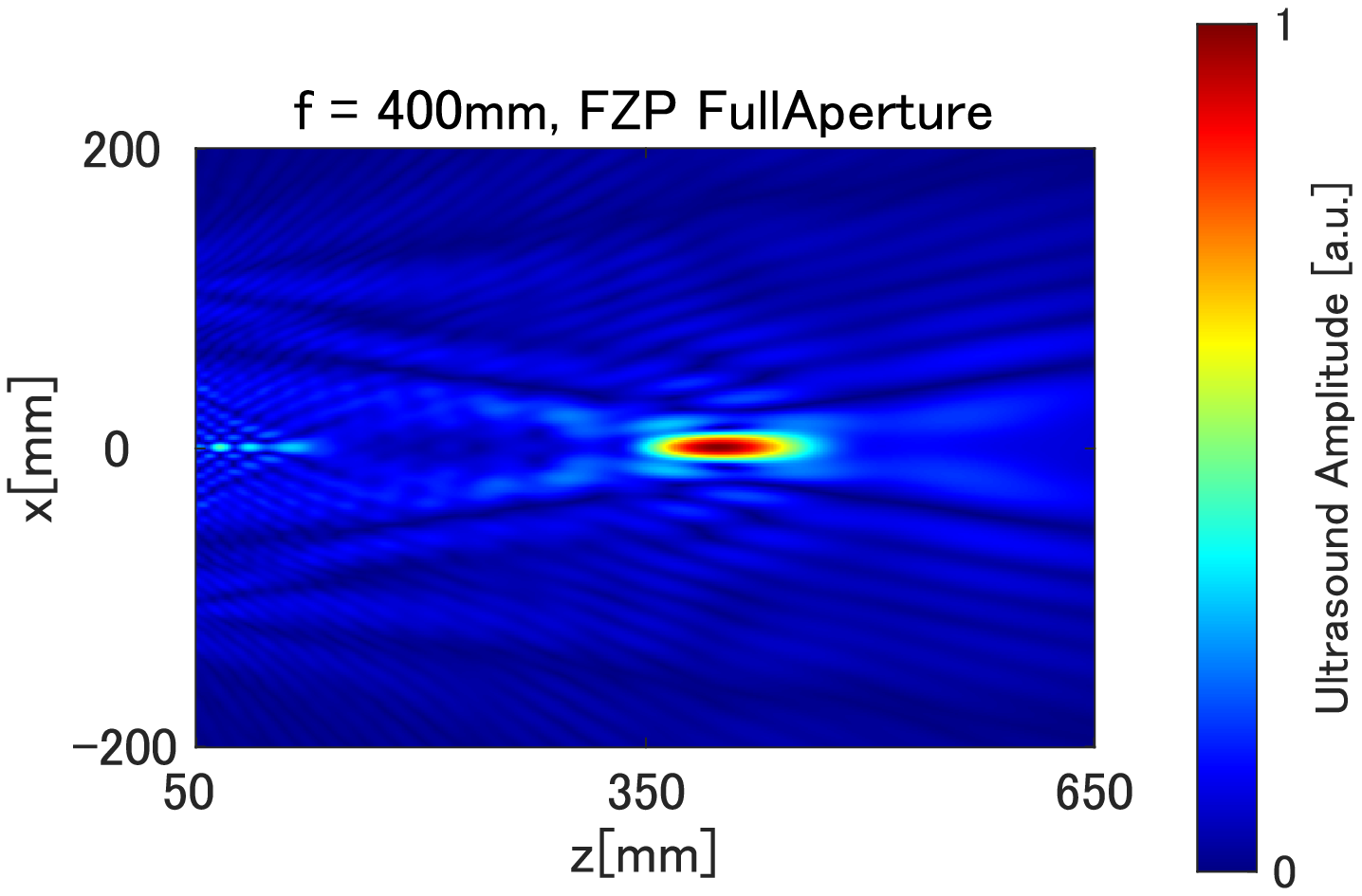}
\caption{\label{fig:drivingpattern1}Calculated FZP patterns (left column), normalized ultrasound amplitude fields by a continuous planar wave source under the FZPs in numerical simulations in the focal plane paralell to the $xy$-plane (Middle column), and that in the $xz$-plane (right column), for the focal depth of 150~mm and 400~mm, respectively. The coordinate system is as defined in the Fig. \ref{fig:setup} illustrating the experiment setup.}
\end{figure*}
 
The designing procedure of FZPs follows this principle. The amplitude distribution $P(\bm{r}_t)$ on the FZP is generated from the distribution of the driving phase $\theta(\bm{r}_t)$ of the phased array element, calculated in Eq. (1) with arbitrary spatial distributions. The most commonly used phase-to-amplitude conversion rules are as follows: 1) determine $\alpha$ so that the phase at the point on the irradiation plane closest to the focus is zero, 2) calculate the remainder of the driving phase divided by $2\pi$, and 3) set the amplitude of the element to ON ($P(\bm{r}_t) = 1$) when the remainder is from zero to $\pi$ and set the amplitude OFF ($P(\bm{r}_t) = 0$) otherwise:
 \begin{align}
P(\bm{r}_t) = \left\{
\begin{array}{cc}
1, & 2n\pi \leq \theta(\bm{r}_t) < (2n + 1)\pi\\
0, & (2n + 1)\pi \leq \theta(\bm{r}_t) < 2(n+1)\pi.
\end{array}
\right., n = 0,1,2,\ldots 
\end{align}
The above rule one is considered as effective in creating a strong sound field, because each element cannot be regarded as completely nondirectional in practice, and its ultrasound emission to the front is the strongest. In the following parts of the paper, we describe our investigations on the spatial properties of the ultrasound field generated according to this method via numerical and real environment experiments.

\section{Numerical Experiments}
\subsection{Calculation of acoustic wave convergence with amplitude FZP}
First, we evaluated the focusing performance of FZPs attached to an in-phase planer wave source. In real-environment experiments described in the following section, we utilized airborne ultrasonic phased arrays with each element driven in-phase as a plane wave source. Therefore, the size of the plane wave source in the numerical simulations was set to 370 mm $\times$ 290 mm, which is approximately equivalent to that of the actual phased arrays. 
We assumed that each point of the amplitude on the FZP was a omnidirectional point source.
The free-field bounary condition and no dissipation of medium in the calculation domain of ultrasound propagation are also assumed.
Under these conditions, the ultrasound field $P_f(\bm{r})$ was calculated using the wave superposition principle:
\begin{align}
P_f(\bm{r}) = \sum_{t = 1}^N P(\bm{r}_t)\frac{e^{-\mathrm{j}k ||\bm{r}- \bm{r}_t||}}{||\bm{r}- \bm{r}_t||},
\end{align}
where $\mathrm{j} = \sqrt{-1}$ denotes the imaginary unit and $N$ denotes the number of source points.
Here $t = 1,\ldots,N$ denotes the index of source locations.
All simulations are performed using the above equation.

Two types of amplitude FZPs were designed in line with the above design criteria using Eq. (2) to converge sound around a focal point located apart from the plane wave source by 150~mm and 40~mm, respectively. The spatial resolution of the wave field calculation area was also set to 1 mm. The ultrasonic frequency was set to 40 kHz, which is widely utilized in current mid-air ultrasound applications.
The FZP patterns were calculated with a spatial resolution of 1 mm, less than 1/4 of the wavelength.
Figure \ref{fig:drivingpattern1} illustrates a simulation result of the ultrasonic sound field including the focal point for each FZP. 
The left column of the figures indicates the values of $P(\bm{r}_t)$.
We adopted MATLAB for all numerical simulations in this study.
The sound field generated by the designed FZPs indicates successful formation of an ultrasound focus for each case.
Concentric acoustic emissions outside the focal region were observed in both FZPs in the $xy$-plane amplitude simulations.
\begin{figure}[t]
\includegraphics[width = 3in]{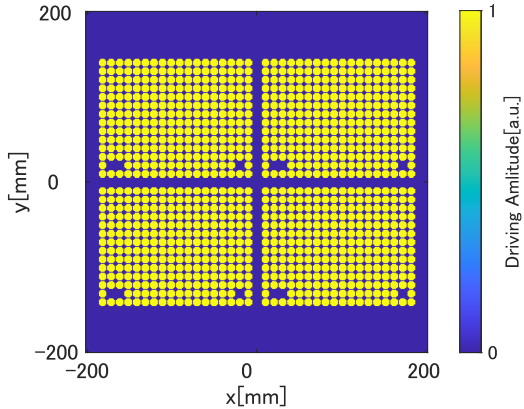}
\caption{\label{fig:arrangement}Transducer arragement in the numerical simulations and measurement experiments. Yellow regions indicate ultrasound emitting areas.}
\end{figure}
\begin{figure*}[t]
\includegraphics[width=2in]{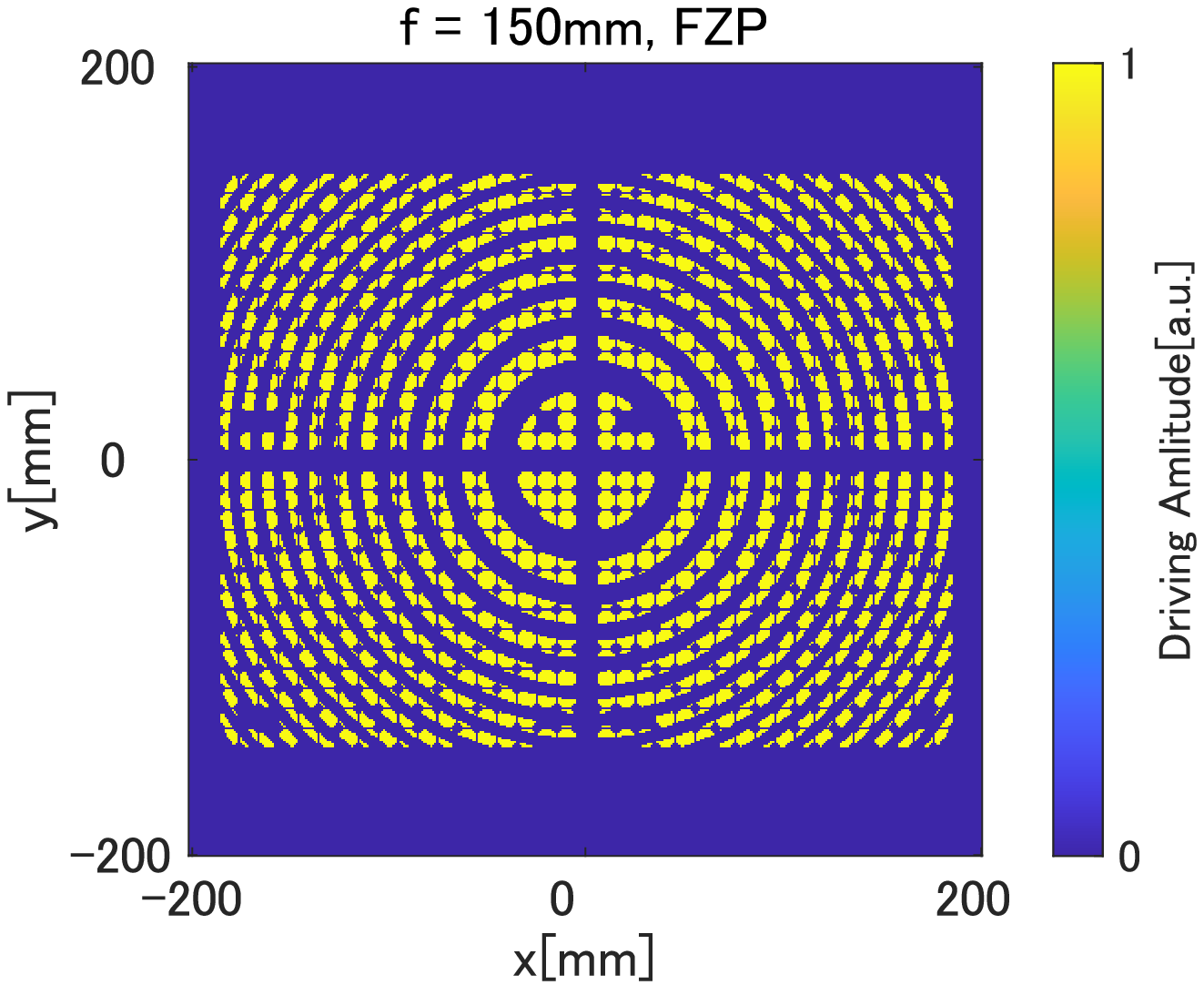}
\includegraphics[width=2in]{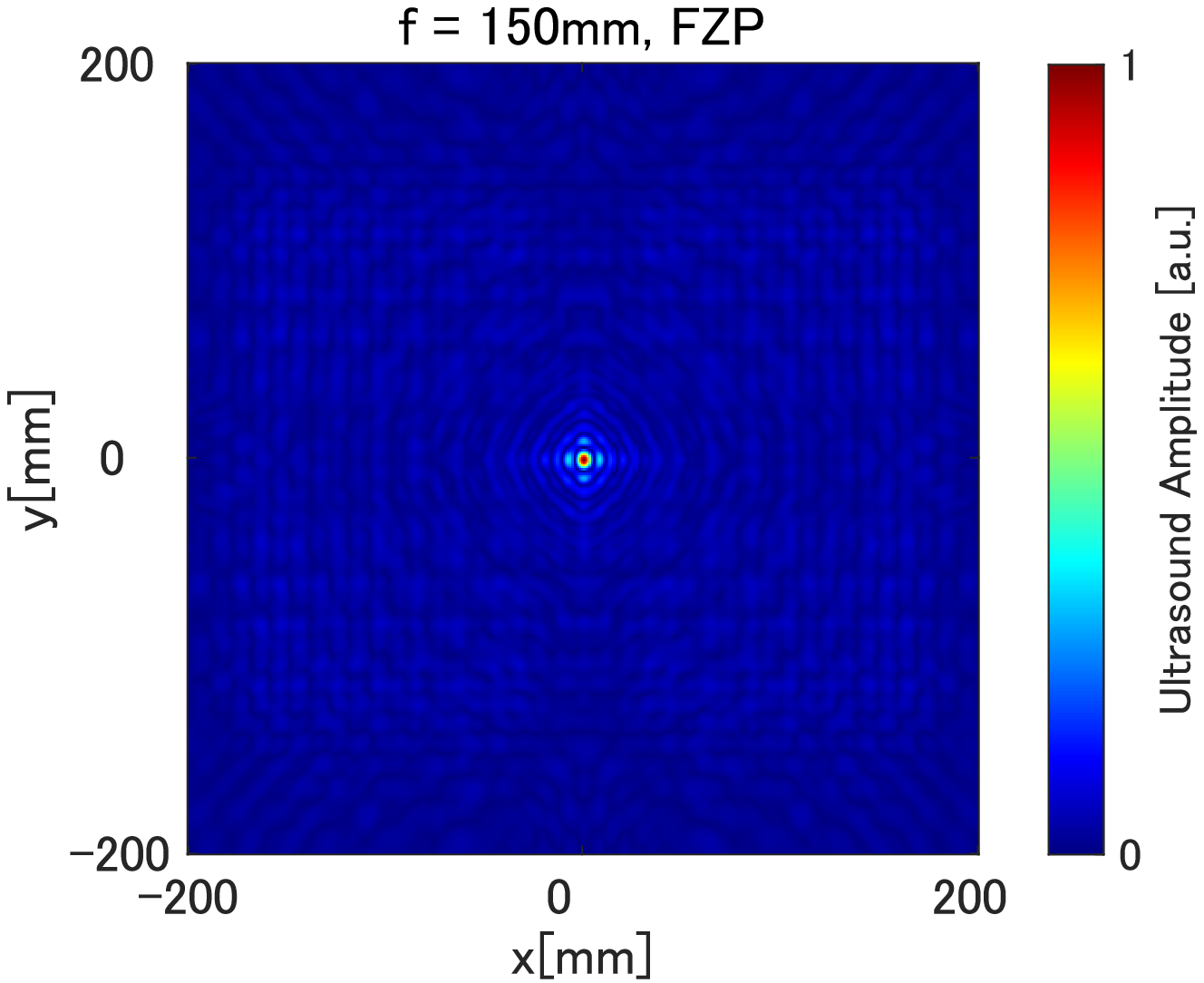}
\includegraphics[width=2in]{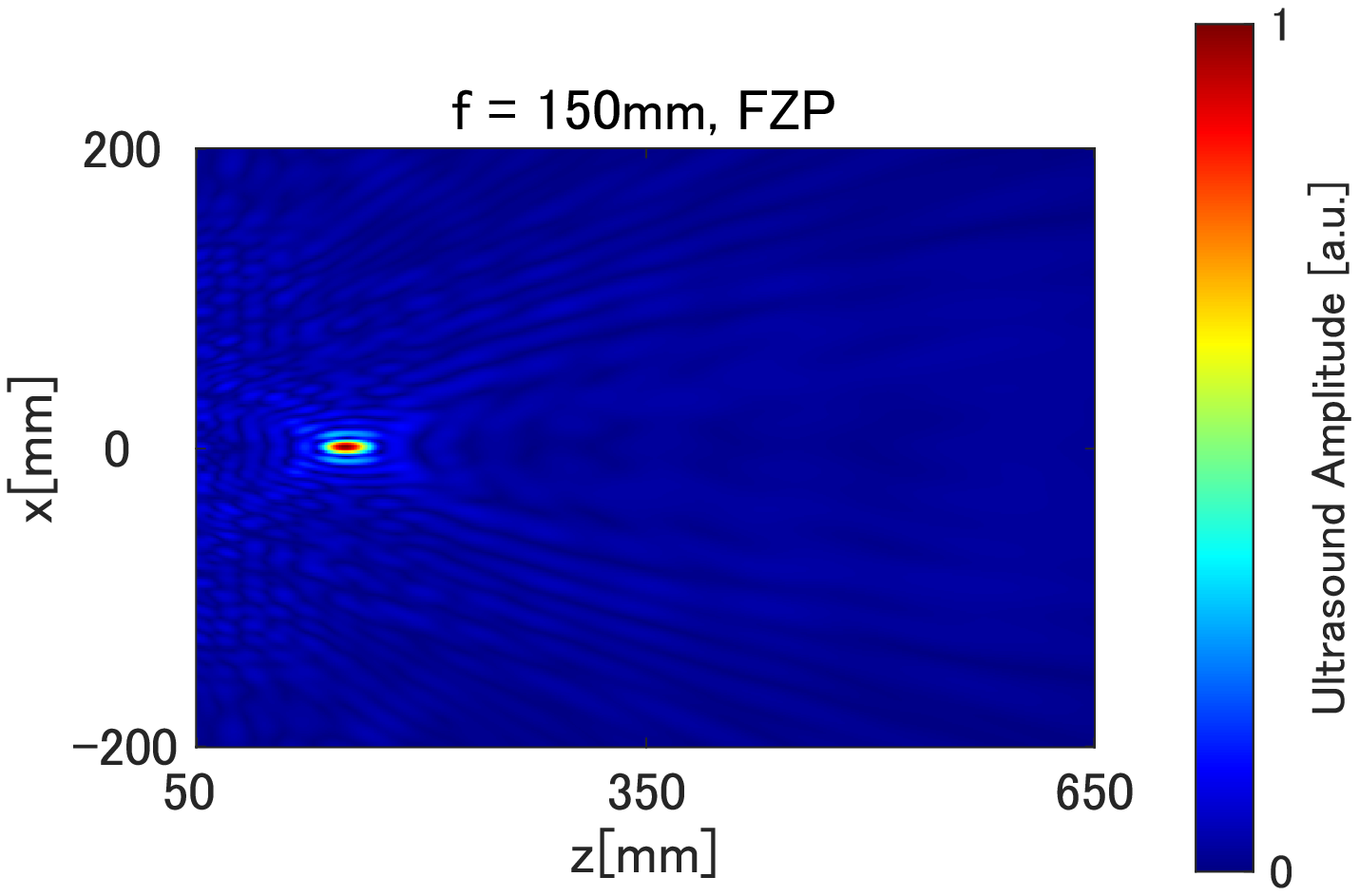}
\includegraphics[width=2in]{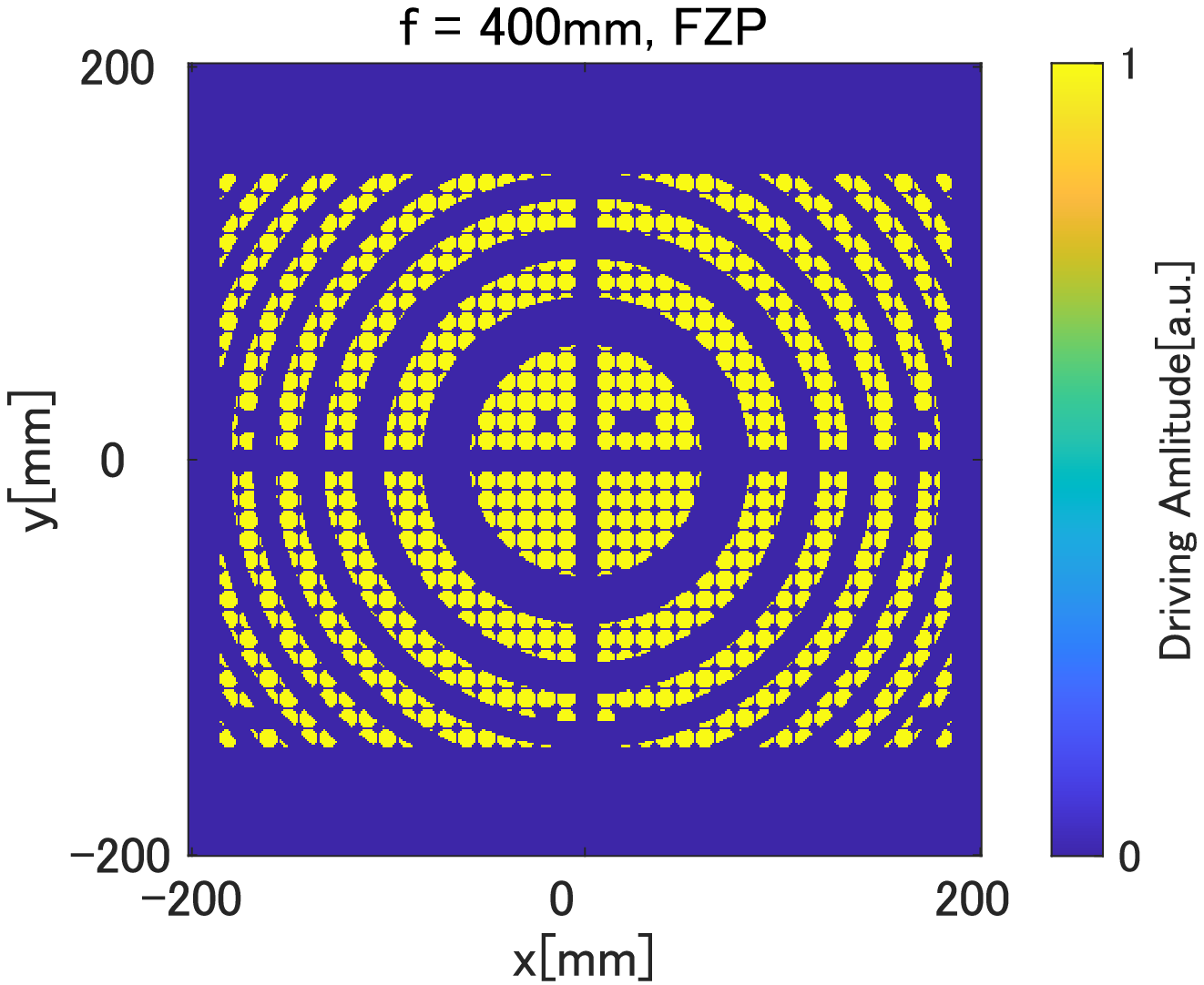}
\includegraphics[width=2in]{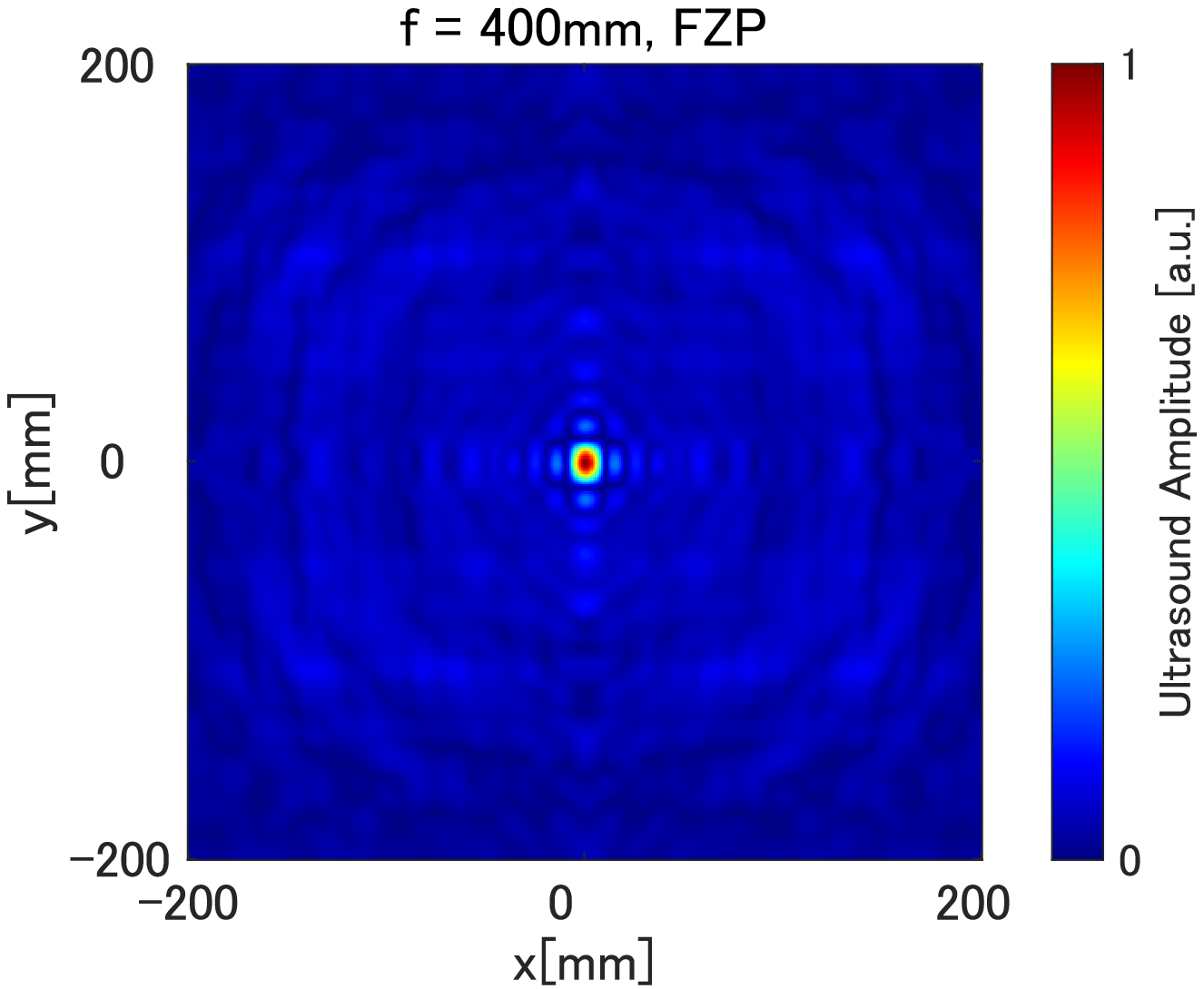}
\includegraphics[width=2in]{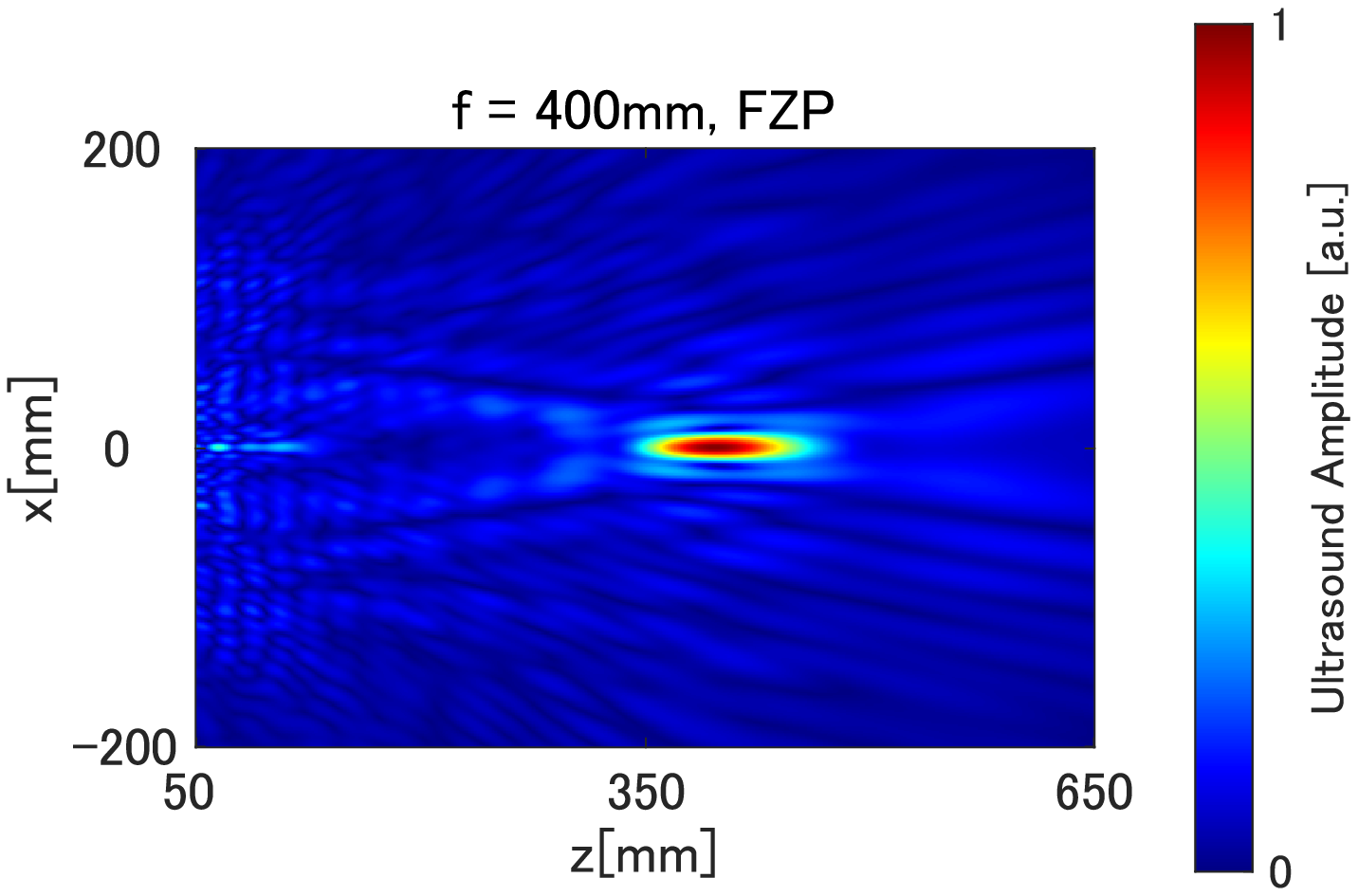}
\caption{\label{fig:drivingpattern2}Simulation results for the case where FZPs are located on the transducer array driven with synchronized uniform phase delays. Calculated amplitude patterns (left column), normalized amplitude fields in numerical simulations in the focal plane parallel to the $xy$-plane (middle column), and that in the $xz$-plane (right column), for the focal depths of 150~mm and 400~mm, respectively.}
\end{figure*}
\begin{figure*}[t]
\includegraphics[width=2in]{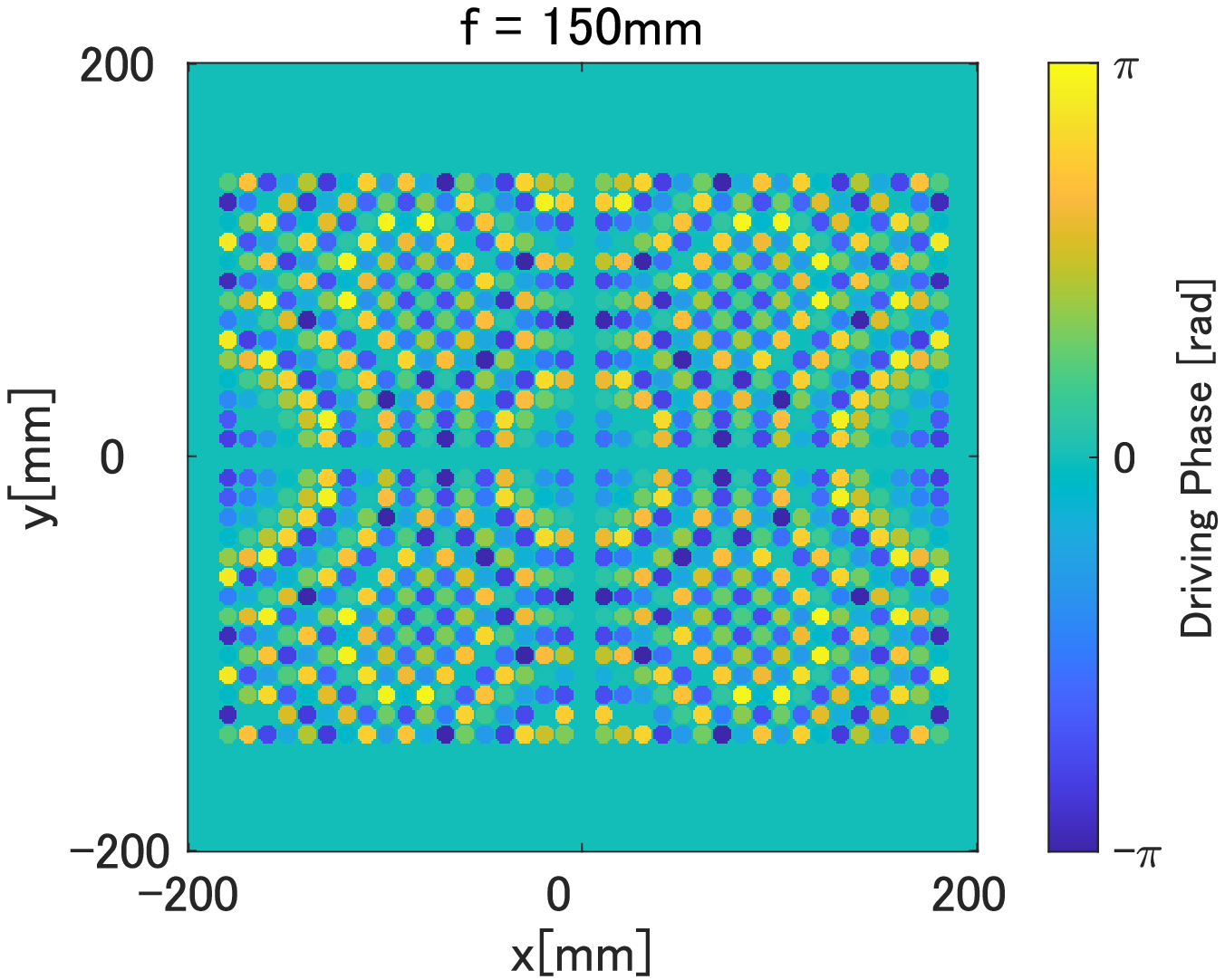}
\includegraphics[width=2in]{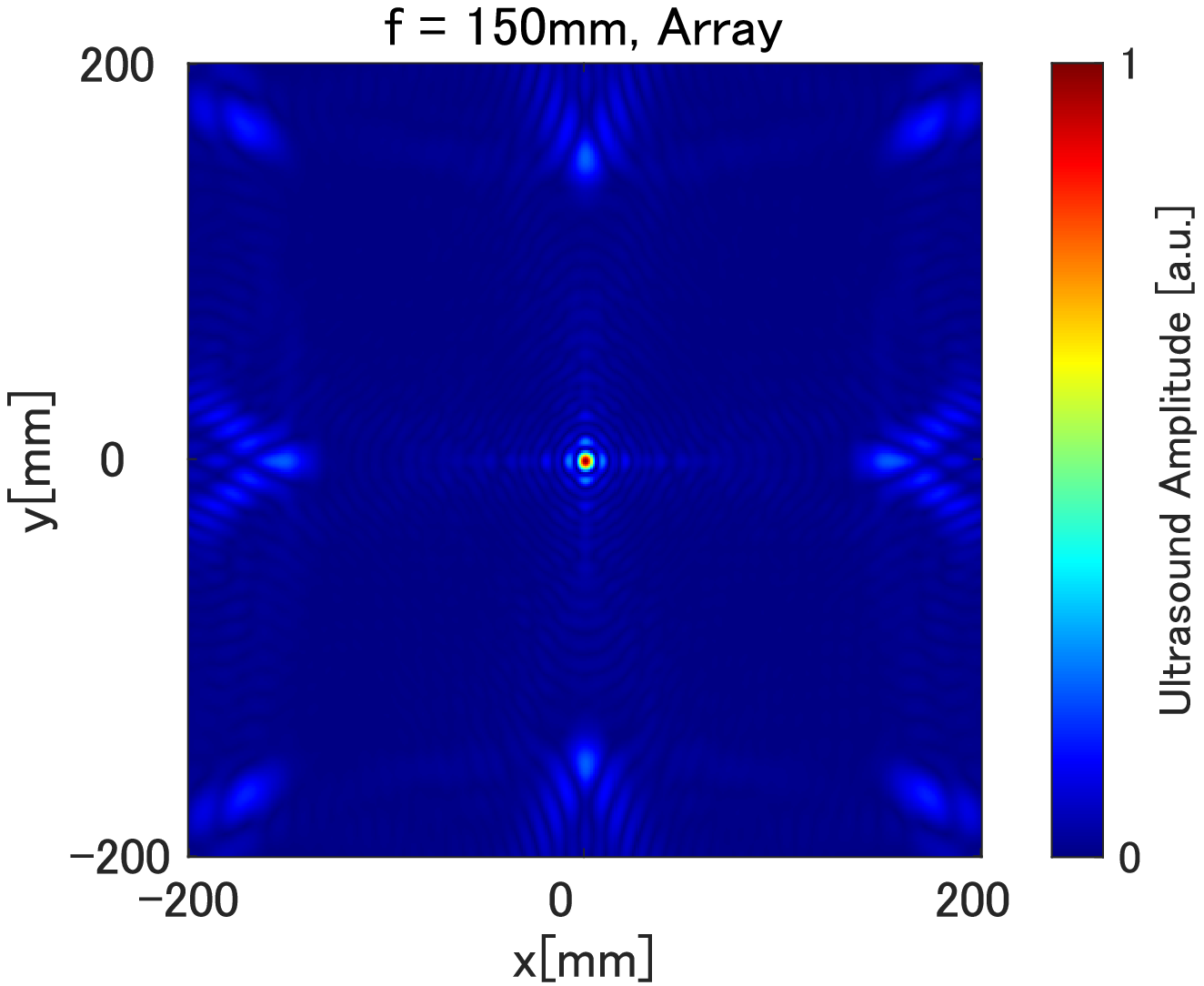}
\includegraphics[width=2in]{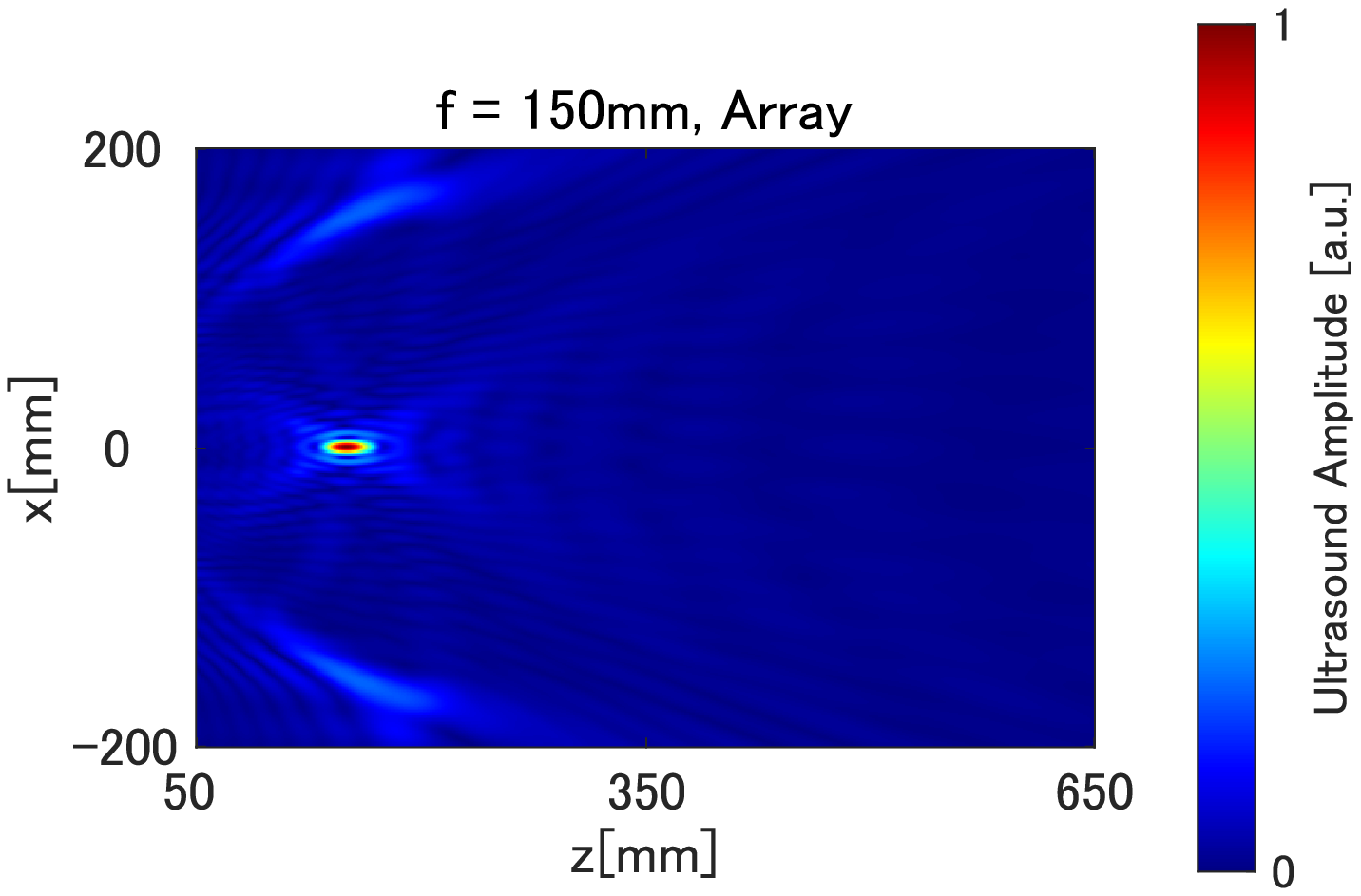}
\includegraphics[width=2in]{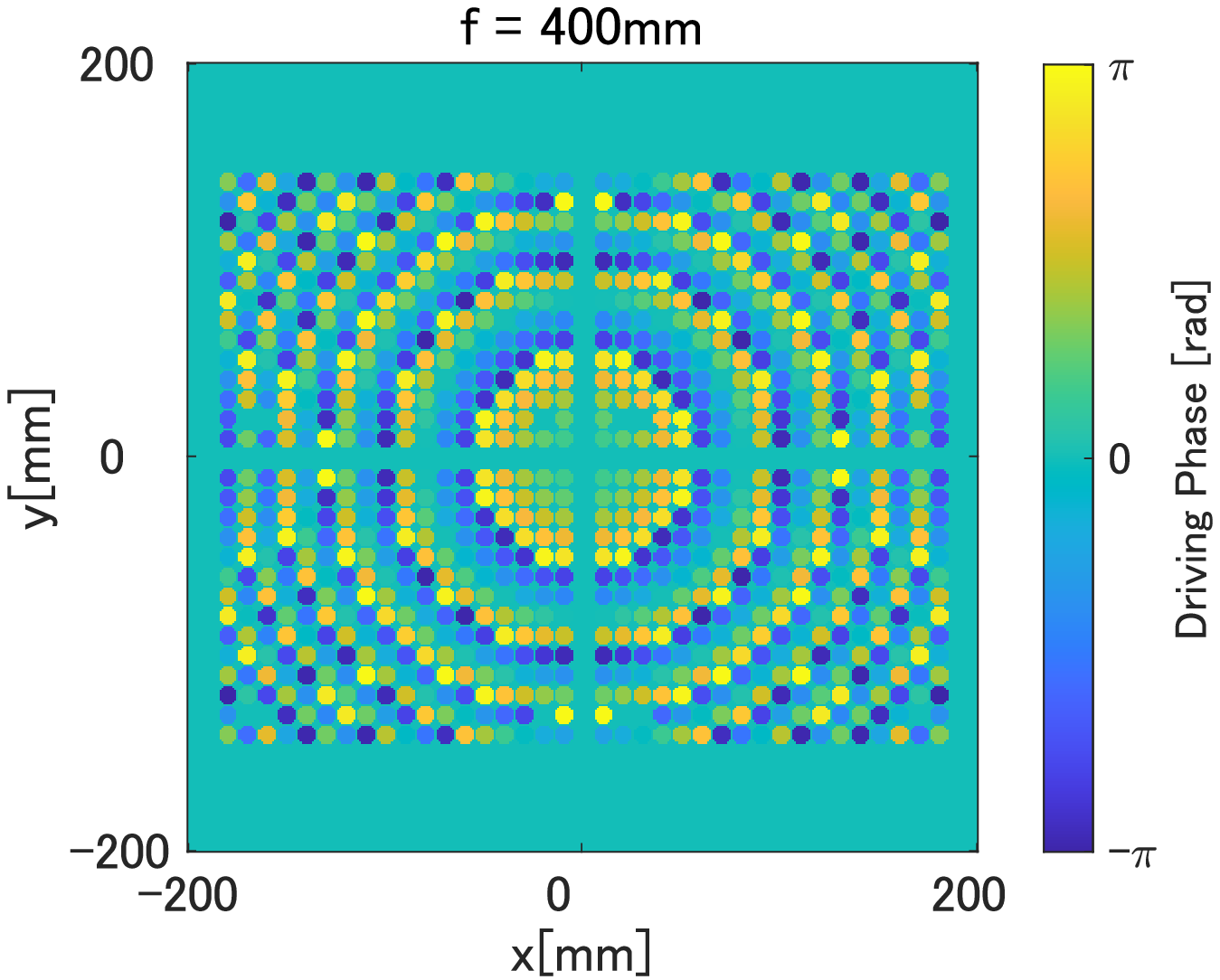}
\includegraphics[width=2in]{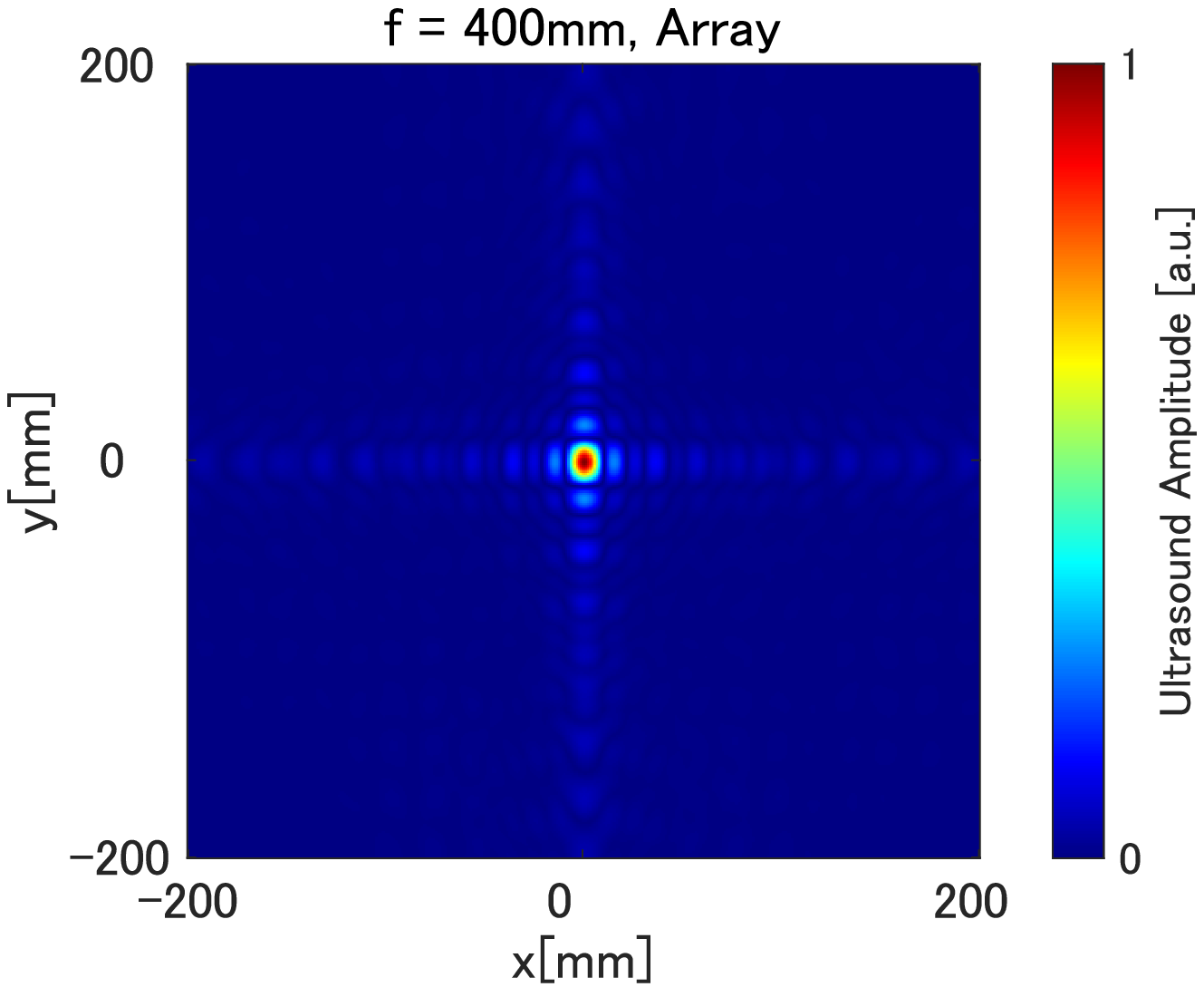}
\includegraphics[width=2in]{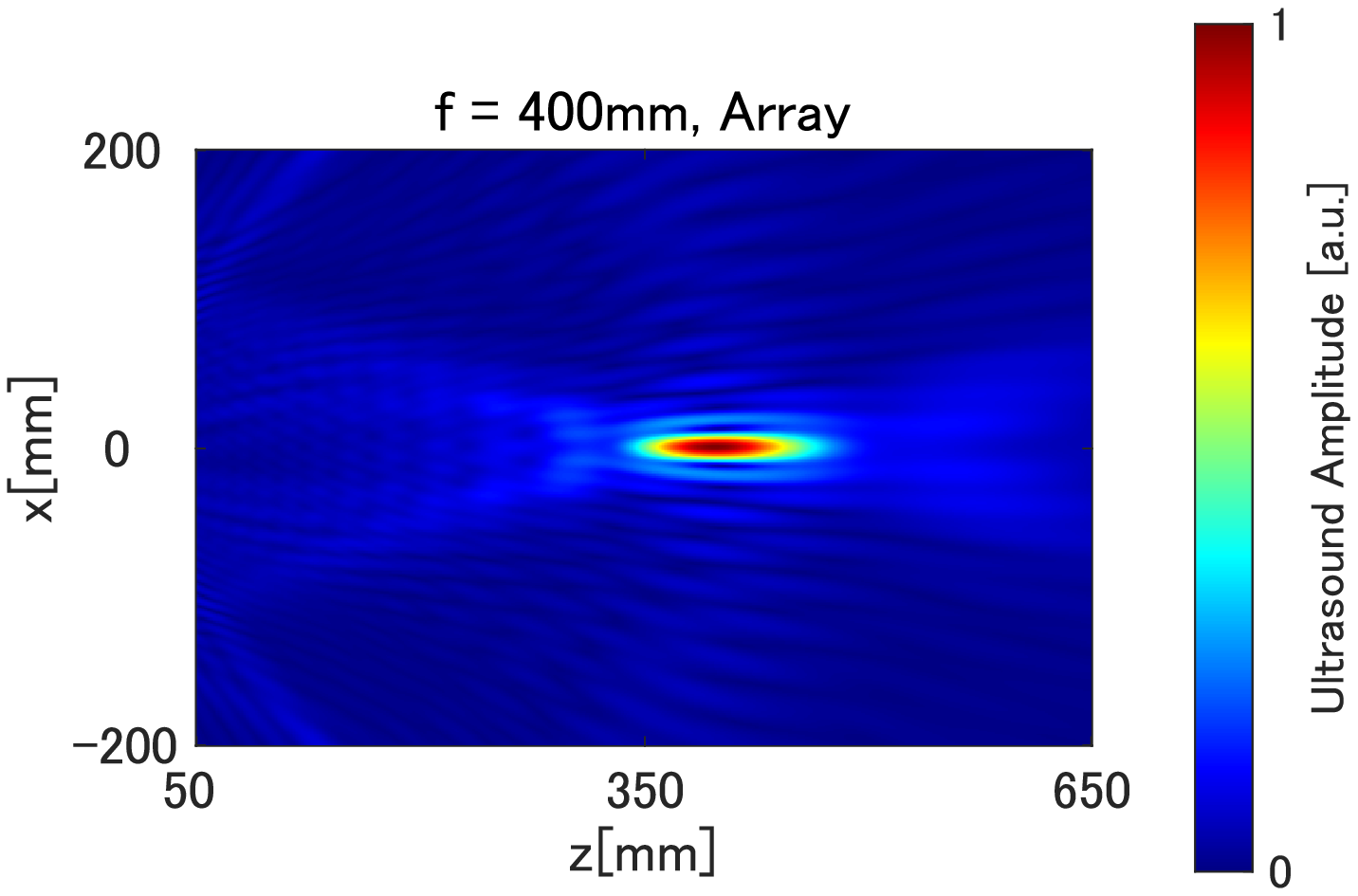}
\caption{\label{fig:drivingpattern3}Simulation results for the case where the transducer array is driven with individual element phase delays for forming an ultrasound focus. Calculated amplitude patterns (left column), normalized amplitude fields in numerical simulations in the focal plane parallel to the $xy$-plane (middle column) and that in the $xz$-plane (right column), for the focal depths of 150~mm and 400~mm.}
\end{figure*}

\subsection{Focus formation by FZP using a plane wave source comprising multiple ultrasonic transducers}
As mentioned in the introduction, we consider forming the vibrating surface by arranging cylindrical aerial ultrasonic transducers in a two-dimensional grid, which are commercially available and have been utilized for several preceding studies.
In this second numerical simulation, we assumed that each transducer could be modeled as a set of infinitesimal point sources uniformly distributed on its vibrating surface of 10 mm diameter, which is equivalent to that of the transducers utilized in the following experiments.
The arrangement of the transducer in the simulation was set identical to the real devices employed in the real environment experiment. The FZP pattern was superimposed onto the transducer array in which the driving signals of all transducers set were identical. As with the previous simulations, the spatial distribution of the source FZP plane was set to 1 mm.
For the fidelity of simulations compared with the real transducer arrangement on the phased arrays, periodic gaps were created between the cylinders and some regions were set where transducers were not mounted, that corresponded to the screws in the actual devices utilized in the experiments.
The transducer arrangement in the $xy$-plane in the numerical and real-world experiments is depicted in Fig. \ref{fig:arrangement}.
\begin{figure*}[t]
\includegraphics[width=2in]{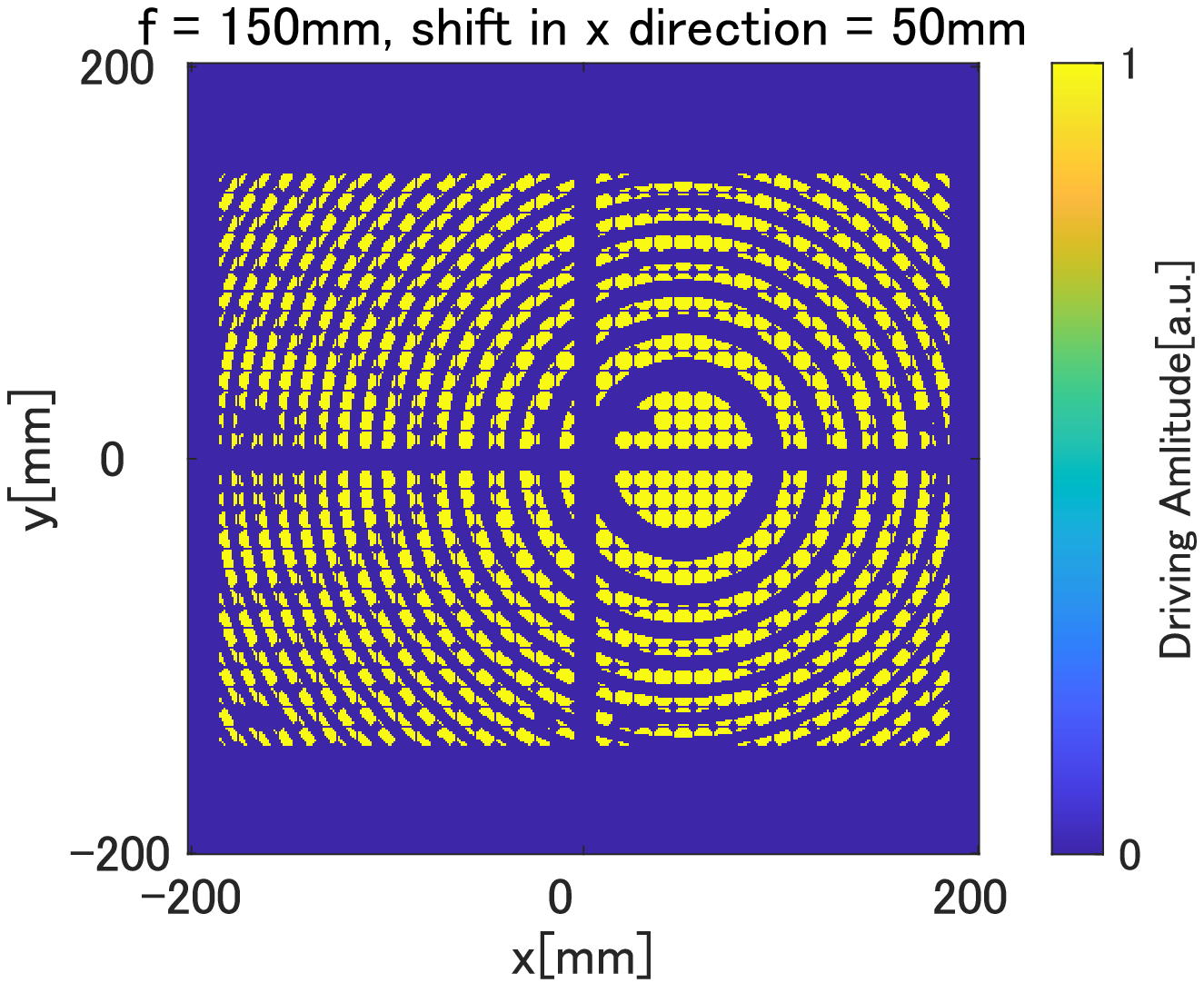}
\includegraphics[width=2in]{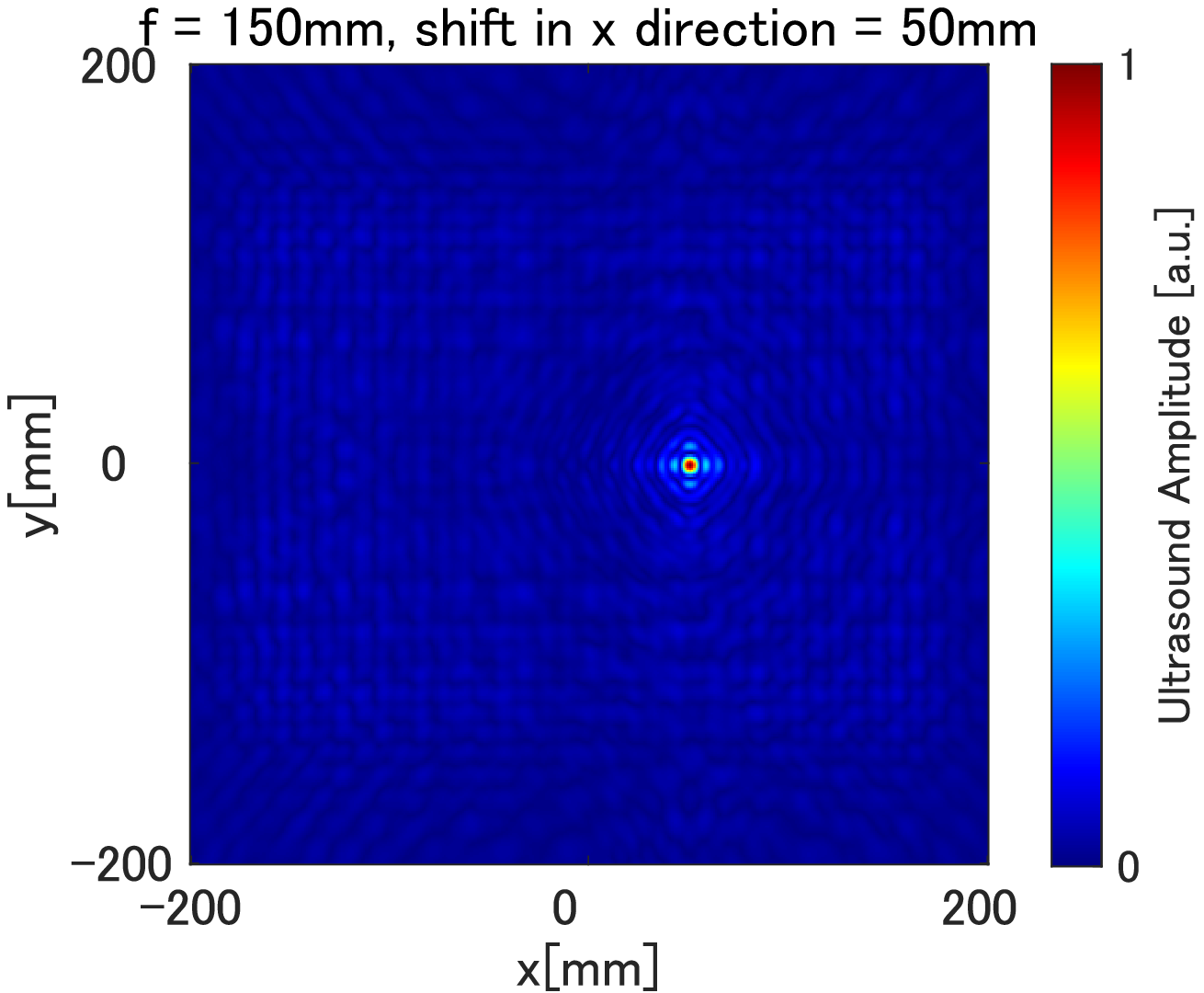}
\includegraphics[width=2.5in]{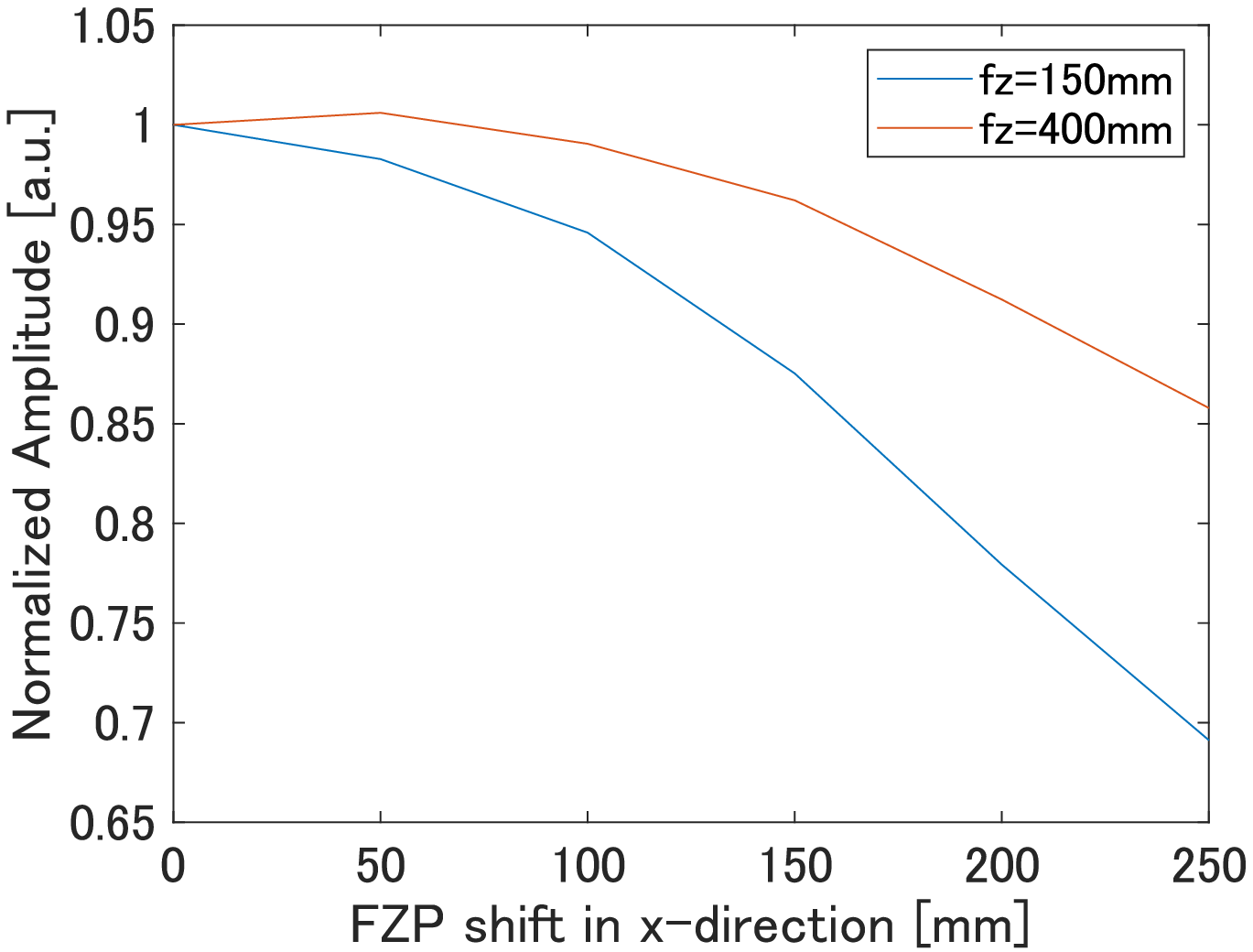}
\caption{\label{fig:shiftsim}Examples of simulation results for the case where FZPs shifted in $x$-direction are located on the transducer array driven with synchronized uniform phase delays. Calculated amplitude pattern (left), normalized amplitude field in the focal plane parallel to the $xy$-plane (middle), and relative focal amplitudes normalized by the case with no FZP shift (right).}
\end{figure*}
\begin{figure*}[t]
\centerline{\includegraphics[width=5in]{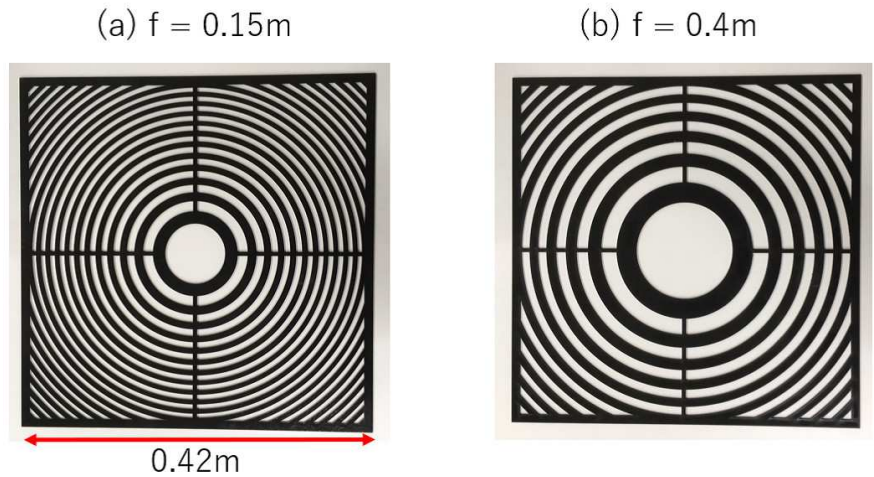}}
\caption{\label{fig:fzp}Fabricated FZPs with focal depths of (a) 150~mm and (b) 400~mm.}
\baselineskip=11pt
\raggedright
\end{figure*}
 \begin{figure}[t]
\centerline{\includegraphics[width=3in]{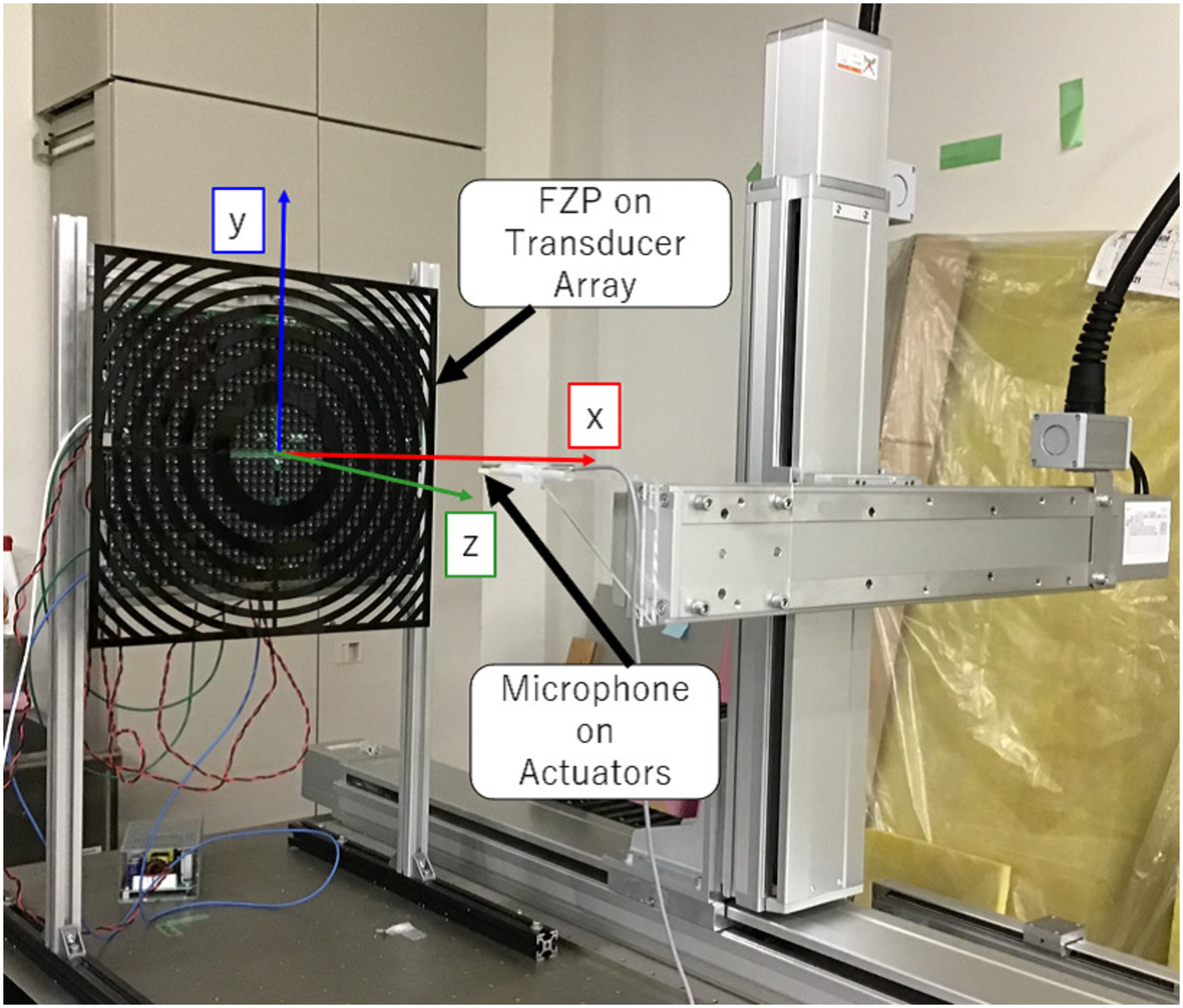}}
\caption{\label{fig:setup} Coordinate system in the experiments and experimental setup with four units of ultrasound phased arrays as planar wave sources, covered by the FZP.}
\baselineskip=11pt
\raggedright
\end{figure}
Figure \ref{fig:drivingpattern2} illustrates the calculated source amplitude distributions and generated sound fields.
The left column of the figures shows the amplitude patterns of the ultrasound source $P(\bm{r}_t)$, comprising multiple ultrasound transducers partially covered by the FZPs.
In this simulation, all emissions were in the same phase and FZP was modeled as complete ultrasound mask with no thickness.
The results demonstrate that adequate focusing was achieved with a periodically perforated emission plane comprising a set of in-phase-driven transducers.
Weak grid-like amplitude patterns that were superimposed on the focus in the $xy$-plane simulation are observed.
These patterns were not observed in the amplitude distributions generated from the previous simulations, where no periodical source gaps were considered.
Therefore, generation of these patterns can be ascribed to the periodical defects in the emission plane of the wave source under the FZPs.
Apart from that, the generated patterns with these two simulations have great similarity to one another.

Next, we evaluated the focal formation by the conventional phase controlling method of transducer arrays, for the relative assessment of the focusing performances of the FZP-based methods.
The arrangement and amplitudes of the transducers were set identical to those in the previous simulation.
The output phase of each transducer was calculated in line with Eq. (1) with $\alpha = 0$, that is, $P(\bm{r}_t)$ in Eq. (3) was determined as $P(\bm{r}_t) = e^{\mathrm{j}\theta(\bm{r}_t)} = e^{\mathrm{j}k||\bm{r}_t - \bm{r}_f||}$.
The output amplitude of the sources was all identical.

Figure \ref{fig:drivingpattern3} illustrates the calculation results of the amplitude distribution by the phase-controlled transducer arrays.
With the focal depth of 150~mm, prominent grating lobes are observed at a distance from the focal point, instead of the widespread granular amplitude patterns observed in the case with FZPs.
This is owing to the fact that the spatial resolution of the phase distribution control of the radiating surface depends on the size of the transducer elements, and therefore cannot achieve a phase distribution finer than half a wavelength.
The spatial distributions of FZPs we create in the experiment are finer than half the wavelength, and this mitigated the generation of the grating lobes.
In the case where the focal depth is 400~mm, the grating lobes are located further from the focus, and cannot be seen in the region of simulation.

The phased array is expected to be more efficient than FZPs in forming a focal point because the radiating surface is not shielded unlike the FZPs. As indicated in the next chapter, numerical simulation and real-environment experiments both indicate that the ultrasound intensity at the focus was reduced by using FZPs than performing phase control over all transducers, when the output of each element was the same.

\subsection{Lateral movement of FZPs over transducers driven in phase}
We numerically investigated how the lateral movement of the FZPs over the transducers affects the focusing.
In the simulations, we calculated the focal amplitudes generated by the in-phase transducers and the FZP over them laterally shifted in $x$-direction by 0, 50, 100, 150, 200, and 250~mm.
Figure \ref{fig:shiftsim} depicts an example of amplitude pattern and its corresponding generated ultrasound field for the case with an FZP with the focal depth of 150~mm that is shifted by 50~mm in the $x$-direction. 
A lateral shift of the generated focus that corresponded to the lateral FZP shift is observed.
The figure also includes a graph that shows the relative focal amplitude with respect to the lateral shift of the FZPs.
The simulation results show that for the lateral shifts less than 10~mm do not significantly affect the focal amplitude, whereas the shift greater than the half of the aperture dimension drastically reduce the focal amplitude. 
It is also observed that the focal amplitude attenuation per a unit length shift of FZPs is smaller for the greater focal depth.

\section{physical experiments}
\subsection{Ultrasound focusing by the binary hologram}
We fabricated two types of FZPs made of acrylic plate with 2 mm of thickness, which have 150~mm and 400~mm of focal depth, respectively (Fig. \ref{fig:fzp}).
The acrylic sheets were cut out by a laser cutter (VD-60100, COMMAX, Co., Ltd., Japan) based on CAD data with the geometric positioning and size of each component defined with spatial quantization of 1~mm. 

In the experiment, we utilized custom-made 40 kHz phased arrays \cite{autd3} with their all transducers driven in-phase as a wave source, on which the fabricated FZP was placed to form an ultrasound focus. The custom-made phased array unit utilized in the experiment contained 249 40 kHz ultrasound transducers (T4010A1, Nippon Ceramic, Co., Ltd., Japan) arranged in a two-dimensional lattice. We employed four units of the phased arrays forming an ultrasound emitting aperture of 374.0~mm $\times$ 292.8~mm in the experiment, which corresponds to the spatial configuration of the numerical experiments. For ultrasound scanning measurement, a standard microphone system (1/8-in. microphone, type 4138-A-015; pre-amplifier, type 2670; condition amplifier, type 2690-A; all products of Hottinger, Br\"{u}el and Kj\ae r, Denmark) was utilized. The microphone was mounted on the tip of three-dimensional linear actuators (type ICSB3, product of IAI, Japan). The microphone on the actuators scanned the sound field to capture the acoustic pressure distributions in designated regions. The entire experimental setup and coordinate system in the experiments are illustrated in Fig. \ref{fig:setup}.
 
The scanning was completed with spatial interval of 5~mm for the $x$- and $y$- axes, and 10~mm for the $z$- axis illustrated in Fig. \ref{fig:setup}.
The ultrasound output of the arrays was adequately weakened, compared with its possible maximum for avoiding measurement saturation of the microphone.
Across all the measurements, the output intensity of the transducers were set identical.
For each measurement, the center of the coordinate system in the $xy$-plane was slightly adjusted so that it yielded the maximum observed pressure within the range less than 2~mm after manually setting the coordinate origin on the center of the arrays.
In this manual calibration, the origin of the $z$-axis was set to the surface of the phased array transducers.
While setting the transducer output power unchanged, we also sequentially measured the sound field created by the phase-controlled transducers at corresponding focal depths for the comparison of focusing performances among the two methods.
\begin{figure*}[t]
\includegraphics[width=2.5in]{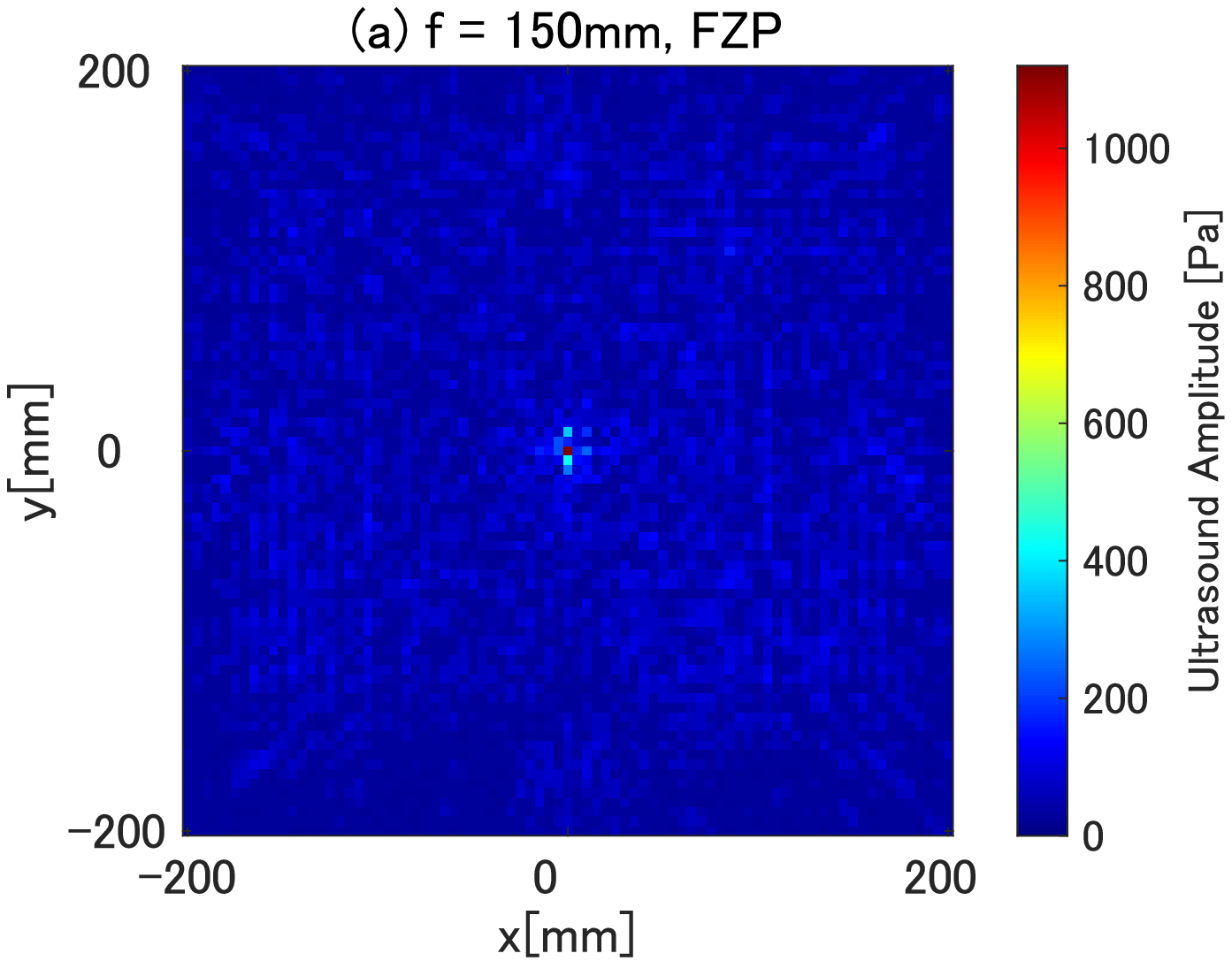}
\includegraphics[width=2.5in]{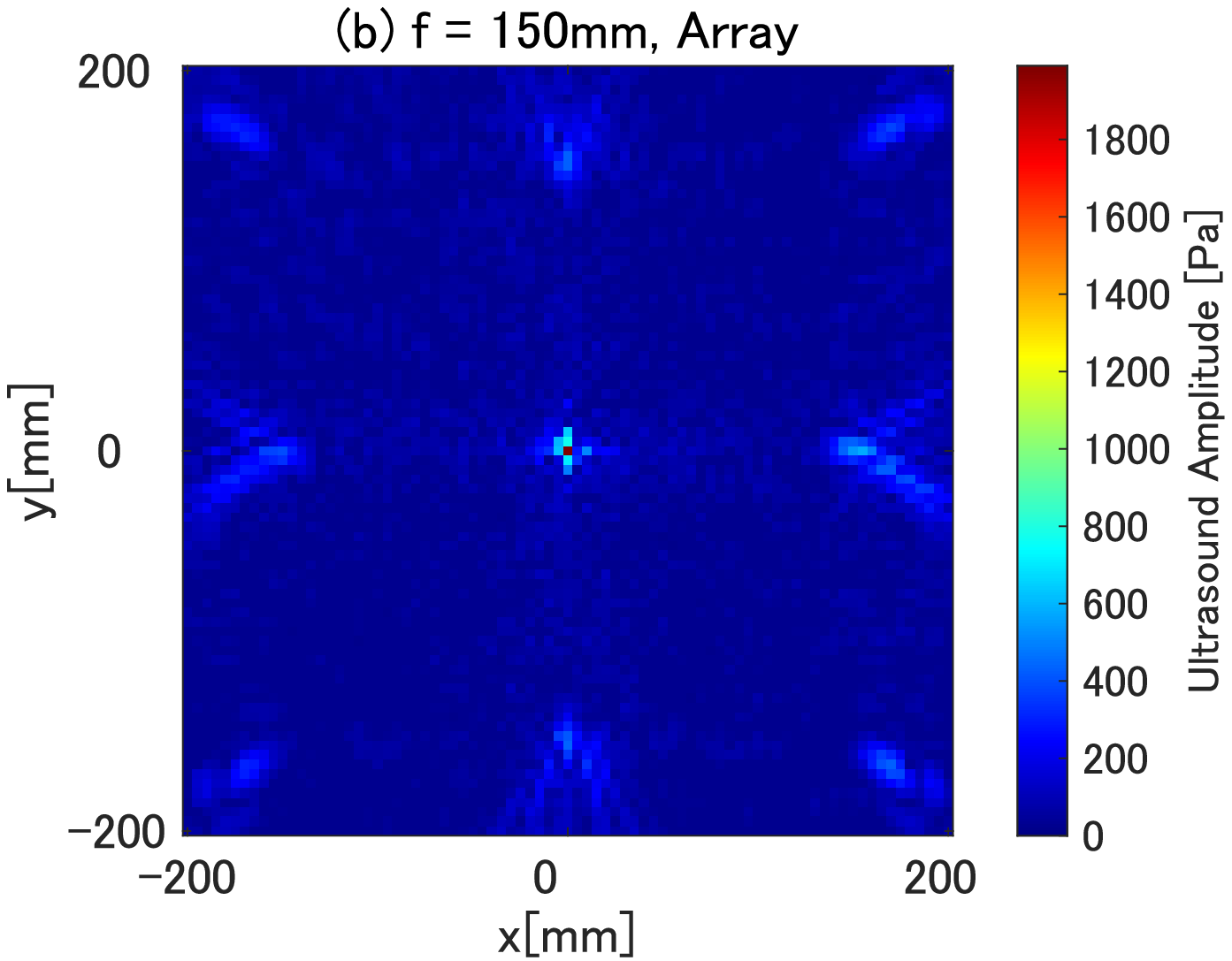}
\includegraphics[width=2.5in]{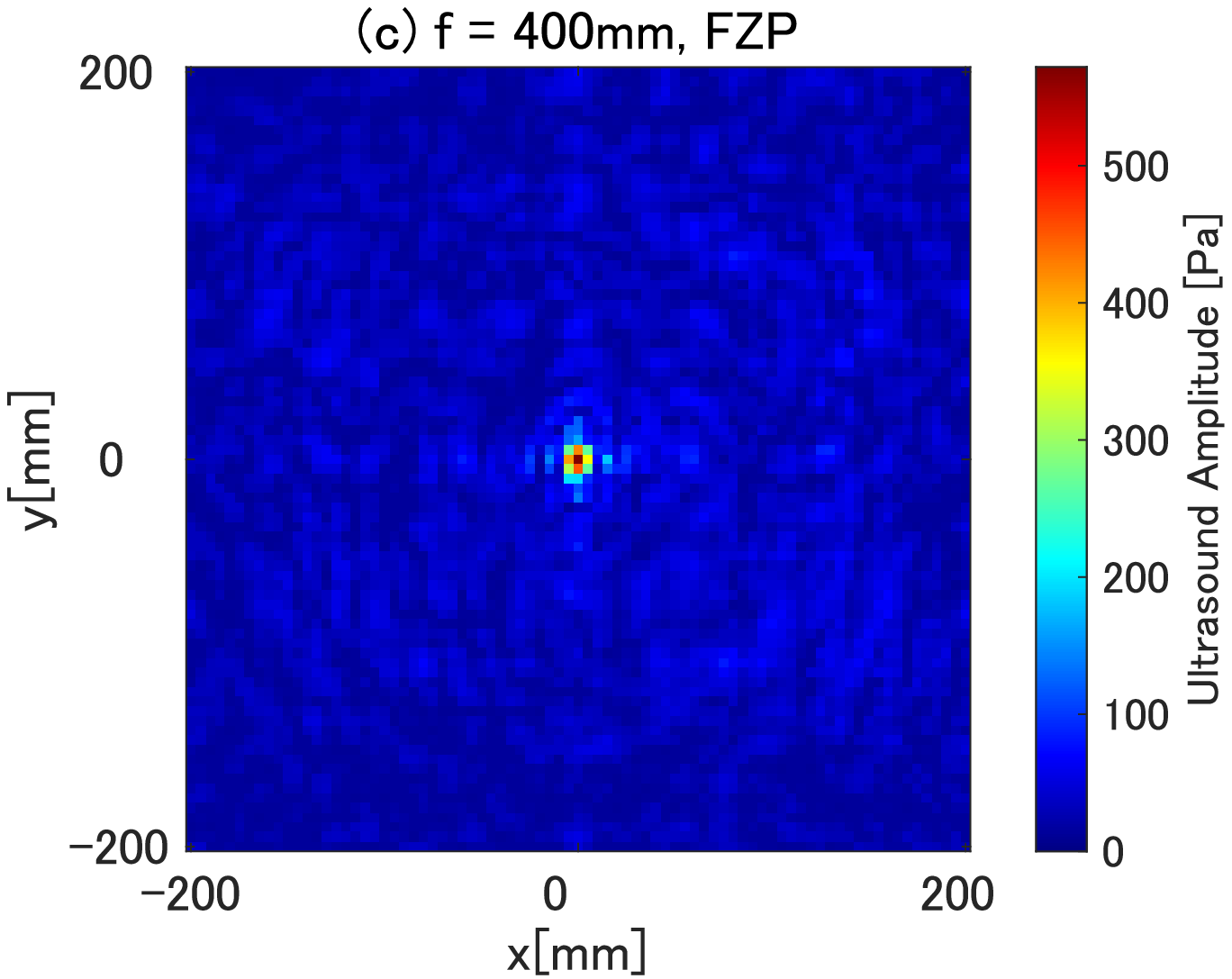}
\includegraphics[width=2.5in]{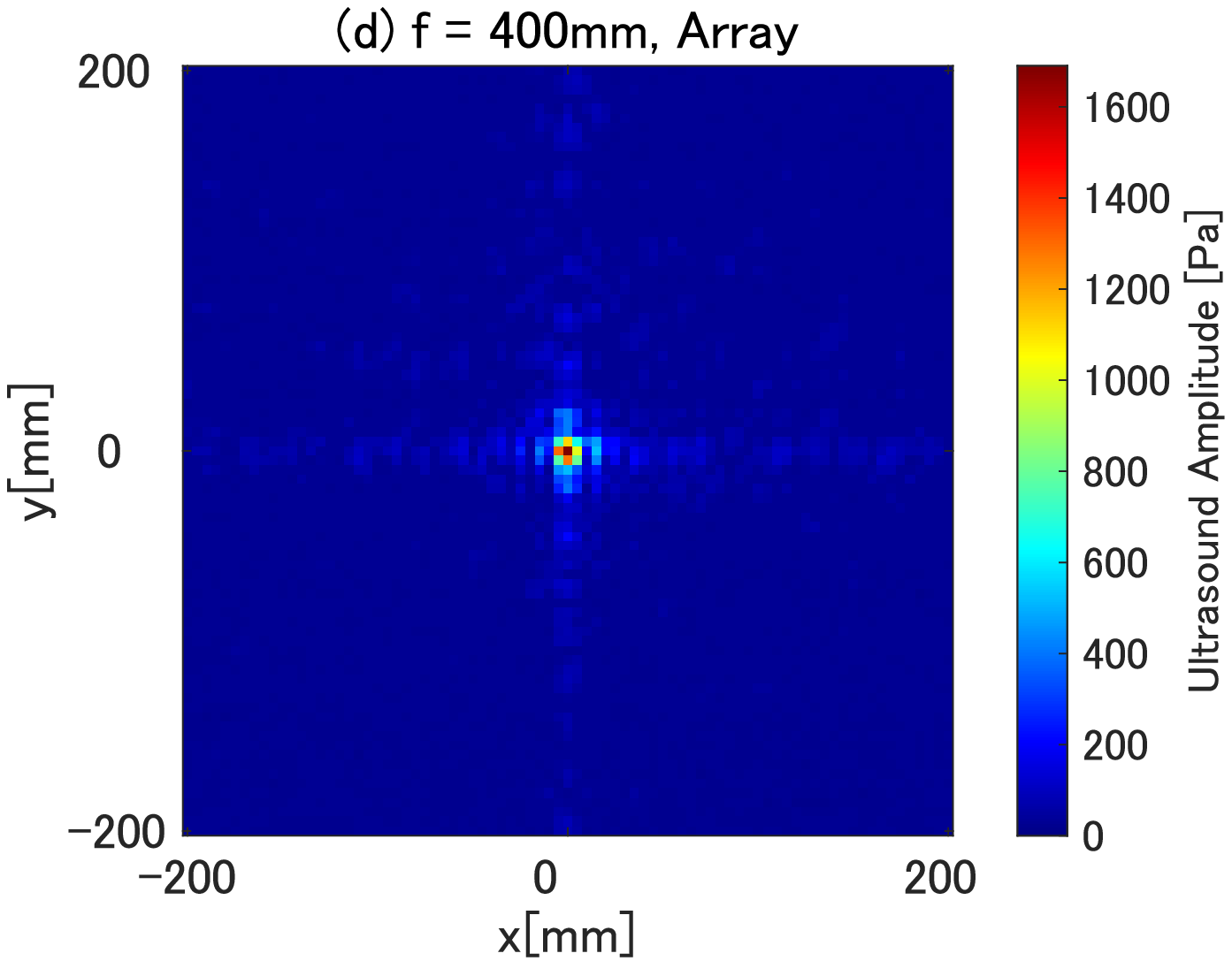}
\includegraphics[width=2.5in]{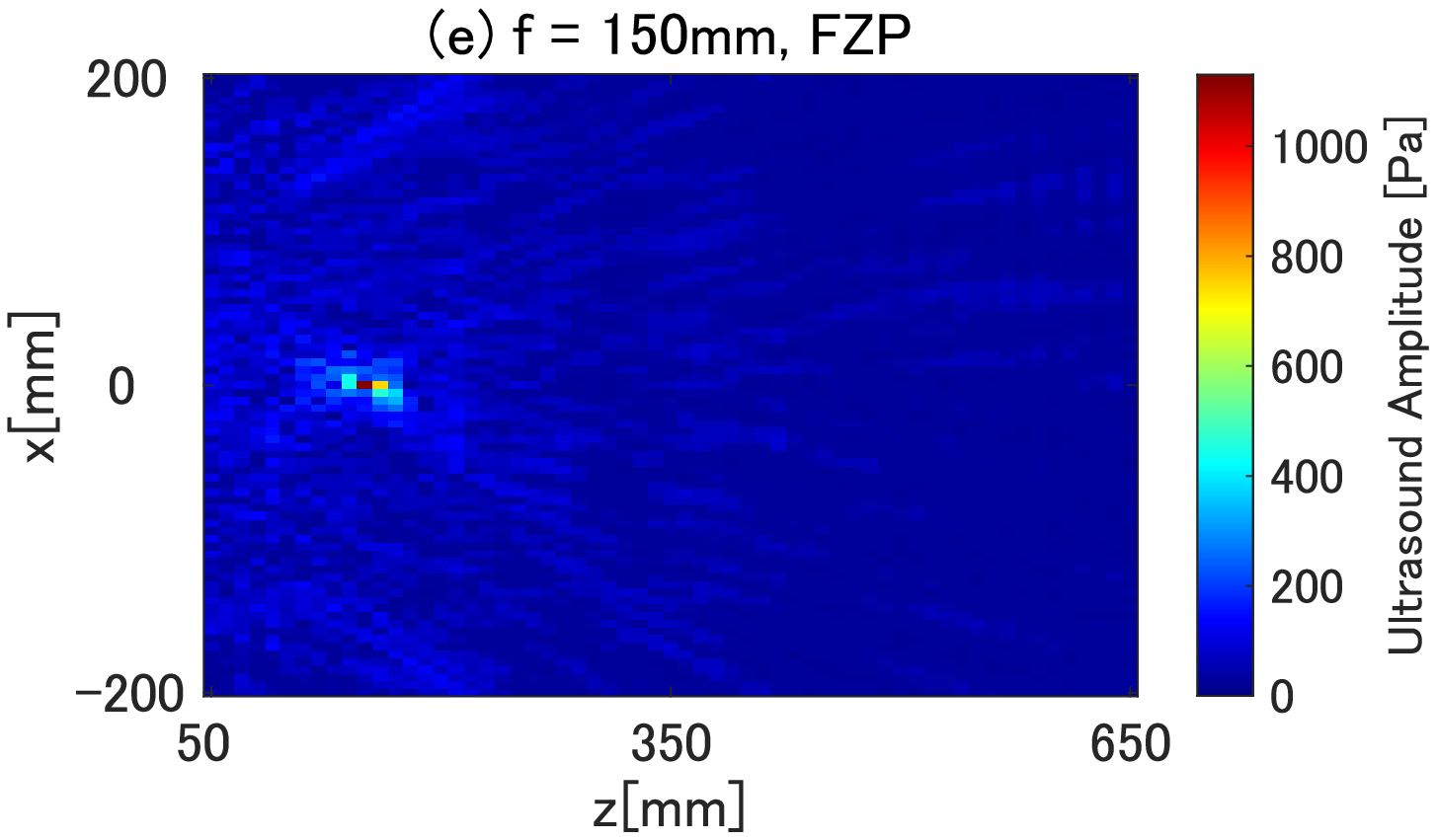}
\includegraphics[width=2.5in]{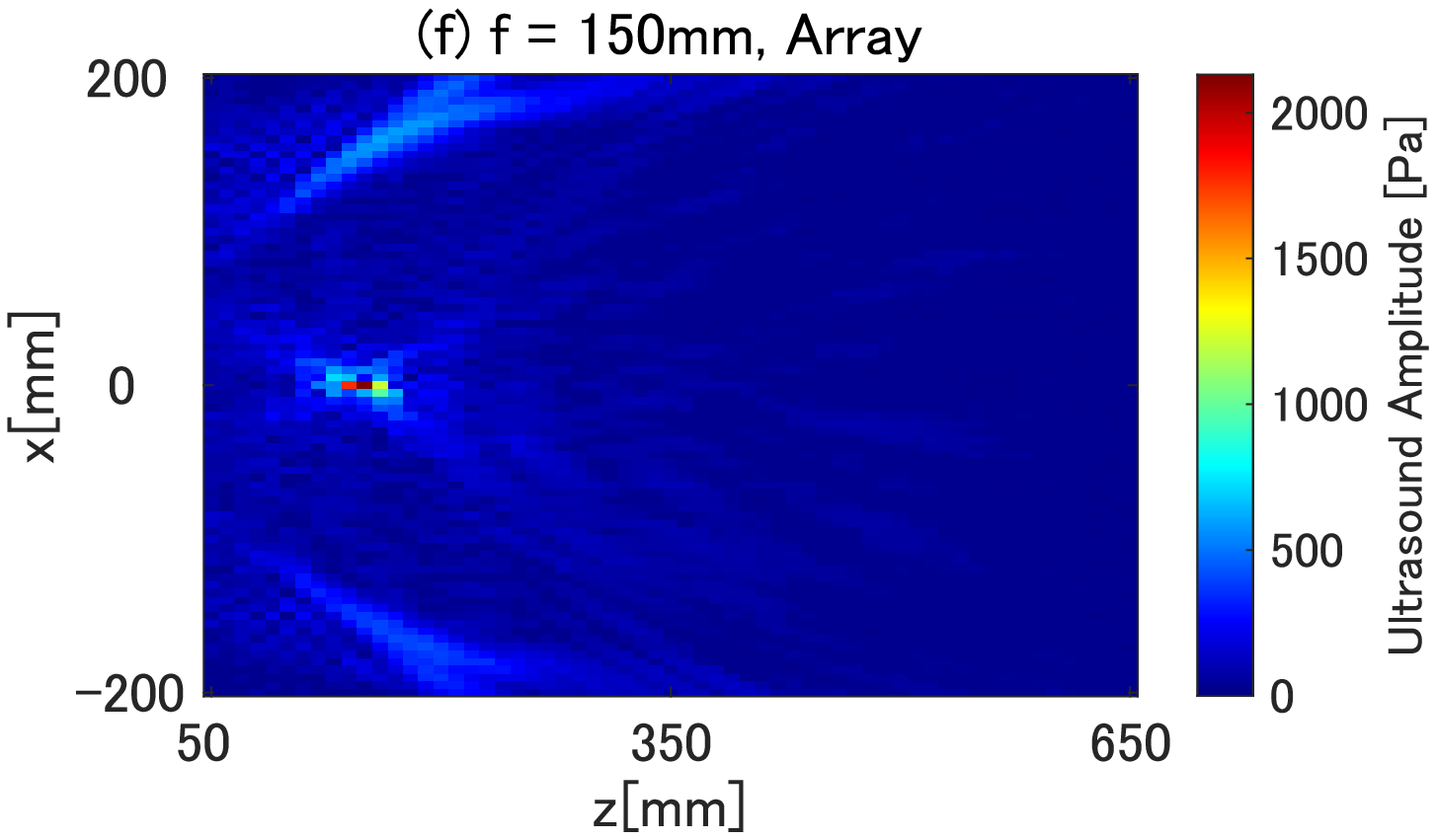}
\includegraphics[width=2.5in]{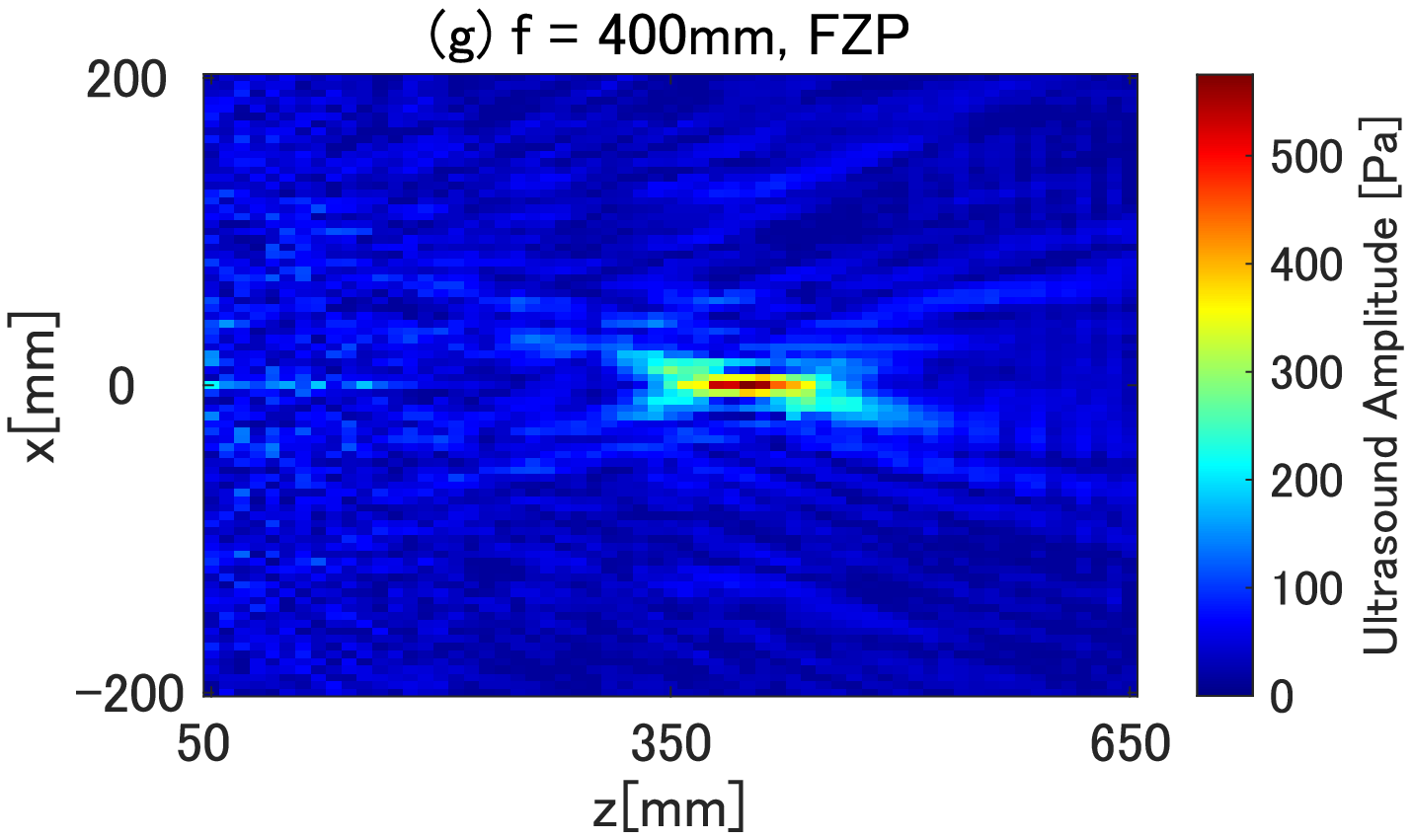}
\includegraphics[width=2.5in]{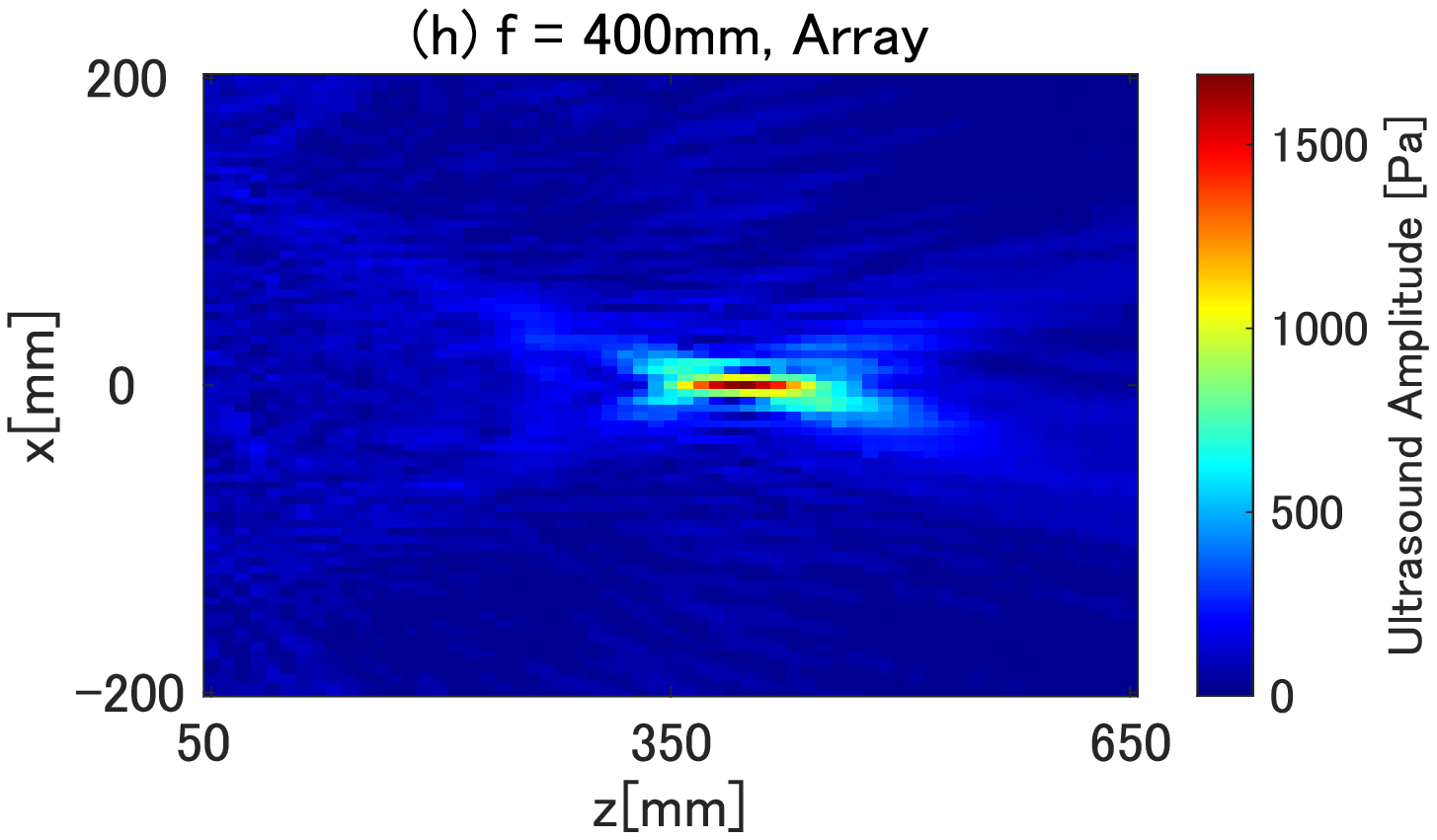}
\caption{\label{fig:measurements}Measured ultrasound amplitude distribution when focus was formed by the FZPs (left column) or the phase controlling of transducer arrays (right column), for the focal depth of 150~mm ((a), (b), (e), and (f)) and 400~mm ((c), (d), (g), and (h)). }
\baselineskip=11pt
\end{figure*}

\begin{figure*}[t]
\includegraphics[width=3in]{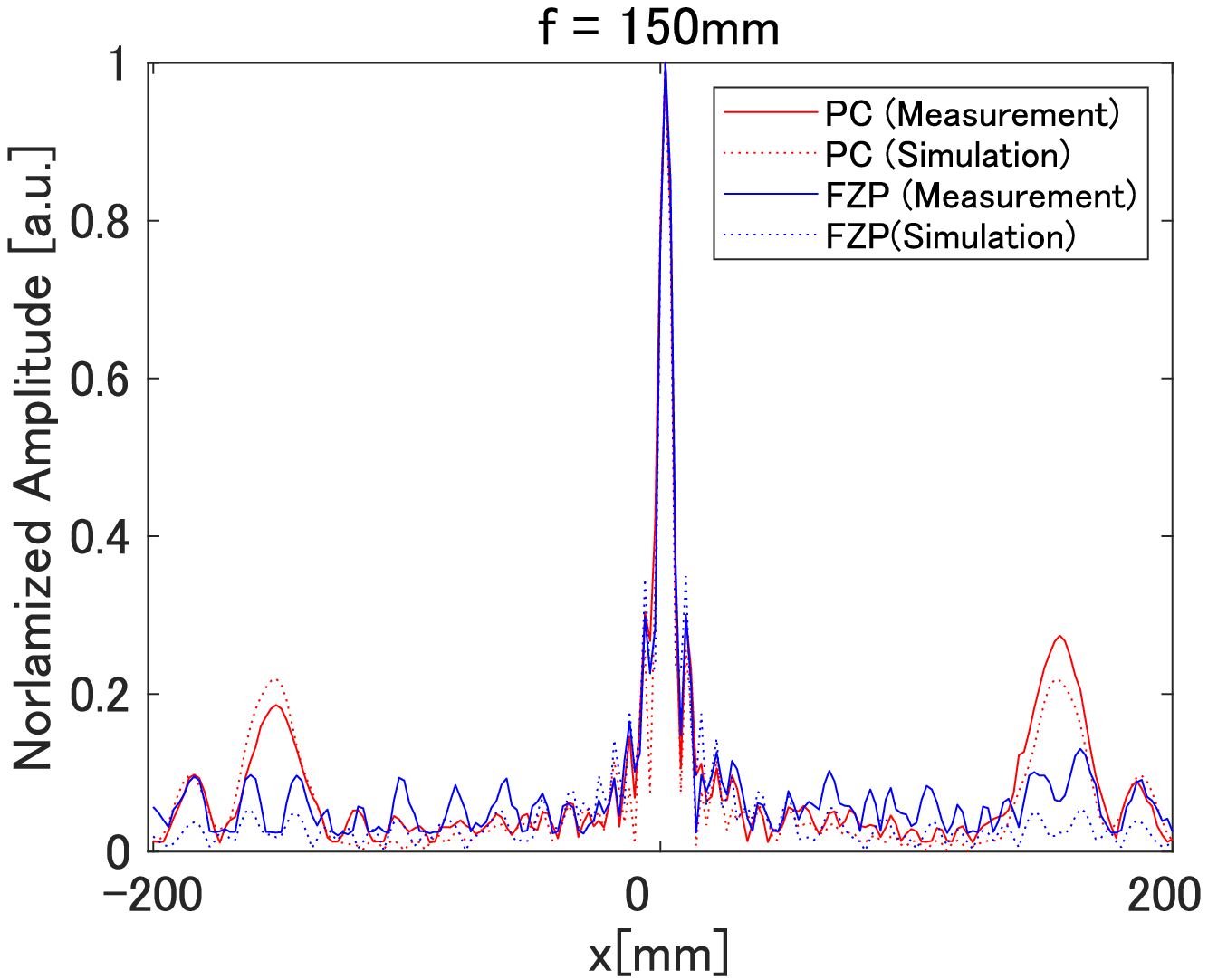}
\includegraphics[width=3in]{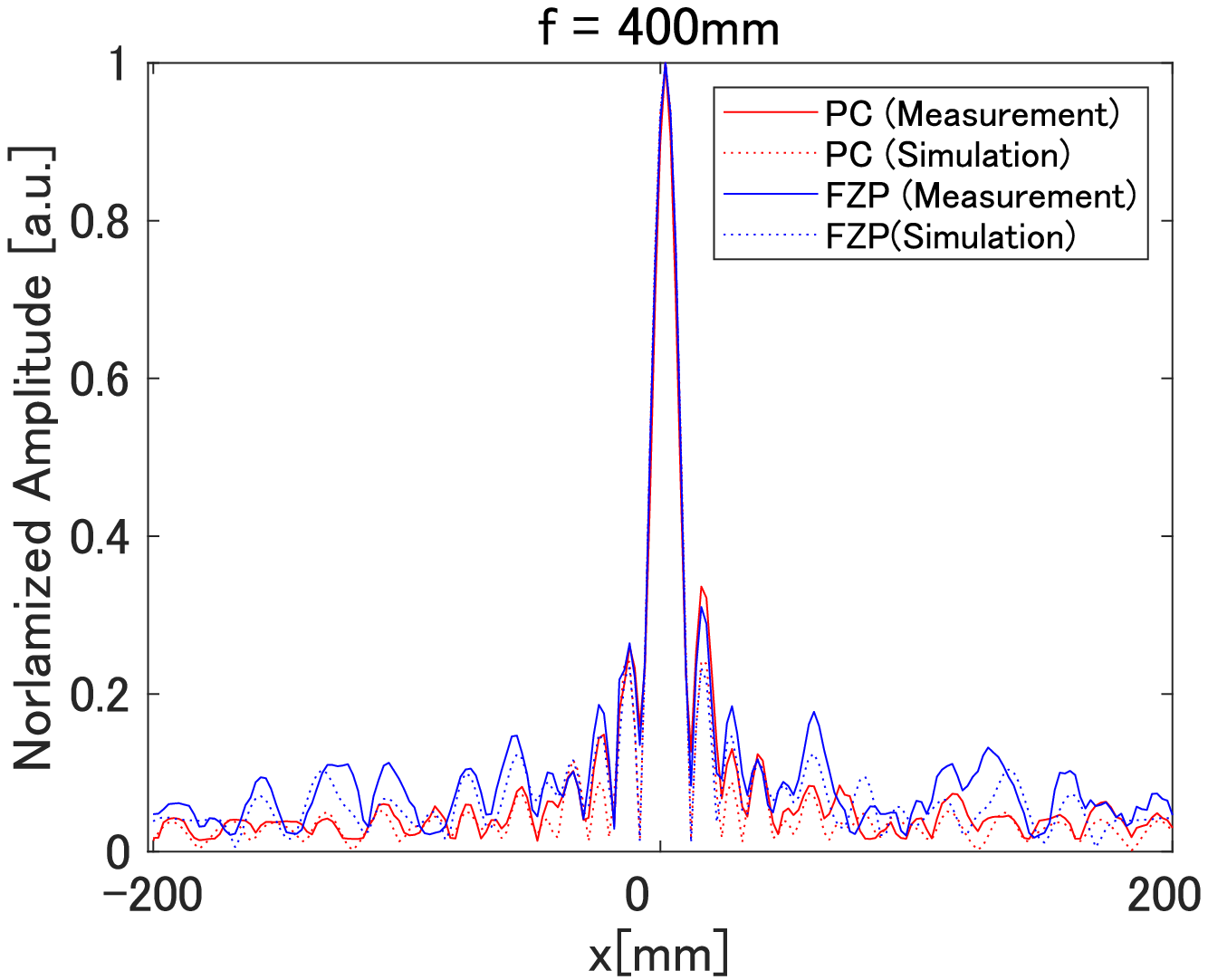}
\includegraphics[width=3in]{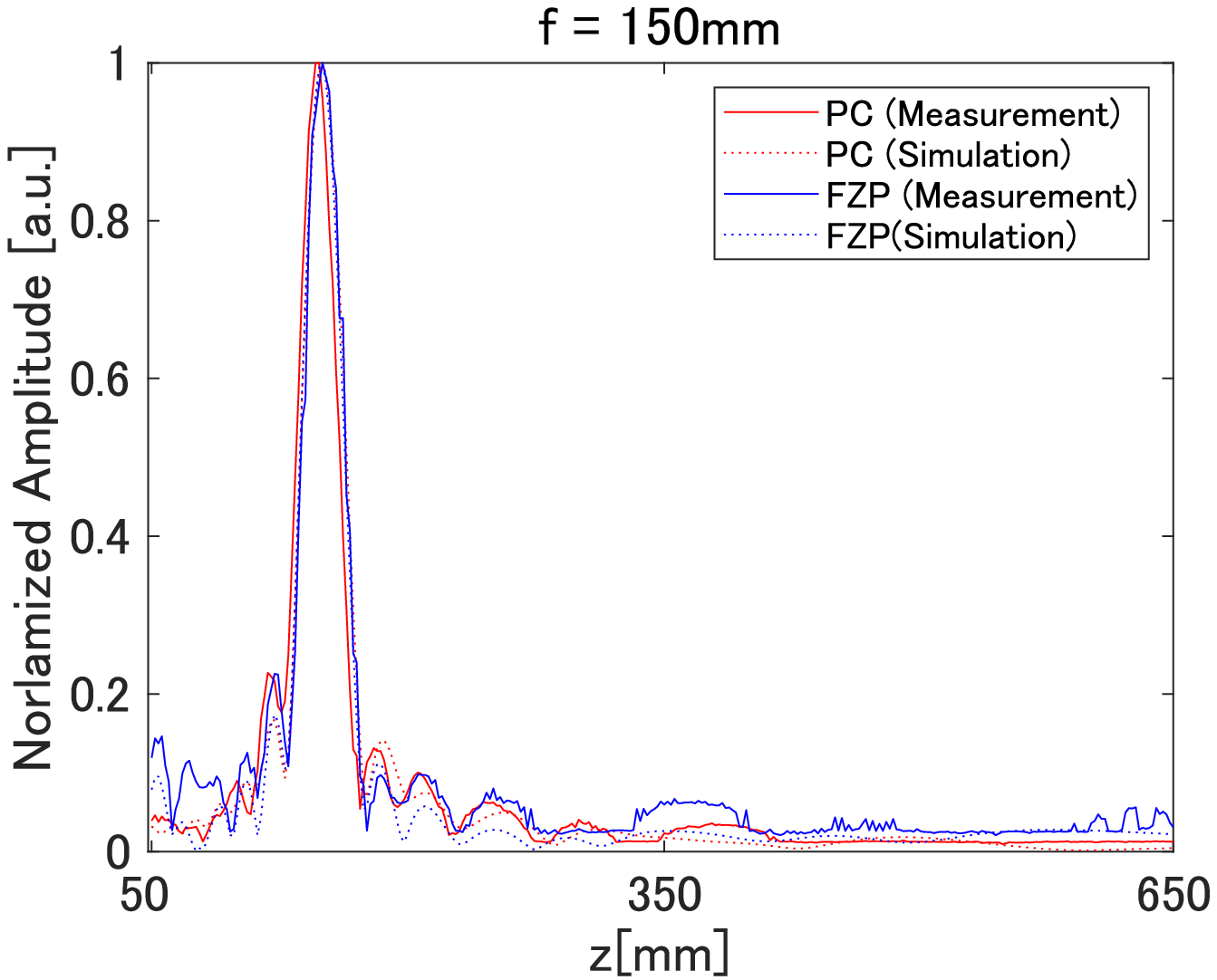}
\includegraphics[width=3in]{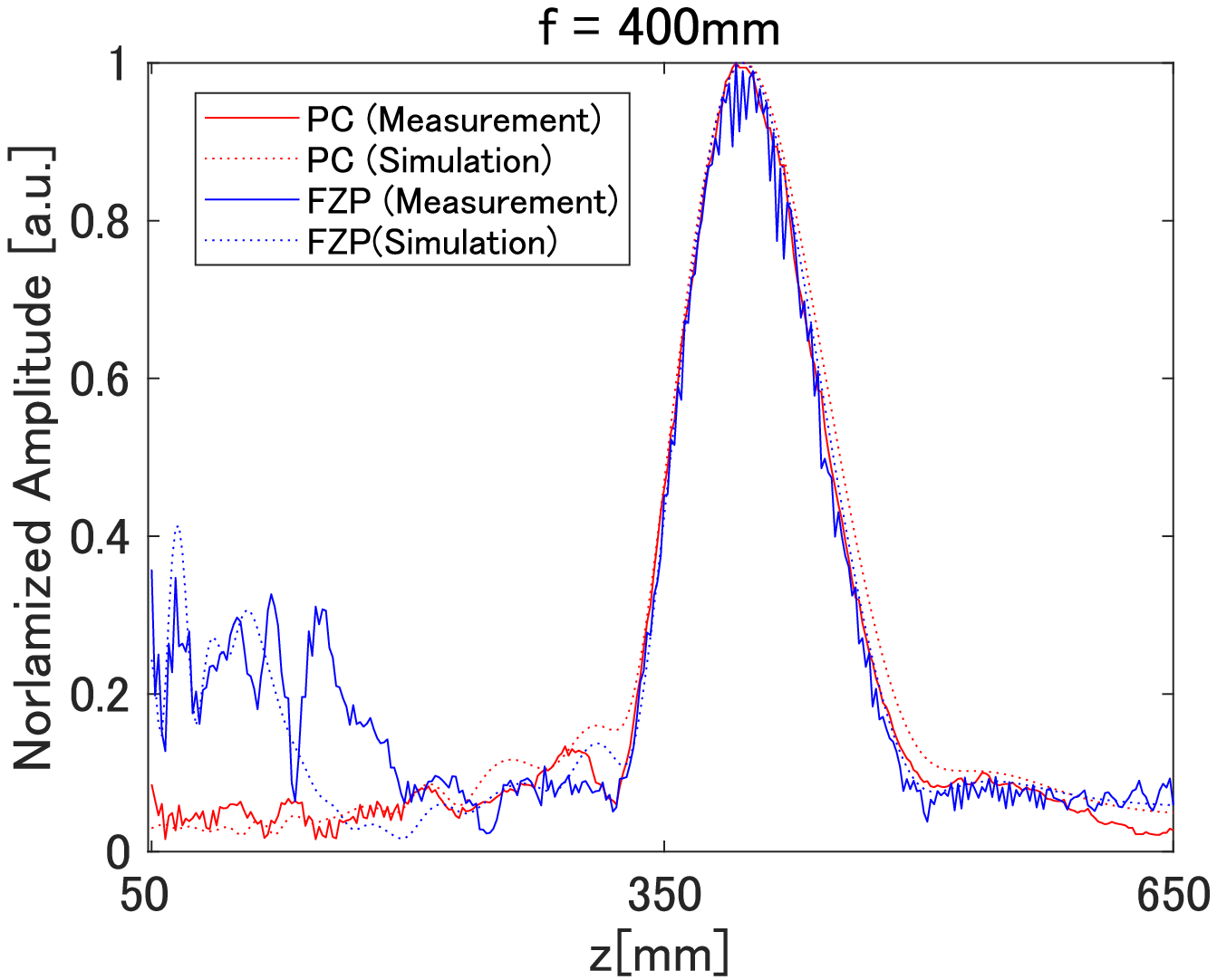}
\caption{\label{fig:measurementsline}Measured and simulated acoustic amplitude distributions normalized by the maximum amplitude of each graph, for the focal depth of 150~mm (left column) and 400~mm (right column), on the line parallel to the $x$-axis at the focal depth (upper row) and on the $z$-axis (lower row). "PC" indicates the case with phase-controlled transducers.}
\baselineskip=11pt
\raggedright
\end{figure*}

Figure \ref{fig:measurements} illustrates the results of the sound field measurements with the FZPs.
As in the simulation experiment, it was confirmed that the generation of grating lobes was suppressed by the FZP with focal depth of 150~mm. 
Figure \ref{fig:measurementsline} illustrates spatially finer measurement results with 2~mm scanning intervals along the $x$- and $z$- axes with amplitude distributions obtained by the numerical simulation, normalized by the output obtained with the maximum value of each case. 
The graphs indicate that the spatial profile of the amplitude distributions indicate good agreement with those predicted by numerical simulations.
The sizes of the focus with both cases with the FZP and phase-controlled transducers were comparable.

\begin{table}[t]
\begin{tabular}{|l|l|l|}
\hline
Focal Depth & \begin{tabular}[c]{@{}l@{}}Amplitude Ratio\\ (Simulation)\end{tabular} & \begin{tabular}[c]{@{}l@{}}Amplitude Ratio\\ (Measurement)\end{tabular} \\ \hline
150 mm      & 59.1\%                                                                 & 52.1\%                                                                  \\ \hline
400 mm      & 38.2\%                                                                 & 33.3\%                                                                  \\ \hline
\end{tabular}
\caption{Focal amplitude ratio of the FZP case to the phase-controlled transducer case. }
\label{tab:table1}
\end{table}
At the same time, The focal amplitudes with the FZPs clearly reduced when compared with that generated by the phase-controlled transducers (Table 1).
Although the focal power efficiency was reduced, compared with the conventional phased array method, we consider it was still valid for most of the current mid-air ultrasound use. 
We confirmed that aerial vibrotactile presentation, one of such prevalent applications, can be achieved.
By applying amplitude modulation at 150 Hz as is often done in airborne ultrasound presentation systems, distinct pinpoint tactile stimuli could be felt by placing bare hands over the ultrasound focus generated by the FZPs for the possible maximum ultrasound output from the transducers.
From this result, it is expected that FZP-based ultrasound manipulation can be applied in other scenarios that have been demonstrated by precedent studies.

\subsection{The mobility of the focus in binary holograms and multi-focusing}
\begin{figure}[t]
\includegraphics[width=3in]{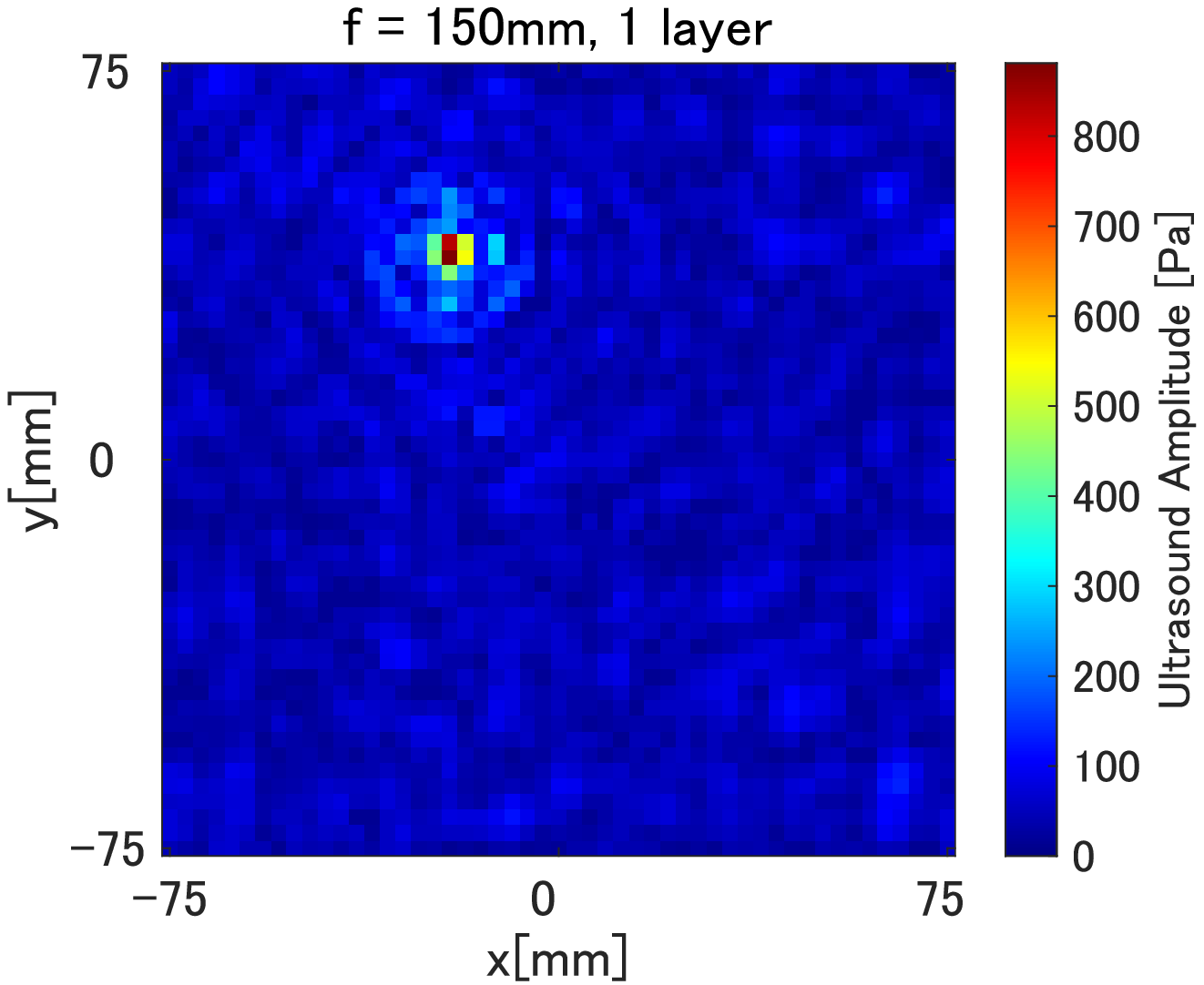}
\includegraphics[width=3in]{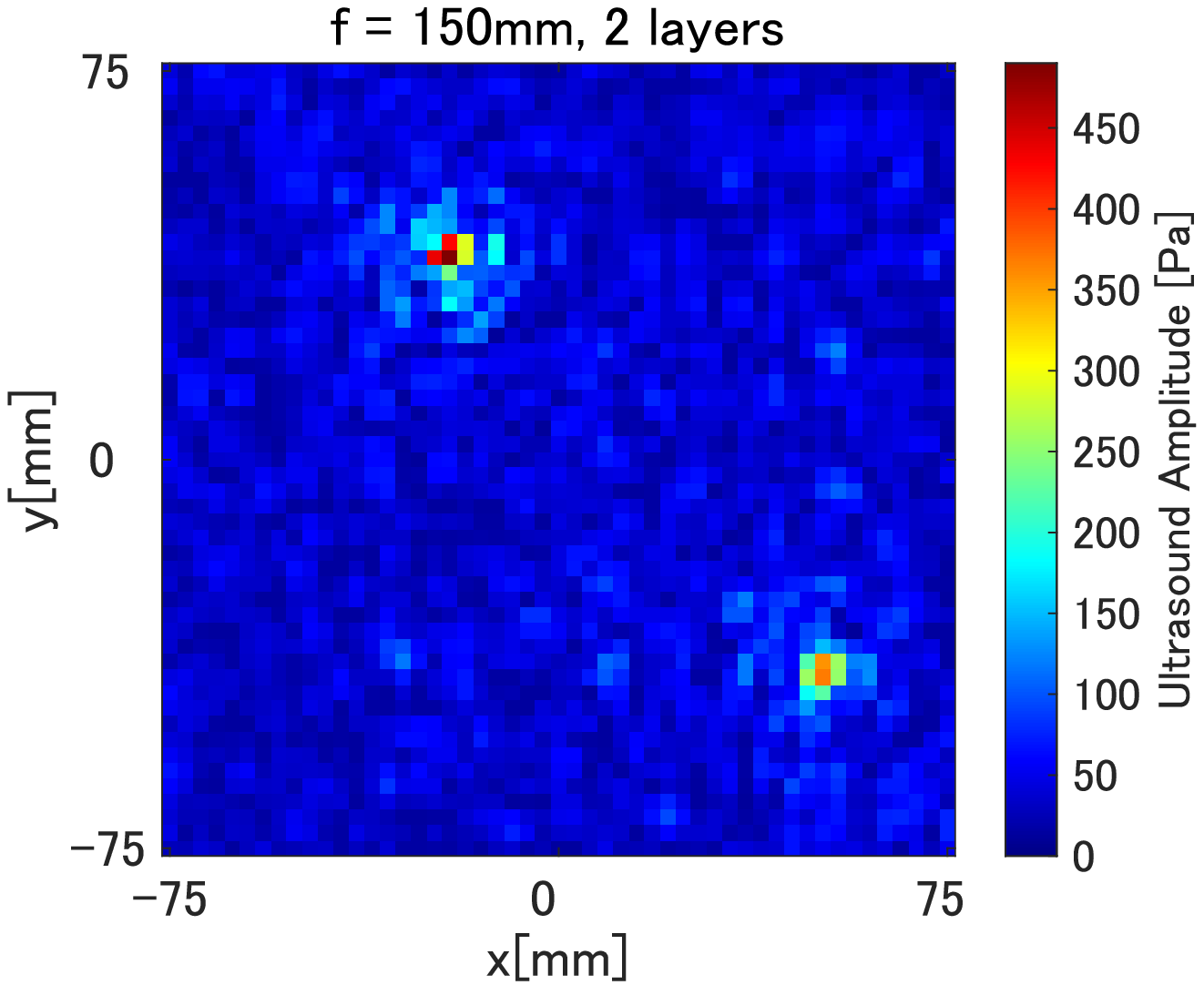}
\caption{\label{fig:measurements2}Measured acoustic amplitude distribution with shifted single FZP (Ueft figure) and two layers of FZPs (lower figure).}
\baselineskip=11pt
\raggedright
\end{figure}

Furthermore, we placed an FZP with focal depth of 150~mm on the in-phase transducer array.
We shifted the center of the FZPs apart from the center of the arrays as the previous numerical simulations.
We performed the sound field measurement as in the previous experiments and observed that the focus moved to the same position as the center of the FZP.
Then, we placed another 150~mm-focal-depth FZP on that FZP and observed that two ultrasound foci were simultaneously generated at positions corresponding to the centers of both FZPs (Fig.\ref{fig:measurements2}). 

As observed in the numerical simulations, the experiment also demonstrates that the focal movement by only translational movement of FZP is realized, which is much easier than moving the entire sound sources toward desired focal positions. 
Simultaneously, the power of foci were approximately halved from that with a single focus case.
Regarding the case with two foci, the relative power of unwanted acoustic emission around the focal region was stronger than that observed by the single-layer-FZP case.
For this case, vibrotactile stimulation was faintly felt on the palm with 150~Hz amplitude modulation applied to the possible maximum ultrasound output from the transducers.

\section{Discussions}
\subsection{Focusing performance for varied focal depths} 
In both simulations and measurements, it is observed that the main lobe width of the ultrasound foci were comparable between the cases with FZPs and phase-controlled transducers.
Based on this fact, out of the focal amplitude ratio in Table 1, the relative focusing efficacy of the FZP compared to that of phase-controlled transducer array can be roughly estimated because the amplitude distributions around the foci were almost identical for the both cases.
According to the precedent study supposing that the focal sound intensity is in proportion to the square of the focal amplitude \cite{Tarrazo-Serrano2019}, the relative focusing efficacy of FZP with the focusing depth of 150~mm was 34.9~\% in numerical simulation and 27.1~\% in measurement, compared with that of phase-controlled transducers.
With the focusing depth of 400~mm, the relative focusing efficacy of FZP compared to the phase-controlled transducers was 14.6~\% in numerical simulation and 11.1~\% in measurement.
In the preceding studies, the phase-controlling version of the FZP, which causes no ultrasound blockage on the plate was studied \cite{Tarrazo-Serrano2019, Mortezaie1984}.
Such devices have a physical effect similar to the phase-controlled transducer arrays in generating ultrasound foci, showing better focusing efficacy.
In that study, the focusing efficacy of the phase-controlled FZP was approximately four times as great as the amplitude-controlled FZP, which was similar to the results in our experiments with the focusing depth of 150~mm.
At the same time, the past research did not handle the issue of grating lobes because no periodic amplitude gaps in the source plane were taken into account.

Thus, the relative focusing efficacy with a longer focal depth gets smaller, which can be attributed to the fact that the “rings” on the FZP get more distant from one another for a longer focal depth, resulting in a smaller number of rings existing in a finite FZP aperture. 
For the most extreme case where the focus is fairly far from the FZP, there may be only one open circle in the FZP.
In that case, the resulting ultrasound field is equivalent to that with a windowed planar emission, where no longer proper focusing is expected.
In the same situation, a driving phase distribution on the emission plane is realized with the phase-controlled transducer case, which is still capable of forming a focus.
 
The generation of grating lobes by the phase-controlled transducers was suppressed with the use of FZP, at the cost of unintended ultrasound emission around the focal region.
 This is because of the emission pattern with the FZP being finer than the wavelength, unlike that with phase-controlled transducers inevitably becoming coarser owing to the transducer size. 
However, such strong grating lobes are not observed in the case with the focal depth of 400~mm in the measurement area.
The grating lobes exist in a region more apart from the focus with a greater focal depth.
Therefore, in the cases where the grating lobes are so far from the focus that they can be neglected, the phase-controlled scheme outperforms the FZP-based method owing to less unintended acoustic emissions. 
 
The intervals of rings on the FZP also depend on the wavelength.
 As the wavelength decreases, the central circle becomes smaller, and more rings exist in the FZP, which is expected to improve sound collection performance. 
At the same time, it becomes much more difficult to decrease the dimension of the phase-controlled transducers to realize proper focusing.
Therefore, FZP-based focusing scheme will be more suitable for ultrasound emission with a higher frequency.

\subsection{Effect of FZP thickness}
In the numerical experiment, the thickness of the FZPs were not considered.
In practice, direct waves to the focus were expected to be partially blocked off by the thickness of FZPs.
The smaller the elevation angle from the sound source to the focus is, the greater the part of sound waves is blocked, because the sides of the FZP rings serve as walls.
We utilized 2~mm-thick acrylic plates for fabricating FZPs in our experiments, with which the blockage percentage by the FZP walls was estimated to be several percent compared to the case with FZP with no thickness.
The error in focal amplitude ratio of the FZP cases to the phase-controlled transducer cases between the numerical and experimental results may partially be explained by this effect.

With an experiment where two FZPs were stacked to create two individual foci, the amplitudes of the foci were less than that of the single focus created with one FZP.
This is presumably because one FZP blocked the positive contribution to the focus from the other FZP, and the valid area of the planar sound source became considerably smaller.
Another factor for this power reduction may be the total thickness of the two FZP layers being 4~mm, which might have caused more blocking-off effect of direct waves and complicated sound reflections between the layers.
 
\subsection{Variations of driving frequencies}
FZP-based focusing can achieve finer spatial resolution than the conventional phased-array technique.
This resolution gap between the two methods is more prominent for a higher ultrasound frequency, because of the difficulty in downsizing transducers of the corresponding resonant frequency.
However, the spatial resolution of FZP patterning can be readily improved, as there have been a great bunch of minute machining techniques including the laser cutting.
In addition, a higher frequency source results in more rings created in FZPs, which will contribute to a more proper focusing.
Therefore, our method can be effectively implemented as a form of miniature emission plane with a higher frequency source, as well as enlarged mid-air-ultrasound workspaces.
\begin{figure}[t]
\includegraphics[width=3in]{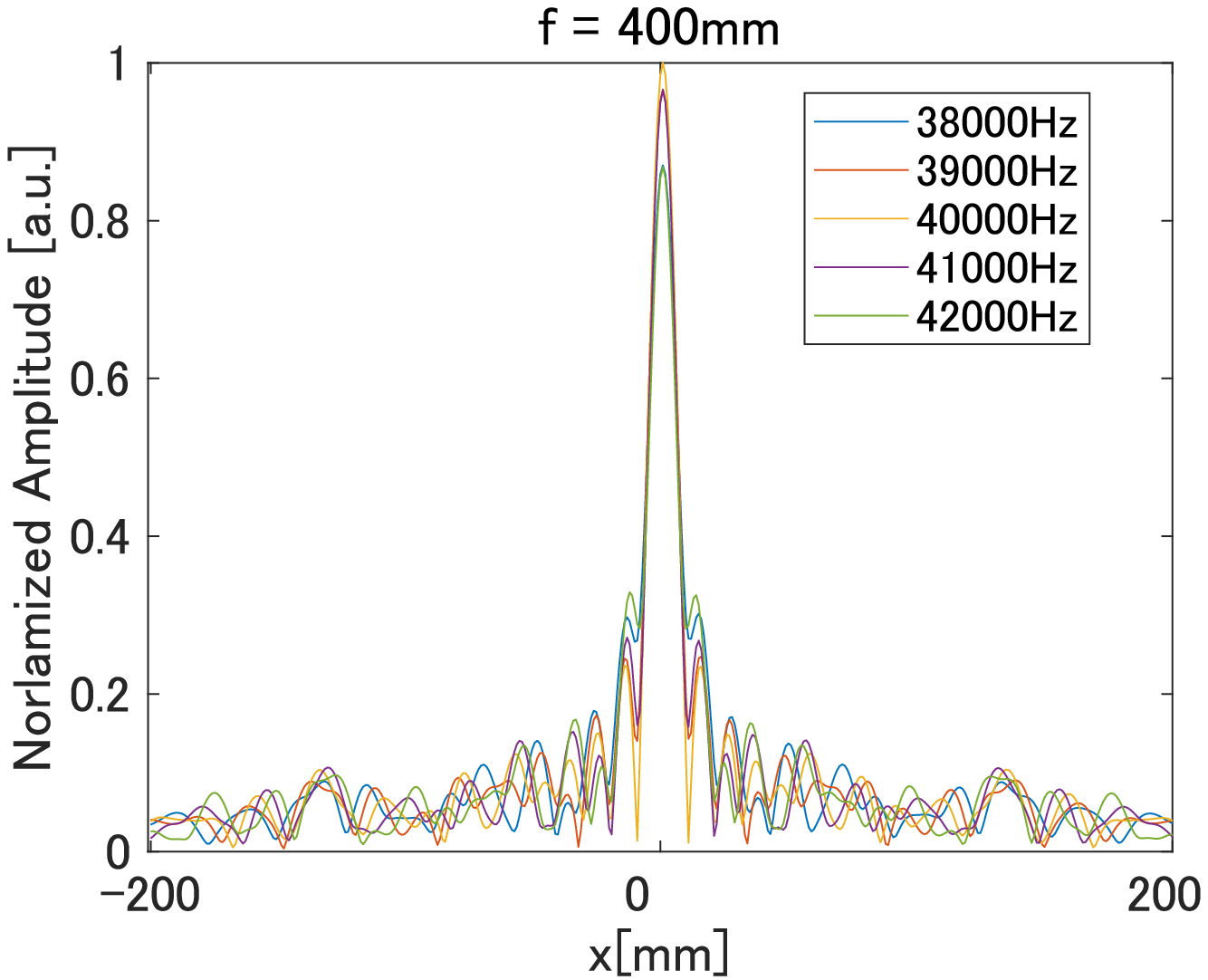}
\includegraphics[width=3in]{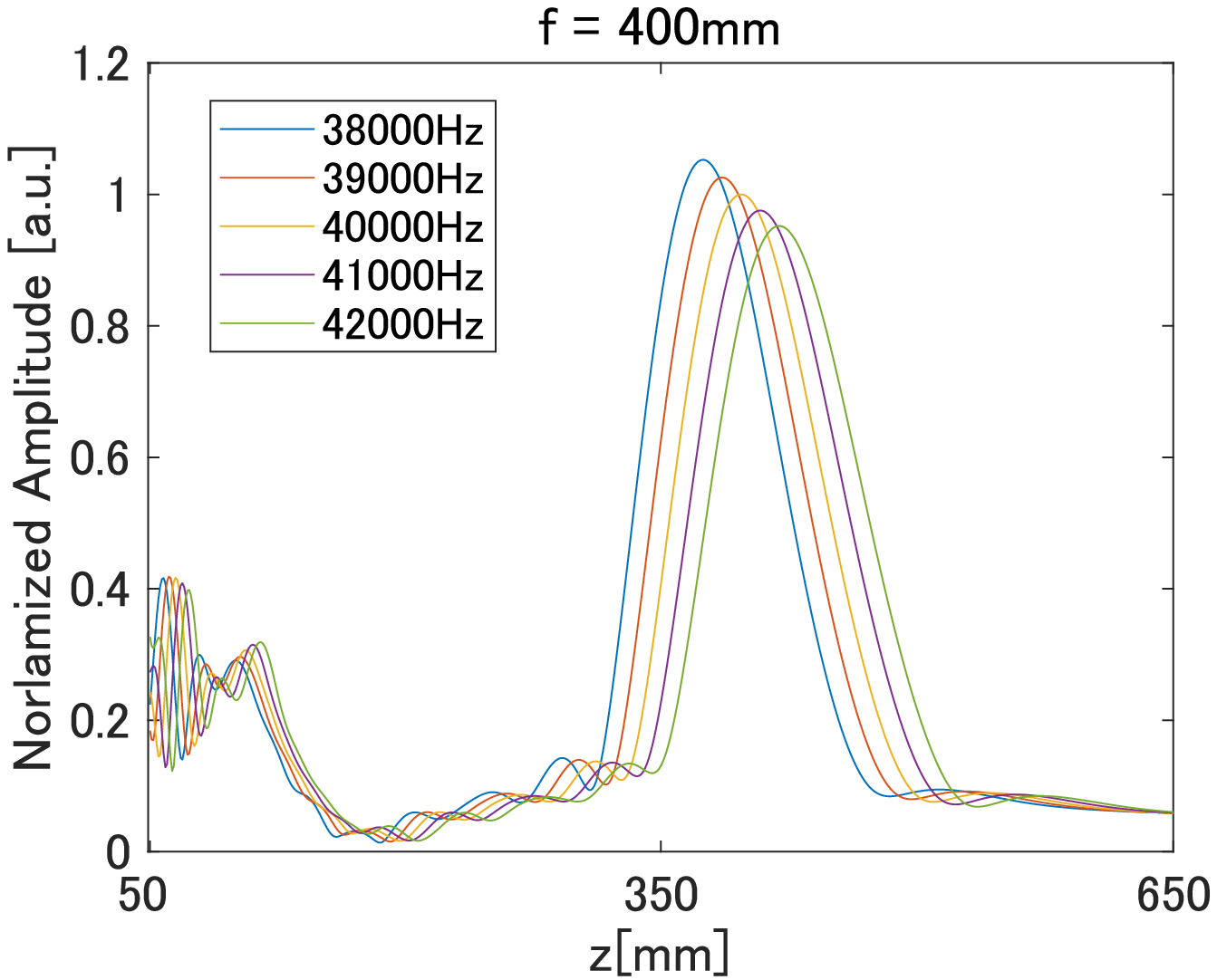}
\caption{\label{fig:freqvary}Numerically calculated amplitude distribution on the line parallel to the $x$-axis at the focal depth (upper) and on the $z$-axis (lower) by varying the driving frequency of the transducers. The graphs correspond to the case with the FZP fabricated for 40~kHz ultrasound focusing at 400~mm apart from the FZP.}
\end{figure}

\section{Frequency mismatches and misalignment between sources and FZPs}
An FZP for generating ultrasound focus is fabricated for its corresponding fixed driving frequency and focal depth.
Therefore, the frequency of the ultrasound sources should match that corresponds to the designing process of the FZP for the intended focusing.
We numerically investigated how the frequency mismatch between the sources and the FZP affects the focusing performance.
Figure \ref{fig:freqvary} shows the normalized ultrasound amplitude around the focal region.
The simulation results indicate that the resulting focal depth is affected by the frequency mismatch.
At the same time, focal amplitude at the intended depth becomes highest when no frequency mismatch between the FZP and the sources occurs.
On the other hand, the amplitude peak value at the focus with shifted depths becomes higher as the driving frequency is lowered.
This is presumably because the lowered driving frequencies cause the foci to be formed closer to the FZP, where the attenuation of ultrasound emissions becomes accordingly lower.

Next, we discuss the effect of vertical misalignment of FZP from the source surface.
As the FZP is located farther from the source surface, the input wave to the FZP layer ideally becomes more similar to the perfect plain waves for the infinitely large source aperture.
In this case, the effect of vertical misalignment just appears as the focal depth shift according to the distance between the source and the FZP when they are located parallel to each other.
However, when the source aperture is limited, a smaller portion of emitted wave travels through the FZP when they are located more apart from each other.
This can result in lowered focal amplitude in addition to the focal depth shift.
\section{Conclusions}
In this paper, a method of controlling ultrasound fields using an FZP amplitude binary mask on a plane wave source was developed. 
In the experiments, a 2~mm thick acrylic plate cut out by a laser cutting machine was utilized as a binary mask.
We evaluated focusing performances with our proposed method, where ultrasonic convergence occurs to the same degree of spatial resolution as in the case with a conventional method using phase-controlled transducers.
As a favorable side effect, FZP suppressed grating lobes observed with the conventional method.
We also determined that shifting the FZP over a fixed source can move the focus. 
Furthermore, multi-focusing was achieved by using layers of multiple FZPs with their centers corresponding to the focal positions.

Configuration of our method is very simple.
A planar source with its common driving voltage pulses and the FZPs can be made both inexpensive, thin, and readily implemented and upsized.
Our method is suitable for mid-air ultrasound control system with large apertures or systems using ultrasound sources with a higher frequency.
Implementation of large ultrasound aperture will lead to new large-scale mid-air ultrasonic application systems that employ the entire walls and ceilings as ultrasonic emission planes.

Subsequent challenges include electronical control of amplitude distribution over the source.
We consider that this scenario has the possibility to be realized, as it only needs on-off binary control of emission patterns, instead of inter-element phase control with the temporal preciseness less than one-tenth of the ultrasound period. We believe that such amplitude controlling mechanism requires less effort to be realized than that required to upsize the phased array by simply consolidating a large number of array units with a tremendous number of transducers individually phase-controlled.



\begin{acknowledgments}
This study was supported by JST, PRESTO Grant Number JPMJPR21R9, Japan.
\end{acknowledgments}

\section*{Author Declarations}
The authors have no conflicts of interest.

\section*{Author Contributions}
Masatake Kitano: Conceptualization (equal); Formal Analysis (lead); Investigation (equal); Methodology (supportive); Writing -- original draft (lead); 
Keisuke Hasegawa: Conceptualization (equal); Formal Analysis (supportive); Investigation (equal); Methodology (lead); Writing -- review and editing (lead); Supervision (lead); Funding Acquisition (lead); Resources(lead); Project Administration(lead);

\section*{Data Availability}
The data that support the findings of this study are available from the corresponding author upon reasonable request.

\bibliography{sampbib}

\begin{thebibliography}{41}%
\makeatletter
\providecommand \@ifxundefined [1]{%
 \@ifx{#1\undefined}
}%
\providecommand \@ifnum [1]{%
 \ifnum #1\expandafter \@firstoftwo
 \else \expandafter \@secondoftwo
 \fi
}%
\providecommand \@ifx [1]{%
 \ifx #1\expandafter \@firstoftwo
 \else \expandafter \@secondoftwo
 \fi
}%
\providecommand \natexlab [1]{#1}%
\providecommand \enquote  [1]{``#1''}%
\providecommand \bibnamefont  [1]{#1}%
\providecommand \bibfnamefont [1]{#1}%
\providecommand \citenamefont [1]{#1}%
\providecommand \href@noop [0]{\@secondoftwo}%
\providecommand \href [0]{\begingroup \@sanitize@url \@href}%
\providecommand \@href[1]{\@@startlink{#1}\@@href}%
\providecommand \@@href[1]{\endgroup#1\@@endlink}%
\providecommand \@sanitize@url [0]{\catcode `\\12\catcode `\$12\catcode
  `\&12\catcode `\#12\catcode `\^12\catcode `\_12\catcode `\%12\relax}%
\providecommand \@@startlink[1]{}%
\providecommand \@@endlink[0]{}%
\providecommand \url  [0]{\begingroup\@sanitize@url \@url }%
\providecommand \@url [1]{\endgroup\@href {#1}{\urlprefix }}%
\providecommand \urlprefix  [0]{URL }%
\providecommand \Eprint [0]{\href }%
\providecommand \doibase [0]{http://dx.doi.org/}%
\providecommand \selectlanguage [0]{\@gobble}%
\providecommand \bibinfo  [0]{\@secondoftwo}%
\providecommand \bibfield  [0]{\@secondoftwo}%
\providecommand \translation [1]{[#1]}%
\providecommand \BibitemOpen [0]{}%
\providecommand \bibitemStop [0]{}%
\providecommand \bibitemNoStop [0]{.\EOS\space}%
\providecommand \EOS [0]{\spacefactor3000\relax}%
\providecommand \BibitemShut  [1]{\csname bibitem#1\endcsname}%
\let\auto@bib@innerbib\@empty
\bibitem [{\citenamefont {Strutt}(1884)}]{Strutt1884}%
  \BibitemOpen
  \bibfield  {author} {\bibinfo {author} {\bibfnamefont {J.~W.}\ \bibnamefont
  {Strutt}},\ }\bibfield  {title} {\enquote {\bibinfo {title} {I. on the
  circulation of air observed in kundt’s tubes, and on some allied acoustical
  problems},}\ }\href {\doibase 10.1098/rstl.1884.0002} {\bibfield  {journal}
  {\bibinfo  {journal} {Philosophical Transactions of the Royal Society of
  London}\ }\textbf {\bibinfo {volume} {175}},\ \bibinfo {pages} {1--21}
  (\bibinfo {year} {1884})},\ \Eprint
  {http://arxiv.org/abs/https://royalsocietypublishing.org/doi/pdf/10.1098/rstl.1884.0002}
  {https://royalsocietypublishing.org/doi/pdf/10.1098/rstl.1884.0002}
  \BibitemShut {NoStop}%
\bibitem [{\citenamefont {King}(1934)}]{King1934}%
  \BibitemOpen
  \bibfield  {author} {\bibinfo {author} {\bibfnamefont {L.~V.}\ \bibnamefont
  {King}},\ }\bibfield  {title} {\enquote {\bibinfo {title} {On the acoustic
  radiation pressure on spheres},}\ }\href {\doibase 10.1098/rspa.1934.0215}
  {\bibfield  {journal} {\bibinfo  {journal} {Proceedings of the Royal Society
  of London. Series A - Mathematical and Physical Sciences}\ }\textbf {\bibinfo
  {volume} {147}},\ \bibinfo {pages} {212--240} (\bibinfo {year} {1934})},\
  \Eprint
  {http://arxiv.org/abs/https://royalsocietypublishing.org/doi/pdf/10.1098/rspa.1934.0215}
  {https://royalsocietypublishing.org/doi/pdf/10.1098/rspa.1934.0215}
  \BibitemShut {NoStop}%
\bibitem [{\citenamefont {Westervelt}(1951)}]{Westervelt1951}%
  \BibitemOpen
  \bibfield  {author} {\bibinfo {author} {\bibfnamefont {P.~J.}\ \bibnamefont
  {Westervelt}},\ }\bibfield  {title} {\enquote {\bibinfo {title} {The theory
  of steady forces caused by sound waves},}\ }\href {\doibase
  10.1121/1.1906764} {\bibfield  {journal} {\bibinfo  {journal} {The Journal of
  the Acoustical Society of America}\ }\textbf {\bibinfo {volume} {23}},\
  \bibinfo {pages} {312--315} (\bibinfo {year} {1951})},\ \Eprint
  {http://arxiv.org/abs/https://doi.org/10.1121/1.1906764}
  {https://doi.org/10.1121/1.1906764} \BibitemShut {NoStop}%
\bibitem [{\citenamefont {Westervelt}(1953)}]{Westervelt1953}%
  \BibitemOpen
  \bibfield  {author} {\bibinfo {author} {\bibfnamefont {P.~J.}\ \bibnamefont
  {Westervelt}},\ }\bibfield  {title} {\enquote {\bibinfo {title} {The theory
  of steady rotational flow generated by a sound field},}\ }\href {\doibase
  10.1121/1.1907009} {\bibfield  {journal} {\bibinfo  {journal} {The Journal of
  the Acoustical Society of America}\ }\textbf {\bibinfo {volume} {25}},\
  \bibinfo {pages} {60--67} (\bibinfo {year} {1953})},\ \Eprint
  {http://arxiv.org/abs/https://doi.org/10.1121/1.1907009}
  {https://doi.org/10.1121/1.1907009} \BibitemShut {NoStop}%
\bibitem [{\citenamefont {Hoshi}\ \emph {et~al.}(2010)\citenamefont {Hoshi},
  \citenamefont {Takahashi}, \citenamefont {Iwamoto},\ and\ \citenamefont
  {Shinoda}}]{Hoshi2010}%
  \BibitemOpen
  \bibfield  {author} {\bibinfo {author} {\bibfnamefont {T.}~\bibnamefont
  {Hoshi}}, \bibinfo {author} {\bibfnamefont {M.}~\bibnamefont {Takahashi}},
  \bibinfo {author} {\bibfnamefont {T.}~\bibnamefont {Iwamoto}}, \ and\
  \bibinfo {author} {\bibfnamefont {H.}~\bibnamefont {Shinoda}},\ }\bibfield
  {title} {\enquote {\bibinfo {title} {Noncontact tactile display based on
  radiation pressure of airborne ultrasound},}\ }\href {\doibase
  10.1109/TOH.2010.4} {\bibfield  {journal} {\bibinfo  {journal} {IEEE
  Transactions on Haptics}\ }\textbf {\bibinfo {volume} {3}},\ \bibinfo {pages}
  {155--165} (\bibinfo {year} {2010})}\BibitemShut {NoStop}%
\bibitem [{\citenamefont {Takahashi}, \citenamefont {Hasegawa},\ and\
  \citenamefont {Shinoda}(2020)}]{Takahashi2020}%
  \BibitemOpen
  \bibfield  {author} {\bibinfo {author} {\bibfnamefont {R.}~\bibnamefont
  {Takahashi}}, \bibinfo {author} {\bibfnamefont {K.}~\bibnamefont {Hasegawa}},
  \ and\ \bibinfo {author} {\bibfnamefont {H.}~\bibnamefont {Shinoda}},\
  }\bibfield  {title} {\enquote {\bibinfo {title} {Tactile stimulation by
  repetitive lateral movement of midair ultrasound focus},}\ }\href {\doibase
  10.1109/TOH.2019.2946136} {\bibfield  {journal} {\bibinfo  {journal} {IEEE
  Transactions on Haptics}\ }\textbf {\bibinfo {volume} {13}},\ \bibinfo
  {pages} {334--342} (\bibinfo {year} {2020})}\BibitemShut {NoStop}%
\bibitem [{\citenamefont {Frier}\ \emph {et~al.}(2018)\citenamefont {Frier},
  \citenamefont {Ablart}, \citenamefont {Chilles}, \citenamefont {Long},
  \citenamefont {Giordano}, \citenamefont {Obrist},\ and\ \citenamefont
  {Subramanian}}]{Frier2018}%
  \BibitemOpen
  \bibfield  {author} {\bibinfo {author} {\bibfnamefont {W.}~\bibnamefont
  {Frier}}, \bibinfo {author} {\bibfnamefont {D.}~\bibnamefont {Ablart}},
  \bibinfo {author} {\bibfnamefont {J.}~\bibnamefont {Chilles}}, \bibinfo
  {author} {\bibfnamefont {B.}~\bibnamefont {Long}}, \bibinfo {author}
  {\bibfnamefont {M.}~\bibnamefont {Giordano}}, \bibinfo {author}
  {\bibfnamefont {M.}~\bibnamefont {Obrist}}, \ and\ \bibinfo {author}
  {\bibfnamefont {S.}~\bibnamefont {Subramanian}},\ }\bibfield  {title}
  {\enquote {\bibinfo {title} {Using spatiotemporal modulation to draw tactile
  patterns in mid-air},}\ }in\ \href@noop {} {\emph {\bibinfo {booktitle}
  {Haptics: Science, Technology, and Applications}}},\ \bibinfo {editor}
  {edited by\ \bibinfo {editor} {\bibfnamefont {D.}~\bibnamefont
  {Prattichizzo}}, \bibinfo {editor} {\bibfnamefont {H.}~\bibnamefont
  {Shinoda}}, \bibinfo {editor} {\bibfnamefont {H.~Z.}\ \bibnamefont {Tan}},
  \bibinfo {editor} {\bibfnamefont {E.}~\bibnamefont {Ruffaldi}}, \ and\
  \bibinfo {editor} {\bibfnamefont {A.}~\bibnamefont {Frisoli}}}\ (\bibinfo
  {publisher} {Springer International Publishing},\ \bibinfo {address} {Cham},\
  \bibinfo {year} {2018})\ pp.\ \bibinfo {pages} {270--281}\BibitemShut
  {NoStop}%
\bibitem [{\citenamefont {Hajas}\ \emph {et~al.}(2020)\citenamefont {Hajas},
  \citenamefont {Pittera}, \citenamefont {Nasce}, \citenamefont {Georgiou},\
  and\ \citenamefont {Obrist}}]{Hajas2020}%
  \BibitemOpen
  \bibfield  {author} {\bibinfo {author} {\bibfnamefont {D.}~\bibnamefont
  {Hajas}}, \bibinfo {author} {\bibfnamefont {D.}~\bibnamefont {Pittera}},
  \bibinfo {author} {\bibfnamefont {A.}~\bibnamefont {Nasce}}, \bibinfo
  {author} {\bibfnamefont {O.}~\bibnamefont {Georgiou}}, \ and\ \bibinfo
  {author} {\bibfnamefont {M.}~\bibnamefont {Obrist}},\ }\bibfield  {title}
  {\enquote {\bibinfo {title} {Mid-air haptic rendering of 2d geometric shapes
  with a dynamic tactile pointer},}\ }\href {\doibase 10.1109/TOH.2020.2966445}
  {\bibfield  {journal} {\bibinfo  {journal} {IEEE Transactions on Haptics}\
  }\textbf {\bibinfo {volume} {13}},\ \bibinfo {pages} {806--817} (\bibinfo
  {year} {2020})}\BibitemShut {NoStop}%
\bibitem [{\citenamefont {Morales}\ \emph {et~al.}(2019)\citenamefont
  {Morales}, \citenamefont {Marzo}, \citenamefont {Subramanian},\ and\
  \citenamefont {Mart\'{\i}nez}}]{Morales2019}%
  \BibitemOpen
  \bibfield  {author} {\bibinfo {author} {\bibfnamefont {R.}~\bibnamefont
  {Morales}}, \bibinfo {author} {\bibfnamefont {A.}~\bibnamefont {Marzo}},
  \bibinfo {author} {\bibfnamefont {S.}~\bibnamefont {Subramanian}}, \ and\
  \bibinfo {author} {\bibfnamefont {D.}~\bibnamefont {Mart\'{\i}nez}},\
  }\bibfield  {title} {\enquote {\bibinfo {title} {Leviprops: Animating
  levitated optimized fabric structures using holographic acoustic tweezers},}\
  }in\ \href {\doibase 10.1145/3332165.3347882} {\emph {\bibinfo {booktitle}
  {Proceedings of the 32nd Annual ACM Symposium on User Interface Software and
  Technology}}},\ \bibinfo {series and number} {UIST '19}\ (\bibinfo
  {publisher} {Association for Computing Machinery},\ \bibinfo {address} {New
  York, NY, USA},\ \bibinfo {year} {2019})\ p.\ \bibinfo {pages}
  {651–661}\BibitemShut {NoStop}%
\bibitem [{\citenamefont {Marzo}\ \emph {et~al.}(2017)\citenamefont {Marzo},
  \citenamefont {Ghobrial}, \citenamefont {Cox}, \citenamefont {Caleap},
  \citenamefont {Croxford},\ and\ \citenamefont {Drinkwater}}]{Marzo2017}%
  \BibitemOpen
  \bibfield  {author} {\bibinfo {author} {\bibfnamefont {A.}~\bibnamefont
  {Marzo}}, \bibinfo {author} {\bibfnamefont {A.}~\bibnamefont {Ghobrial}},
  \bibinfo {author} {\bibfnamefont {L.}~\bibnamefont {Cox}}, \bibinfo {author}
  {\bibfnamefont {M.}~\bibnamefont {Caleap}}, \bibinfo {author} {\bibfnamefont
  {A.}~\bibnamefont {Croxford}}, \ and\ \bibinfo {author} {\bibfnamefont
  {B.~W.}\ \bibnamefont {Drinkwater}},\ }\bibfield  {title} {\enquote {\bibinfo
  {title} {Realization of compact tractor beams using acoustic delay-lines},}\
  }\href {\doibase 10.1063/1.4972407} {\bibfield  {journal} {\bibinfo
  {journal} {Applied Physics Letters}\ }\textbf {\bibinfo {volume} {110}},\
  \bibinfo {pages} {014102} (\bibinfo {year} {2017})},\ \Eprint
  {http://arxiv.org/abs/https://doi.org/10.1063/1.4972407}
  {https://doi.org/10.1063/1.4972407} \BibitemShut {NoStop}%
\bibitem [{\citenamefont {Inoue}\ \emph {et~al.}(2019)\citenamefont {Inoue},
  \citenamefont {Mogami}, \citenamefont {Ichiyama}, \citenamefont {Noda},
  \citenamefont {Makino},\ and\ \citenamefont {Shinoda}}]{Inoue2019}%
  \BibitemOpen
  \bibfield  {author} {\bibinfo {author} {\bibfnamefont {S.}~\bibnamefont
  {Inoue}}, \bibinfo {author} {\bibfnamefont {S.}~\bibnamefont {Mogami}},
  \bibinfo {author} {\bibfnamefont {T.}~\bibnamefont {Ichiyama}}, \bibinfo
  {author} {\bibfnamefont {A.}~\bibnamefont {Noda}}, \bibinfo {author}
  {\bibfnamefont {Y.}~\bibnamefont {Makino}}, \ and\ \bibinfo {author}
  {\bibfnamefont {H.}~\bibnamefont {Shinoda}},\ }\bibfield  {title} {\enquote
  {\bibinfo {title} {Acoustical boundary hologram for macroscopic rigid-body
  levitation},}\ }\href {\doibase 10.1121/1.5087130} {\bibfield  {journal}
  {\bibinfo  {journal} {The Journal of the Acoustical Society of America}\
  }\textbf {\bibinfo {volume} {145}},\ \bibinfo {pages} {328--337} (\bibinfo
  {year} {2019})},\ \Eprint
  {http://arxiv.org/abs/https://doi.org/10.1121/1.5087130}
  {https://doi.org/10.1121/1.5087130} \BibitemShut {NoStop}%
\bibitem [{\citenamefont {Hirayama}\ \emph {et~al.}(2019)\citenamefont
  {Hirayama}, \citenamefont {Martinez~Plasencia}, \citenamefont {Masuda},\ and\
  \citenamefont {Subramanian}}]{Hirayama2019}%
  \BibitemOpen
  \bibfield  {author} {\bibinfo {author} {\bibfnamefont {R.}~\bibnamefont
  {Hirayama}}, \bibinfo {author} {\bibfnamefont {D.}~\bibnamefont
  {Martinez~Plasencia}}, \bibinfo {author} {\bibfnamefont {N.}~\bibnamefont
  {Masuda}}, \ and\ \bibinfo {author} {\bibfnamefont {S.}~\bibnamefont
  {Subramanian}},\ }\bibfield  {title} {\enquote {\bibinfo {title} {A
  volumetric display for visual, tactile and audio presentation using acoustic
  trapping},}\ }\href {\doibase 10.1038/s41586-019-1739-5} {\bibfield
  {journal} {\bibinfo  {journal} {Nature}\ }\textbf {\bibinfo {volume} {575}},\
  \bibinfo {pages} {320--323} (\bibinfo {year} {2019})}\BibitemShut {NoStop}%
\bibitem [{\citenamefont {Hirayama}\ \emph {et~al.}(2022)\citenamefont
  {Hirayama}, \citenamefont {Christopoulos}, \citenamefont {Plasencia},\ and\
  \citenamefont {Subramanian}}]{Hirayama2022}%
  \BibitemOpen
  \bibfield  {author} {\bibinfo {author} {\bibfnamefont {R.}~\bibnamefont
  {Hirayama}}, \bibinfo {author} {\bibfnamefont {G.}~\bibnamefont
  {Christopoulos}}, \bibinfo {author} {\bibfnamefont {D.~M.}\ \bibnamefont
  {Plasencia}}, \ and\ \bibinfo {author} {\bibfnamefont {S.}~\bibnamefont
  {Subramanian}},\ }\bibfield  {title} {\enquote {\bibinfo {title} {High-speed
  acoustic holography with arbitrary scattering objects},}\ }\href {\doibase
  10.1126/sciadv.abn7614} {\bibfield  {journal} {\bibinfo  {journal} {Science
  Advances}\ }\textbf {\bibinfo {volume} {8}},\ \bibinfo {pages} {eabn7614}
  (\bibinfo {year} {2022})},\ \Eprint
  {http://arxiv.org/abs/https://www.science.org/doi/pdf/10.1126/sciadv.abn7614}
  {https://www.science.org/doi/pdf/10.1126/sciadv.abn7614} \BibitemShut
  {NoStop}%
\bibitem [{\citenamefont {Hasegawa}\ \emph {et~al.}(2017)\citenamefont
  {Hasegawa}, \citenamefont {Qiu}, \citenamefont {Noda}, \citenamefont
  {Inoue},\ and\ \citenamefont {Shinoda}}]{Hasegawa2017}%
  \BibitemOpen
  \bibfield  {author} {\bibinfo {author} {\bibfnamefont {K.}~\bibnamefont
  {Hasegawa}}, \bibinfo {author} {\bibfnamefont {L.}~\bibnamefont {Qiu}},
  \bibinfo {author} {\bibfnamefont {A.}~\bibnamefont {Noda}}, \bibinfo {author}
  {\bibfnamefont {S.}~\bibnamefont {Inoue}}, \ and\ \bibinfo {author}
  {\bibfnamefont {H.}~\bibnamefont {Shinoda}},\ }\bibfield  {title} {\enquote
  {\bibinfo {title} {Electronically steerable ultrasound-driven long narrow air
  stream},}\ }\href {\doibase 10.1063/1.4985159} {\bibfield  {journal}
  {\bibinfo  {journal} {Applied Physics Letters}\ }\textbf {\bibinfo {volume}
  {111}},\ \bibinfo {pages} {064104} (\bibinfo {year} {2017})},\ \Eprint
  {http://arxiv.org/abs/https://doi.org/10.1063/1.4985159}
  {https://doi.org/10.1063/1.4985159} \BibitemShut {NoStop}%
\bibitem [{\citenamefont {Hasegawa}, \citenamefont {Qiu},\ and\ \citenamefont
  {Shinoda}(2018)}]{Hasegawa2018}%
  \BibitemOpen
  \bibfield  {author} {\bibinfo {author} {\bibfnamefont {K.}~\bibnamefont
  {Hasegawa}}, \bibinfo {author} {\bibfnamefont {L.}~\bibnamefont {Qiu}}, \
  and\ \bibinfo {author} {\bibfnamefont {H.}~\bibnamefont {Shinoda}},\
  }\bibfield  {title} {\enquote {\bibinfo {title} {Midair ultrasound fragrance
  rendering},}\ }\href {\doibase 10.1109/TVCG.2018.2794118} {\bibfield
  {journal} {\bibinfo  {journal} {IEEE Transactions on Visualization and
  Computer Graphics}\ }\textbf {\bibinfo {volume} {24}},\ \bibinfo {pages}
  {1477--1485} (\bibinfo {year} {2018})}\BibitemShut {NoStop}%
\bibitem [{\citenamefont {Hasegawa}, \citenamefont {Yuki},\ and\ \citenamefont
  {Shinoda}(2019)}]{Hasegawa2019}%
  \BibitemOpen
  \bibfield  {author} {\bibinfo {author} {\bibfnamefont {K.}~\bibnamefont
  {Hasegawa}}, \bibinfo {author} {\bibfnamefont {H.}~\bibnamefont {Yuki}}, \
  and\ \bibinfo {author} {\bibfnamefont {H.}~\bibnamefont {Shinoda}},\
  }\bibfield  {title} {\enquote {\bibinfo {title} {Curved acceleration path of
  ultrasound-driven air flow},}\ }\href {\doibase 10.1063/1.5052423} {\bibfield
   {journal} {\bibinfo  {journal} {Journal of Applied Physics}\ }\textbf
  {\bibinfo {volume} {125}},\ \bibinfo {pages} {054902} (\bibinfo {year}
  {2019})},\ \Eprint {http://arxiv.org/abs/https://doi.org/10.1063/1.5052423}
  {https://doi.org/10.1063/1.5052423} \BibitemShut {NoStop}%
\bibitem [{\citenamefont {Prat-Camps}\ \emph {et~al.}(2020)\citenamefont
  {Prat-Camps}, \citenamefont {Christopoulos}, \citenamefont {Hardwick},\ and\
  \citenamefont {Subramanian}}]{Camps2020}%
  \BibitemOpen
  \bibfield  {author} {\bibinfo {author} {\bibfnamefont {J.}~\bibnamefont
  {Prat-Camps}}, \bibinfo {author} {\bibfnamefont {G.}~\bibnamefont
  {Christopoulos}}, \bibinfo {author} {\bibfnamefont {J.}~\bibnamefont
  {Hardwick}}, \ and\ \bibinfo {author} {\bibfnamefont {S.}~\bibnamefont
  {Subramanian}},\ }\bibfield  {title} {\enquote {\bibinfo {title} {A manually
  reconfigurable reflective spatial sound modulator for ultrasonic waves in
  air},}\ }\href {\doibase https://doi.org/10.1002/admt.202000041} {\bibfield
  {journal} {\bibinfo  {journal} {Advanced Materials Technologies}\ }\textbf
  {\bibinfo {volume} {5}},\ \bibinfo {pages} {2000041} (\bibinfo {year}
  {2020})},\ \Eprint
  {http://arxiv.org/abs/https://onlinelibrary.wiley.com/doi/pdf/10.1002/admt.202000041}
  {https://onlinelibrary.wiley.com/doi/pdf/10.1002/admt.202000041} \BibitemShut
  {NoStop}%
\bibitem [{\citenamefont {Melde}\ \emph {et~al.}(2016)\citenamefont {Melde},
  \citenamefont {Mark}, \citenamefont {Qiu},\ and\ \citenamefont
  {Fischer}}]{Melde2016}%
  \BibitemOpen
  \bibfield  {author} {\bibinfo {author} {\bibfnamefont {K.}~\bibnamefont
  {Melde}}, \bibinfo {author} {\bibfnamefont {A.~G.}\ \bibnamefont {Mark}},
  \bibinfo {author} {\bibfnamefont {T.}~\bibnamefont {Qiu}}, \ and\ \bibinfo
  {author} {\bibfnamefont {P.}~\bibnamefont {Fischer}},\ }\bibfield  {title}
  {\enquote {\bibinfo {title} {Holograms for acoustics},}\ }\href {\doibase
  10.1038/nature19755} {\bibfield  {journal} {\bibinfo  {journal} {Nature}\
  }\textbf {\bibinfo {volume} {537}},\ \bibinfo {pages} {518--522} (\bibinfo
  {year} {2016})}\BibitemShut {NoStop}%
\bibitem [{\citenamefont {Jim\'enez-Gamb\'{\i}n}, \citenamefont {Jim\'enez},\
  and\ \citenamefont {Camarena}(2020)}]{Gambin2020}%
  \BibitemOpen
  \bibfield  {author} {\bibinfo {author} {\bibfnamefont {S.}~\bibnamefont
  {Jim\'enez-Gamb\'{\i}n}}, \bibinfo {author} {\bibfnamefont {N.}~\bibnamefont
  {Jim\'enez}}, \ and\ \bibinfo {author} {\bibfnamefont {F.}~\bibnamefont
  {Camarena}},\ }\bibfield  {title} {\enquote {\bibinfo {title} {Transcranial
  focusing of ultrasonic vortices by acoustic holograms},}\ }\href {\doibase
  10.1103/PhysRevApplied.14.054070} {\bibfield  {journal} {\bibinfo  {journal}
  {Phys. Rev. Applied}\ }\textbf {\bibinfo {volume} {14}},\ \bibinfo {pages}
  {054070} (\bibinfo {year} {2020})}\BibitemShut {NoStop}%
\bibitem [{\citenamefont {Memoli}\ \emph {et~al.}(2017)\citenamefont {Memoli},
  \citenamefont {Caleap}, \citenamefont {Asakawa}, \citenamefont {Sahoo},
  \citenamefont {Drinkwater},\ and\ \citenamefont {Subramanian}}]{Memoli2017}%
  \BibitemOpen
  \bibfield  {author} {\bibinfo {author} {\bibfnamefont {G.}~\bibnamefont
  {Memoli}}, \bibinfo {author} {\bibfnamefont {M.}~\bibnamefont {Caleap}},
  \bibinfo {author} {\bibfnamefont {M.}~\bibnamefont {Asakawa}}, \bibinfo
  {author} {\bibfnamefont {D.~R.}\ \bibnamefont {Sahoo}}, \bibinfo {author}
  {\bibfnamefont {B.~W.}\ \bibnamefont {Drinkwater}}, \ and\ \bibinfo {author}
  {\bibfnamefont {S.}~\bibnamefont {Subramanian}},\ }\bibfield  {title}
  {\enquote {\bibinfo {title} {Metamaterial bricks and quantization of
  meta-surfaces},}\ }\href {\doibase 10.1038/ncomms14608} {\bibfield  {journal}
  {\bibinfo  {journal} {Nature Communications}\ }\textbf {\bibinfo {volume}
  {8}},\ \bibinfo {pages} {14608} (\bibinfo {year} {2017})}\BibitemShut
  {NoStop}%
\bibitem [{\citenamefont {Zhao}\ \emph {et~al.}(2018)\citenamefont {Zhao},
  \citenamefont {Chen}, \citenamefont {Wang},\ and\ \citenamefont
  {Zhang}}]{Zhao2018}%
  \BibitemOpen
  \bibfield  {author} {\bibinfo {author} {\bibfnamefont {S.-D.}\ \bibnamefont
  {Zhao}}, \bibinfo {author} {\bibfnamefont {A.-L.}\ \bibnamefont {Chen}},
  \bibinfo {author} {\bibfnamefont {Y.-S.}\ \bibnamefont {Wang}}, \ and\
  \bibinfo {author} {\bibfnamefont {C.}~\bibnamefont {Zhang}},\ }\bibfield
  {title} {\enquote {\bibinfo {title} {Continuously tunable acoustic
  metasurface for transmitted wavefront modulation},}\ }\href {\doibase
  10.1103/PhysRevApplied.10.054066} {\bibfield  {journal} {\bibinfo  {journal}
  {Phys. Rev. Applied}\ }\textbf {\bibinfo {volume} {10}},\ \bibinfo {pages}
  {054066} (\bibinfo {year} {2018})}\BibitemShut {NoStop}%
\bibitem [{\citenamefont {Molerón}, \citenamefont {Serra-Garcia},\ and\
  \citenamefont {Daraio}(2014)}]{Morelon2014}%
  \BibitemOpen
  \bibfield  {author} {\bibinfo {author} {\bibfnamefont {M.}~\bibnamefont
  {Molerón}}, \bibinfo {author} {\bibfnamefont {M.}~\bibnamefont
  {Serra-Garcia}}, \ and\ \bibinfo {author} {\bibfnamefont {C.}~\bibnamefont
  {Daraio}},\ }\bibfield  {title} {\enquote {\bibinfo {title} {Acoustic fresnel
  lenses with extraordinary transmission},}\ }\href {\doibase
  10.1063/1.4896276} {\bibfield  {journal} {\bibinfo  {journal} {Applied
  Physics Letters}\ }\textbf {\bibinfo {volume} {105}},\ \bibinfo {pages}
  {114109} (\bibinfo {year} {2014})},\ \Eprint
  {http://arxiv.org/abs/https://aip.scitation.org/doi/pdf/10.1063/1.4896276}
  {https://aip.scitation.org/doi/pdf/10.1063/1.4896276} \BibitemShut {NoStop}%
\bibitem [{\citenamefont {Li}\ \emph {et~al.}(2014)\citenamefont {Li},
  \citenamefont {Yu}, \citenamefont {Liang}, \citenamefont {Zou}, \citenamefont
  {Li}, \citenamefont {Cheng},\ and\ \citenamefont {Cheng}}]{Li2014}%
  \BibitemOpen
  \bibfield  {author} {\bibinfo {author} {\bibfnamefont {Y.}~\bibnamefont
  {Li}}, \bibinfo {author} {\bibfnamefont {G.}~\bibnamefont {Yu}}, \bibinfo
  {author} {\bibfnamefont {B.}~\bibnamefont {Liang}}, \bibinfo {author}
  {\bibfnamefont {X.}~\bibnamefont {Zou}}, \bibinfo {author} {\bibfnamefont
  {G.}~\bibnamefont {Li}}, \bibinfo {author} {\bibfnamefont {S.}~\bibnamefont
  {Cheng}}, \ and\ \bibinfo {author} {\bibfnamefont {J.}~\bibnamefont
  {Cheng}},\ }\bibfield  {title} {\enquote {\bibinfo {title} {Three-dimensional
  ultrathin planar lenses by acoustic metamaterials},}\ }\href {\doibase
  10.1038/srep06830} {\bibfield  {journal} {\bibinfo  {journal} {Scientific
  Reports}\ }\textbf {\bibinfo {volume} {4}},\ \bibinfo {pages} {6830}
  (\bibinfo {year} {2014})}\BibitemShut {NoStop}%
\bibitem [{\citenamefont {Polychronopoulos}\ and\ \citenamefont
  {Memoli}(2020)}]{Poly2020}%
  \BibitemOpen
  \bibfield  {author} {\bibinfo {author} {\bibfnamefont {S.}~\bibnamefont
  {Polychronopoulos}}\ and\ \bibinfo {author} {\bibfnamefont {G.}~\bibnamefont
  {Memoli}},\ }\bibfield  {title} {\enquote {\bibinfo {title} {Acoustic
  levitation with optimized reflective metamaterials},}\ }\href {\doibase
  10.1038/s41598-020-60978-4} {\bibfield  {journal} {\bibinfo  {journal}
  {Scientific Reports}\ }\textbf {\bibinfo {volume} {10}},\ \bibinfo {pages}
  {4254} (\bibinfo {year} {2020})}\BibitemShut {NoStop}%
\bibitem [{\citenamefont {Sallam}\ \emph {et~al.}(2021)\citenamefont {Sallam},
  \citenamefont {Meesala}, \citenamefont {Hajj},\ and\ \citenamefont
  {Shahab}}]{Sallam2021}%
  \BibitemOpen
  \bibfield  {author} {\bibinfo {author} {\bibfnamefont {A.}~\bibnamefont
  {Sallam}}, \bibinfo {author} {\bibfnamefont {V.~C.}\ \bibnamefont {Meesala}},
  \bibinfo {author} {\bibfnamefont {M.~R.}\ \bibnamefont {Hajj}}, \ and\
  \bibinfo {author} {\bibfnamefont {S.}~\bibnamefont {Shahab}},\ }\bibfield
  {title} {\enquote {\bibinfo {title} {Holographic mirrors for spatial
  ultrasound modulation in contactless acoustic energy transfer systems},}\
  }\href {\doibase 10.1063/5.0065489} {\bibfield  {journal} {\bibinfo
  {journal} {Applied Physics Letters}\ }\textbf {\bibinfo {volume} {119}},\
  \bibinfo {pages} {144101} (\bibinfo {year} {2021})},\ \Eprint
  {http://arxiv.org/abs/https://doi.org/10.1063/5.0065489}
  {https://doi.org/10.1063/5.0065489} \BibitemShut {NoStop}%
\bibitem [{\citenamefont {Zhao}\ \emph {et~al.}(2020)\citenamefont {Zhao},
  \citenamefont {Laredo}, \citenamefont {Ryan}, \citenamefont {Yazdkhasti},
  \citenamefont {Kim}, \citenamefont {Ganye}, \citenamefont {Horiuchi},\ and\
  \citenamefont {Yu}}]{zhao2020}%
  \BibitemOpen
  \bibfield  {author} {\bibinfo {author} {\bibfnamefont {L.}~\bibnamefont
  {Zhao}}, \bibinfo {author} {\bibfnamefont {E.}~\bibnamefont {Laredo}},
  \bibinfo {author} {\bibfnamefont {O.}~\bibnamefont {Ryan}}, \bibinfo {author}
  {\bibfnamefont {A.}~\bibnamefont {Yazdkhasti}}, \bibinfo {author}
  {\bibfnamefont {H.-T.}\ \bibnamefont {Kim}}, \bibinfo {author} {\bibfnamefont
  {R.}~\bibnamefont {Ganye}}, \bibinfo {author} {\bibfnamefont
  {T.}~\bibnamefont {Horiuchi}}, \ and\ \bibinfo {author} {\bibfnamefont
  {M.}~\bibnamefont {Yu}},\ }\bibfield  {title} {\enquote {\bibinfo {title}
  {Ultrasound beam steering with flattened acoustic metamaterial luneburg
  lens},}\ }\href {\doibase 10.1063/1.5140467} {\bibfield  {journal} {\bibinfo
  {journal} {Applied Physics Letters}\ }\textbf {\bibinfo {volume} {116}},\
  \bibinfo {pages} {071902} (\bibinfo {year} {2020})},\ \Eprint
  {http://arxiv.org/abs/https://doi.org/10.1063/1.5140467}
  {https://doi.org/10.1063/1.5140467} \BibitemShut {NoStop}%
\bibitem [{\citenamefont {Brunet}\ \emph {et~al.}(2015)\citenamefont {Brunet},
  \citenamefont {Merlin}, \citenamefont {Mascaro}, \citenamefont {Zimny},
  \citenamefont {Leng}, \citenamefont {Poncelet}, \citenamefont
  {Arist{\'e}gui},\ and\ \citenamefont {Mondain-Monval}}]{Brunet2015}%
  \BibitemOpen
  \bibfield  {author} {\bibinfo {author} {\bibfnamefont {T.}~\bibnamefont
  {Brunet}}, \bibinfo {author} {\bibfnamefont {A.}~\bibnamefont {Merlin}},
  \bibinfo {author} {\bibfnamefont {B.}~\bibnamefont {Mascaro}}, \bibinfo
  {author} {\bibfnamefont {K.}~\bibnamefont {Zimny}}, \bibinfo {author}
  {\bibfnamefont {J.}~\bibnamefont {Leng}}, \bibinfo {author} {\bibfnamefont
  {O.}~\bibnamefont {Poncelet}}, \bibinfo {author} {\bibfnamefont
  {C.}~\bibnamefont {Arist{\'e}gui}}, \ and\ \bibinfo {author} {\bibfnamefont
  {O.}~\bibnamefont {Mondain-Monval}},\ }\bibfield  {title} {\enquote {\bibinfo
  {title} {Soft 3d acoustic metamaterial with negative index},}\ }\href
  {\doibase 10.1038/nmat4164} {\bibfield  {journal} {\bibinfo  {journal}
  {Nature Materials}\ }\textbf {\bibinfo {volume} {14}},\ \bibinfo {pages}
  {384--388} (\bibinfo {year} {2015})}\BibitemShut {NoStop}%
\bibitem [{\citenamefont {Kr\"odel}\ and\ \citenamefont
  {Daraio}(2016)}]{Krodel2016}%
  \BibitemOpen
  \bibfield  {author} {\bibinfo {author} {\bibfnamefont {S.}~\bibnamefont
  {Kr\"odel}}\ and\ \bibinfo {author} {\bibfnamefont {C.}~\bibnamefont
  {Daraio}},\ }\bibfield  {title} {\enquote {\bibinfo {title} {Microlattice
  metamaterials for tailoring ultrasonic transmission with elastoacoustic
  hybridization},}\ }\href {\doibase 10.1103/PhysRevApplied.6.064005}
  {\bibfield  {journal} {\bibinfo  {journal} {Phys. Rev. Applied}\ }\textbf
  {\bibinfo {volume} {6}},\ \bibinfo {pages} {064005} (\bibinfo {year}
  {2016})}\BibitemShut {NoStop}%
\bibitem [{\citenamefont {Schindel}, \citenamefont {Bashford},\ and\
  \citenamefont {Hutchins}(1997)}]{Schindel1997}%
  \BibitemOpen
  \bibfield  {author} {\bibinfo {author} {\bibfnamefont {D.}~\bibnamefont
  {Schindel}}, \bibinfo {author} {\bibfnamefont {A.}~\bibnamefont {Bashford}},
  \ and\ \bibinfo {author} {\bibfnamefont {D.}~\bibnamefont {Hutchins}},\
  }\bibfield  {title} {\enquote {\bibinfo {title} {Focussing of ultrasonic
  waves in air using a micromachined fresnel zone-plate},}\ }\href {\doibase
  https://doi.org/10.1016/S0041-624X(97)00011-5} {\bibfield  {journal}
  {\bibinfo  {journal} {Ultrasonics}\ }\textbf {\bibinfo {volume} {35}},\
  \bibinfo {pages} {275--285} (\bibinfo {year} {1997})}\BibitemShut {NoStop}%
\bibitem [{\citenamefont {Shen}\ \emph {et~al.}(2019)\citenamefont {Shen},
  \citenamefont {Peng}, \citenamefont {Cai}, \citenamefont {Huang},
  \citenamefont {Zhao}, \citenamefont {Qiu}, \citenamefont {Zheng},\ and\
  \citenamefont {Zhu}}]{Shen2019}%
  \BibitemOpen
  \bibfield  {author} {\bibinfo {author} {\bibfnamefont {Y.-X.}\ \bibnamefont
  {Shen}}, \bibinfo {author} {\bibfnamefont {Y.-G.}\ \bibnamefont {Peng}},
  \bibinfo {author} {\bibfnamefont {F.}~\bibnamefont {Cai}}, \bibinfo {author}
  {\bibfnamefont {K.}~\bibnamefont {Huang}}, \bibinfo {author} {\bibfnamefont
  {D.-G.}\ \bibnamefont {Zhao}}, \bibinfo {author} {\bibfnamefont {C.-W.}\
  \bibnamefont {Qiu}}, \bibinfo {author} {\bibfnamefont {H.}~\bibnamefont
  {Zheng}}, \ and\ \bibinfo {author} {\bibfnamefont {X.-F.}\ \bibnamefont
  {Zhu}},\ }\bibfield  {title} {\enquote {\bibinfo {title} {Ultrasonic
  super-oscillation wave-packets with an acoustic meta-lens},}\ }\href
  {\doibase 10.1038/s41467-019-11430-3} {\bibfield  {journal} {\bibinfo
  {journal} {Nature Communications}\ }\textbf {\bibinfo {volume} {10}},\
  \bibinfo {pages} {3411} (\bibinfo {year} {2019})}\BibitemShut {NoStop}%
\bibitem [{\citenamefont {Welter}\ \emph {et~al.}(2011)\citenamefont {Welter},
  \citenamefont {Sathish}, \citenamefont {Christensen}, \citenamefont
  {Brodrick}, \citenamefont {Heebl},\ and\ \citenamefont
  {Cherry}}]{Welter2011}%
  \BibitemOpen
  \bibfield  {author} {\bibinfo {author} {\bibfnamefont {J.~T.}\ \bibnamefont
  {Welter}}, \bibinfo {author} {\bibfnamefont {S.}~\bibnamefont {Sathish}},
  \bibinfo {author} {\bibfnamefont {D.~E.}\ \bibnamefont {Christensen}},
  \bibinfo {author} {\bibfnamefont {P.~G.}\ \bibnamefont {Brodrick}}, \bibinfo
  {author} {\bibfnamefont {J.~D.}\ \bibnamefont {Heebl}}, \ and\ \bibinfo
  {author} {\bibfnamefont {M.~R.}\ \bibnamefont {Cherry}},\ }\bibfield  {title}
  {\enquote {\bibinfo {title} {Focusing of longitudinal ultrasonic waves in air
  with an aperiodic flat lens},}\ }\href {\doibase 10.1121/1.3640841}
  {\bibfield  {journal} {\bibinfo  {journal} {The Journal of the Acoustical
  Society of America}\ }\textbf {\bibinfo {volume} {130}},\ \bibinfo {pages}
  {2789--2796} (\bibinfo {year} {2011})},\ \Eprint
  {http://arxiv.org/abs/https://doi.org/10.1121/1.3640841}
  {https://doi.org/10.1121/1.3640841} \BibitemShut {NoStop}%
\bibitem [{\citenamefont {Brown}\ \emph {et~al.}(2016)\citenamefont {Brown},
  \citenamefont {Jaros}, \citenamefont {Cox},\ and\ \citenamefont
  {Treeby}}]{Brown2016}%
  \BibitemOpen
  \bibfield  {author} {\bibinfo {author} {\bibfnamefont {M.~D.}\ \bibnamefont
  {Brown}}, \bibinfo {author} {\bibfnamefont {J.}~\bibnamefont {Jaros}},
  \bibinfo {author} {\bibfnamefont {B.~T.}\ \bibnamefont {Cox}}, \ and\
  \bibinfo {author} {\bibfnamefont {B.~E.}\ \bibnamefont {Treeby}},\ }\bibfield
   {title} {\enquote {\bibinfo {title} {Control of broadband optically
  generated ultrasound pulses using binary amplitude holograms},}\ }\href
  {\doibase 10.1121/1.4944758} {\bibfield  {journal} {\bibinfo  {journal} {The
  Journal of the Acoustical Society of America}\ }\textbf {\bibinfo {volume}
  {139}},\ \bibinfo {pages} {1637--1647} (\bibinfo {year} {2016})},\ \Eprint
  {http://arxiv.org/abs/https://doi.org/10.1121/1.4944758}
  {https://doi.org/10.1121/1.4944758} \BibitemShut {NoStop}%
\bibitem [{\citenamefont {Korozlu}\ and\ \citenamefont
  {Cicek}(2018)}]{Korozlu2018}%
  \BibitemOpen
  \bibfield  {author} {\bibinfo {author} {\bibfnamefont {N.}~\bibnamefont
  {Korozlu}}\ and\ \bibinfo {author} {\bibfnamefont {A.}~\bibnamefont
  {Cicek}},\ }\bibfield  {title} {\enquote {\bibinfo {title} {Compact acoustic
  lens composed of annular cavities covered by a membrane},}\ }\href {\doibase
  10.1063/1.5043600} {\bibfield  {journal} {\bibinfo  {journal} {Applied
  Physics Letters}\ }\textbf {\bibinfo {volume} {113}},\ \bibinfo {pages}
  {183504} (\bibinfo {year} {2018})},\ \Eprint
  {http://arxiv.org/abs/https://doi.org/10.1063/1.5043600}
  {https://doi.org/10.1063/1.5043600} \BibitemShut {NoStop}%
\bibitem [{\citenamefont {Tarraz{\'o}-Serrano}\ \emph
  {et~al.}(2019)\citenamefont {Tarraz{\'o}-Serrano}, \citenamefont
  {P{\'e}rez-L{\'o}pez}, \citenamefont {Candelas}, \citenamefont {Uris},\ and\
  \citenamefont {Rubio}}]{Tarrazo-Serrano2019}%
  \BibitemOpen
  \bibfield  {author} {\bibinfo {author} {\bibfnamefont {D.}~\bibnamefont
  {Tarraz{\'o}-Serrano}}, \bibinfo {author} {\bibfnamefont {S.}~\bibnamefont
  {P{\'e}rez-L{\'o}pez}}, \bibinfo {author} {\bibfnamefont {P.}~\bibnamefont
  {Candelas}}, \bibinfo {author} {\bibfnamefont {A.}~\bibnamefont {Uris}}, \
  and\ \bibinfo {author} {\bibfnamefont {C.}~\bibnamefont {Rubio}},\ }\bibfield
   {title} {\enquote {\bibinfo {title} {Acoustic focusing enhancement in
  fresnel zone plate lenses},}\ }\href {\doibase 10.1038/s41598-019-43495-x}
  {\bibfield  {journal} {\bibinfo  {journal} {Scientific Reports}\ }\textbf
  {\bibinfo {volume} {9}},\ \bibinfo {pages} {7067} (\bibinfo {year}
  {2019})}\BibitemShut {NoStop}%
\bibitem [{\citenamefont {Hur}\ \emph {et~al.}(2022)\citenamefont {Hur},
  \citenamefont {Choi}, \citenamefont {Yoon}, \citenamefont {Kim},
  \citenamefont {Lee},\ and\ \citenamefont {Kim}}]{Hur2022}%
  \BibitemOpen
  \bibfield  {author} {\bibinfo {author} {\bibfnamefont {S.}~\bibnamefont
  {Hur}}, \bibinfo {author} {\bibfnamefont {H.}~\bibnamefont {Choi}}, \bibinfo
  {author} {\bibfnamefont {G.~H.}\ \bibnamefont {Yoon}}, \bibinfo {author}
  {\bibfnamefont {N.~W.}\ \bibnamefont {Kim}}, \bibinfo {author} {\bibfnamefont
  {D.-G.}\ \bibnamefont {Lee}}, \ and\ \bibinfo {author} {\bibfnamefont
  {Y.~T.}\ \bibnamefont {Kim}},\ }\bibfield  {title} {\enquote {\bibinfo
  {title} {Planar ultrasonic transducer based on a metasurface piezoelectric
  ring array for subwavelength acoustic focusing in water},}\ }\href {\doibase
  10.1038/s41598-022-05547-7} {\bibfield  {journal} {\bibinfo  {journal}
  {Scientific Reports}\ }\textbf {\bibinfo {volume} {12}},\ \bibinfo {pages}
  {1485} (\bibinfo {year} {2022})}\BibitemShut {NoStop}%
\bibitem [{\citenamefont {Allevato}\ \emph {et~al.}(2022)\citenamefont
  {Allevato}, \citenamefont {Rutsch}, \citenamefont {Hinrichs}, \citenamefont
  {Haugwitz}, \citenamefont {M\"{u}ller}, \citenamefont {Pesavento},\ and\
  \citenamefont {Kupnik}}]{Allevato2022}%
  \BibitemOpen
  \bibfield  {author} {\bibinfo {author} {\bibfnamefont {G.}~\bibnamefont
  {Allevato}}, \bibinfo {author} {\bibfnamefont {M.}~\bibnamefont {Rutsch}},
  \bibinfo {author} {\bibfnamefont {J.}~\bibnamefont {Hinrichs}}, \bibinfo
  {author} {\bibfnamefont {C.}~\bibnamefont {Haugwitz}}, \bibinfo {author}
  {\bibfnamefont {R.}~\bibnamefont {M\"{u}ller}}, \bibinfo {author}
  {\bibfnamefont {M.}~\bibnamefont {Pesavento}}, \ and\ \bibinfo {author}
  {\bibfnamefont {M.}~\bibnamefont {Kupnik}},\ }\bibfield  {title} {\enquote
  {\bibinfo {title} {Air-coupled ultrasonic spiral phased array for
  high-precision beamforming and imaging},}\ }\href {\doibase
  10.1109/OJUFFC.2022.3142710} {\bibfield  {journal} {\bibinfo  {journal} {IEEE
  Open Journal of Ultrasonics, Ferroelectrics, and Frequency Control}\ }\textbf
  {\bibinfo {volume} {2}},\ \bibinfo {pages} {40--54} (\bibinfo {year}
  {2022})}\BibitemShut {NoStop}%
\bibitem [{\citenamefont {Tseng}, \citenamefont {Bedair},\ and\ \citenamefont
  {Lazarus}(2018)}]{Tseng2017}%
  \BibitemOpen
  \bibfield  {author} {\bibinfo {author} {\bibfnamefont {V.~F.-G.}\
  \bibnamefont {Tseng}}, \bibinfo {author} {\bibfnamefont {S.~S.}\ \bibnamefont
  {Bedair}}, \ and\ \bibinfo {author} {\bibfnamefont {N.}~\bibnamefont
  {Lazarus}},\ }\bibfield  {title} {\enquote {\bibinfo {title} {Phased array
  focusing for acoustic wireless power transfer},}\ }\href {\doibase
  10.1109/TUFFC.2017.2771283} {\bibfield  {journal} {\bibinfo  {journal} {IEEE
  Transactions on Ultrasonics, Ferroelectrics, and Frequency Control}\ }\textbf
  {\bibinfo {volume} {65}},\ \bibinfo {pages} {39--49} (\bibinfo {year}
  {2018})}\BibitemShut {NoStop}%
\bibitem [{\citenamefont {Marzo}, \citenamefont {Corkett},\ and\ \citenamefont
  {Drinkwater}(2018)}]{Marzo2017-2}%
  \BibitemOpen
  \bibfield  {author} {\bibinfo {author} {\bibfnamefont {A.}~\bibnamefont
  {Marzo}}, \bibinfo {author} {\bibfnamefont {T.}~\bibnamefont {Corkett}}, \
  and\ \bibinfo {author} {\bibfnamefont {B.~W.}\ \bibnamefont {Drinkwater}},\
  }\bibfield  {title} {\enquote {\bibinfo {title} {Ultraino: An open
  phased-array system for narrowband airborne ultrasound transmission},}\
  }\href {\doibase 10.1109/TUFFC.2017.2769399} {\bibfield  {journal} {\bibinfo
  {journal} {IEEE Transactions on Ultrasonics, Ferroelectrics, and Frequency
  Control}\ }\textbf {\bibinfo {volume} {65}},\ \bibinfo {pages} {102--111}
  (\bibinfo {year} {2018})}\BibitemShut {NoStop}%
\bibitem [{\citenamefont {Rutsch}\ \emph {et~al.}(2015)\citenamefont {Rutsch},
  \citenamefont {Konetzke}, \citenamefont {Unger}, \citenamefont {Hoffmann},
  \citenamefont {Ramadas}, \citenamefont {Dixon},\ and\ \citenamefont
  {Kupnik}}]{Rutsch2015}%
  \BibitemOpen
  \bibfield  {author} {\bibinfo {author} {\bibfnamefont {M.}~\bibnamefont
  {Rutsch}}, \bibinfo {author} {\bibfnamefont {E.}~\bibnamefont {Konetzke}},
  \bibinfo {author} {\bibfnamefont {A.}~\bibnamefont {Unger}}, \bibinfo
  {author} {\bibfnamefont {M.}~\bibnamefont {Hoffmann}}, \bibinfo {author}
  {\bibfnamefont {S.~N.}\ \bibnamefont {Ramadas}}, \bibinfo {author}
  {\bibfnamefont {S.}~\bibnamefont {Dixon}}, \ and\ \bibinfo {author}
  {\bibfnamefont {M.}~\bibnamefont {Kupnik}},\ }\bibfield  {title} {\enquote
  {\bibinfo {title} {Extending the receive performance of phased ultrasonic
  transducer arrays in air down to 40 khz and below},}\ }in\ \href {\doibase
  10.1109/ULTSYM.2015.0095} {\emph {\bibinfo {booktitle} {2015 IEEE
  International Ultrasonics Symposium (IUS)}}}\ (\bibinfo {year} {2015})\ pp.\
  \bibinfo {pages} {1--4}\BibitemShut {NoStop}%
\bibitem [{\citenamefont {Suzuki}\ \emph {et~al.}(2021)\citenamefont {Suzuki},
  \citenamefont {Inoue}, \citenamefont {Fujiwara}, \citenamefont {Makino},\
  and\ \citenamefont {Shinoda}}]{autd3}%
  \BibitemOpen
  \bibfield  {author} {\bibinfo {author} {\bibfnamefont {S.}~\bibnamefont
  {Suzuki}}, \bibinfo {author} {\bibfnamefont {S.}~\bibnamefont {Inoue}},
  \bibinfo {author} {\bibfnamefont {M.}~\bibnamefont {Fujiwara}}, \bibinfo
  {author} {\bibfnamefont {Y.}~\bibnamefont {Makino}}, \ and\ \bibinfo {author}
  {\bibfnamefont {H.}~\bibnamefont {Shinoda}},\ }\bibfield  {title} {\enquote
  {\bibinfo {title} {Autd3: Scalable airborne ultrasound tactile display},}\
  }\href {\doibase 10.1109/TOH.2021.3069976} {\bibfield  {journal} {\bibinfo
  {journal} {IEEE Transactions on Haptics}\ }\textbf {\bibinfo {volume} {14}},\
  \bibinfo {pages} {740--749} (\bibinfo {year} {2021})}\BibitemShut {NoStop}%
\bibitem [{\citenamefont {Mortezaie}\ and\ \citenamefont
  {Wade}(1984)}]{Mortezaie1984}%
  \BibitemOpen
  \bibfield  {author} {\bibinfo {author} {\bibfnamefont {M.}~\bibnamefont
  {Mortezaie}}\ and\ \bibinfo {author} {\bibfnamefont {G.}~\bibnamefont
  {Wade}},\ }\enquote {\bibinfo {title} {Fresnel zone plate and fresnel phase
  plate patterns for acoustic transducers},}\ in\ \href {\doibase
  10.1007/978-1-4613-2779-0_27} {\emph {\bibinfo {booktitle} {Acoustical
  Imaging}}},\ \bibinfo {editor} {edited by\ \bibinfo {editor} {\bibfnamefont
  {M.}~\bibnamefont {Kaveh}}, \bibinfo {editor} {\bibfnamefont {R.~K.}\
  \bibnamefont {Mueller}}, \ and\ \bibinfo {editor} {\bibfnamefont {J.~F.}\
  \bibnamefont {Greenleaf}}}\ (\bibinfo  {publisher} {Springer US},\ \bibinfo
  {address} {Boston, MA},\ \bibinfo {year} {1984})\ pp.\ \bibinfo {pages}
  {345--355}\BibitemShut {NoStop}%
\end{thebibliography}%

\end{document}